\documentclass[aps,pra,eqsecnum,superscriptaddress]{revtex4-2}

\usepackage{amssymb,amsmath}
\usepackage{slashed}
\usepackage{accents}
\usepackage{subdepth}
\usepackage{graphicx}
\usepackage{sansmath}
\usepackage{bm}
\usepackage{soul}
\usepackage{musicography}
\usepackage{yfonts}
\usepackage{upgreek}

\usepackage[breaklinks=true]{hyperref}
\hypersetup{
	colorlinks   = true, 
	urlcolor     = blue, 
	linkcolor    = red, 
	citecolor   = green 
}

\DeclareMathOperator{\tr}{tr}

\newcommand{\sD}{\mathcal{D}}

\newcommand{\sH}{\mathcal{H}}

\newcommand{\sN}{\mathcal{N}}

\newcommand{\ket}[1]{\vert#1\rangle}
\newcommand{\bra}[1]{\langle#1\vert}

\newcommand{\proj}[1]{| #1\rangle\!\langle #1 |}

\newcommand{\iprod}[2]{\langle#1\vert#2\rangle}
\newcommand{\oprod}[2]{\vert#1\rangle\langle#2\vert}

\newcommand{\smallfrac}[2]{{\textstyle{\frac#1#2}}}

\newcommand{\plus}{{\mathord{+}}}
\newcommand{\minus}{{\mathord{-}}}
\renewcommand{\Im}{\mathrm{Im}}
\renewcommand{\Re}{\mathrm{Re}}

\newcommand{\inv}{{\,\text{-}\hspace{-1pt}1}}
\newcommand{\ad}{\mathrm{ad}}

\newcommand{\Z}{\mathcal{Z}}

\renewcommand{\L}{\mathcal{L}}

\newcommand{\D}{\mathcal{D}}

\newcommand{\E}{\mathcal{E}}

\newcommand{\Rinv}[1]{\underleftarrow{#1}}
\newcommand{\Odot}{{\mathcal{O}\hspace{-0.575em}\raisebox{0.5pt}{$\boldsymbol{\cdot}$}\hspace{.275em}}}

\newcommand{\R}{\mathbb R}
\newcommand{\C}{\mathbb C}

\newcommand{\Hb}{\mathcal H}

\newcommand{\GL}{\mathrm{GL}}

\newcommand{\csch}{\mathrm{csch}\,}

\newcommand{\Ho}{H_{\textrm{o}}}

\newcommand{\lowerintsub}[1]{{\raisebox{-1.5pt}{$\scriptstyle#1$}}}

\usepackage{color}

\begin{document}
	
	\title{Simultaneous Momentum and Position Measurement\\and the Instrumental Weyl-Heisenberg Group}
	
	\author{Christopher S.~Jackson}
	\email{omgphysics@gmail.com}
    \noaffiliation
	
	\author{Carlton M.~Caves}
	\email{ccaves@unm.edu}
	\affiliation{Center for Quantum Information and Control,\\University of New Mexico, Albuquerque, New Mexico 87131-0001}
	
	\date{\today}
	
\begin{abstract}
	The canonical commutation relation, $[Q,P]=i\hbar$, stands at the foundation of quantum theory and the original Hilbert space.
	The interpretation of $P$ and $Q$ as observables has always relied on the analogies that exist between the unitary transformations of Hilbert space and the canonical (a.k.a.~contact) transformations of classical phase space.
	Now that the theory of quantum measurement is essentially complete (this took a while), it is possible to revisit the canonical commutation relation in a way that sets the foundation of quantum theory not on unitary transformations, but on positive transformations.
	This paper shows how the concept of simultaneous measurement leads to a fundamental differential geometric problem whose solution shows us the following:
	The simultaneous $P \& Q$ measurement (SPQM) defines a universal measuring instrument, which takes the shape of a 7-dimensional manifold, a universal covering group we call the Instrumental Weyl-Heisenberg Group, $\mathrm{IWH}$.
	The group $\mathrm{IWH}$ connects the identity to classical phase space in unexpected ways that are significant enough that the positive-operator-valued measure (POVM) offers a complete alternative to energy quantization.
	Five of the dimensions define processes that can be easily recognized and understood.
	The other two dimensions, the normalization and phase in the center of $\mathrm{IWH}$, are less familiar.
	The normalization, in particular, requires special handling in order to describe and understand the SPQM instrument.
\end{abstract}
	
	\maketitle

\vfill\pagebreak

\tableofcontents

\vfill\pagebreak

\section{Introduction}

After World War II, theoretical quantum physics became dominated by the design of quantum field theory.
There were three branches of physics that stemmed from this: High-Energy, Condensed-Matter, and Atomic-Molecular-Optical (AMO.)
Although incredibly developed as predictive methods, quantum field theory in all three of these branches has left some very basic ideas of quantum observation underdeveloped.
That this is indeed still the case is evident in how the coherent-state resolutions of the identity or ``overcomplete bases''~\cite{Klauder1960a,Perelomov1986a,WMZhang1990a,Brif1999a} are usually finessed:
\begin{equation}\label{amp}
	1^{\text{(amp)}}_{Z} = Z\int_{\mathbb{C}}\frac{d^2\alpha}{\pi} \proj{\sqrt{Z}\alpha}
\end{equation}
for bosonic amplitudes (where $Z$ is an arbitrary complex scalar)~\cite{Klauder1960a,Glauber1963a,Sudarshan1963a,Glauber1963b,Glauber1965a}, and
\begin{equation}\label{spin}
	1^{\text{(spin)}}_{j}= (2j+1)\int_{S^2}\frac{d\mu(\hat n)}{4\pi} \proj{j,\hat{n}}
\end{equation}
for fermionic spins (where $2j$ is an integer)~\cite{Perelomov1986a,WMZhang1990a,Brif1999a,Massar1995a}.
These overcomplete bases are both key to establishing functional integration~\cite{Wiener1921b,Wiener1921a,Wiener1924a,Feynman1942a,Feynman1948a,Kac1947a,Kac1959a,Klauder1960a,Feynman2010a,Bourbaki2004a,Chaichian2001a} and to understanding the observation of energy quanta~\cite{TaMatsubara1955a}.
Yet these overcomplete bases do not fit the most basic idea of a Hermitian eigenbasis and as such are often not considered a serious form of observation.
By itself, this inattention to how one could observe in these overcomplete bases leaves a missing piece at the foundation of quantum mechanics and quantum field theory.

Meanwhile, an understanding of the overcomplete bases as a bona fide means of observation has been slowly coming to light.
First, the idea of observables matured into the general mathematical theories of instruments and operations~\cite{Ludwig1983a,Ludwig1985a,Schwinger1959a,Wigner1963a,Jauch1967a,Davies1970a,Davies1976a,Lindblad1976a,Kraus1983a,Peres1993a,Nielsen2000a,Wiseman2009a}.
In this more contemporary language, the overcomplete bases of equations~\ref{amp} and~\ref{spin} are called \emph{positive-operator-valued measures} or POVMs and are understood to be \emph{informationally complete}, meaning the distribution observed by any one of these POVMs is enough to reconstruct the quantum state.
With this theoretical technology, it has further been discovered that the overcomplete bases correspond to various forms of continual (or continuous) simultaneous observation.
The standard coherent-state POVM (equation~\ref{amp}) was discovered to be the effect of simultaneously observing both quadratures of a leaky cavity, a form of observation we will call \emph{Goetsch-Graham-Wiseman (GGW) heterodyne detection\/}~\cite{Barchielli1991a,Goetsch1994a,Wiseman1994b,Wiseman1996a,CSJackson2022a}.
Then the spin-coherent POVM of equation~\ref{spin} was discovered to be the effect of simultaneously observing the three orthogonal spin components, a form of observation we call the \emph{spin-isotropic measurement\/}~\cite{Shojaee2018a,CSJackson2021a,CSJackson2023a}.

Before proceeding, we caution that this paper uses a mathematical apparatus not familiar to most physicists and quantum scientists.  This apparatus is introduced here naturally, as it becomes both desirable and necessary.  Readers who are made uncomfortable by this apparatus are urged to consult the companion paper~\cite{CSJackson2023a}, which attempts to persuade the reader that the unfamiliar mathematical concepts and techniques are essential tools---a new way of thinking and doing---and then introduces these tools as gently as possible.

GGW heterodyne detection and the spin-isotropic measurement work in a very similar way, but they are different in one very important respect.
While GGW heterodyne detection assumes energy-conserving system-meter interactions, the spin-isotropic measurement assumes Hermitian meter displacements,
	$-iH^\text{(iso)} dt/\hbar = -\sqrt{\kappa\,dt}\,J_k \otimes 2\sigma \partial_q$,
where $J_k$ is an orthogonal spin component of the system, $q$ is the meter register, $\sigma$ is the width of the meter pointer, and $\kappa$ is the measurement rate.
In both cases, the measuring instrument consists of Kraus operators defined by a time-ordered exponential over the duration~$T$ of the measurement.
For GGW heterodyne detection, the Kraus operators are~\cite{CSJackson2022a}
\begin{equation}\label{introGGW}
	L^{(\text{GGW})}[dw_{[0,T)}]
	= \mathcal{T}\exp\!\left(\int_0^{T-dt} -2a^\dag a \,\kappa\,dt + 2a \sqrt{\kappa}\,dw_t^*\right)\,,
\end{equation}
where $a=(Q+iP)/\sqrt{2\hbar}$ is the usual complex-amplitude operator and $dw_t=(dW_t^q+idW_t^p)/\sqrt2$ is the registered complex Wiener path.
For spin-isotropic measurement, the Kraus operators are~\cite{CSJackson2021a,CSJackson2023a}
\begin{equation}\label{introIso}
L^{(\text{iso})}[d\vec{W}_{[0,T)}]
= \mathcal{T}\exp\!\left(\int_0^{T-dt} -\vec{J}^{\,2} \,\kappa\,dt + \vec{J}\cdot \!\sqrt{\kappa}\,d\vec{W}_t\right)\,,
\end{equation}
where $\vec{J}=(J_x,J_y,J_z)$ is the triple of orthogonal spin-component observables and $d\vec{W}_t=(dW^x_t,dW^y_t,dW^z_t)$ is the registered 3-vector of Wiener paths.
The most striking feature about the instruments defined by equations \ref{introGGW} and \ref{introIso} is that they can be integrated \emph{universally}|that is, independent of matrix representation.  The difference between the two cases can now be summarized as the following: integrating equation~\ref{introIso} defines a 7-dimensional manifold that requires the theory of symmetric spaces, whereas integrating equation~\ref{introGGW} defines a 3-dimensional manifold that is much more straightforward.

This paper is an analysis of the quadrature analog of the spin-isotropic measurement, a form of observation we call the \emph{Simultaneous $P$\&$Q$ Measurement} or \emph{SPQM}.
The name SPQM is our homage to Alberto Barchielli~\cite{BarchielliSPQMtribute,Beard2015a}, who appears to be the first to have considered and analyzed this problem~\cite{Barchielli1982a}.
SPQM generates a measuring instrument with Kraus operators~\cite{CSJackson2023a}
\begin{equation}\label{introSPQM}
	\boxed{
		\vphantom{\Bigg(}
		\hspace{15pt}
		L^{(\text{SPQM})}\left[dw_{[0,T)}\right]
		= \mathcal{T}\exp\!\left(\int_0^{T-dt}\hspace{-5pt}-2\Ho \,\kappa dt + P \sqrt{\kappa}\,dW^p_t + Q \sqrt{\kappa}\,dW^q_t\right)\,,
		\hspace{15pt}
		\vphantom{\Bigg)}
	}
\end{equation}
where $2\Ho \equiv P^2+Q^2$ and $P$ and $Q$ are (dimensionless) canonical momentum and position (or the conserved quadrature components of a harmonic oscillator).
This time-ordered exponential defines another fundamental 7-dimensional manifold, the universal covering group, which for SPQM we call the \emph{Instrumental Weyl-Heisenberg Group}, $G=\mathrm{IWH}$.
The universal covering group is defined by a map $\gamma$ with the universal property that for any Hilbert space $\Hb$ carrying the paths of Kraus operators $L^{(\text{SPQM})}:\C^{T/dt}\longrightarrow\GL(\Hb)$, there exists a unique representation $R$ such that
\begin{equation}
	L^{(\text{SPQM})} = R\circ \gamma
	\hspace{20pt}
	\text{where}
	\hspace{20pt}
	\C^{T/dt}
	\xrightarrow{\;\gamma\;}
	\mathrm{IWH}
	\xrightarrow{\;R\;}
	\GL(\Hb)\,.
\end{equation}
This universal way of considering $G=\mathrm{IWH}$ essentially amounts to suspending the choice of $\Hb$ and therefore $\hbar$, but it is very important to appreciate that the measuring instrument is, in fact, fundamentally independent of the Hilbert space and therefore $\hbar$.
The same is true for the spin-isotropic measurement, except that different irreducible representations don't amount to choices of $\hbar$, but rather to choices of the total angular momentum number $j$.

The universal covering group $\mathrm{IWH}$ can be navigated in much the same way as can be done for semisimple Lie groups, with the use of right-invariant vector fields and decompositions similar to those of Cartan and Harish-Chandra.
In particular, the sample-paths defined by $x(t) = \gamma[dw_{[0,t)}]$ diffuse according to a Fokker-Planck-Kolmogorov equation,
\begin{equation}\label{diffuse}
	\boxed{
		\vphantom{\Bigg(}
		\hspace{15pt}
		\frac{1}{\kappa}\frac{\partial}{\partial t} D_{t}(x)=\left(2\Rinv{\Ho}+\frac12\Rinv{P}\Rinv{P}+\frac12\Rinv{Q}\Rinv{Q}\right)\![D_t](x)\,,
		\hspace{15pt}
		\vphantom{\Bigg)}
	}
\end{equation}
where
\begin{equation}\label{KODFintro}
\boxed{
	\vphantom{\Bigg(}
	\hspace{15pt}
	D_T(x)\equiv\int \D\mu\!\left[dw_{[0,T)}\right]\delta\Big(x,\gamma\!\left[dw_{[0,T)}\right]\Big)
	\hspace{10pt}
	\vphantom{\Bigg)}
}
\end{equation}
is the \emph{Kraus-operator distribution function} of the SPQM instrument with respect to the Haar measure~\cite{Haar1933a,vonNeumann1999a,Nachbin1965a,Bourbaki2004a} of $\mathrm{IWH}$ and $\Rinv{\Ho}$, $\Rinv{Q}$, and $\Rinv{P}$ are right-invariant derivatives tangent to $\mathrm{IWH}$.  We regard ``Kraus-operator distribution function,'' ``Kraus-operator distribution,'' and ``Kraus-operator density'' as interchangeable, despite subtle differences some might attribute to these usages.  We abbreviate Kraus-operator distribution function as~\hbox{KOD} to invite the reader to use any of these terms.  The KOD can be considered the \emph{universal unraveling} of the total (or unconditional) operation (a completely positive, trace-preserving superoperator),
\begin{align}\label{overallZT}
\boxed{
	\vphantom{\Bigg(}
	\hspace{15pt}
	\Z^{(\text{SPQM})}_T
	\equiv \int \D\mu\!\left[dw_{[0,T)}\right]\Odot\Big(L^{(\text{SPQM})}\!\left[dw_{[0,T)}\right]\!\Big)
	=\int_{\lowerintsub{\mathrm{IWH}}} \hspace{-12pt} d^7\!\mu(x) \,D_T(x) \,\Odot\big(R(x)\big)\,,
	\hspace{10pt}
	\vphantom{\Bigg)}
}
\end{align}
where $\D\mu\!\left[dw_{[0,T)}\right]$ is the Wiener path measure, $d^7\!\mu(x)$ is the Haar measure, and $\Odot(L)\equiv L\!\odot L^\dagger$.  The technology of right-invariant differentiation~\cite{Knapp1988a,Frankel2012a,Kitaev2018a} will not be familiar to most quantum physicists and information scientists.  Introducing SPQM, $\mathrm{IWH}$, the concept of right-invariant motion, and the KOD is the subject of section~\ref{Moving}.

Section~\ref{Coord} translates the co\"ordinate-independent formulation of section~\ref{Moving} to forms that physicists are more likely to recognize.  Indeed,
what the aforementioned decompositions do is to co\"ordinate the points of $\mathrm{IWH}$~\cite{Gilmore2002a,Helgason1978a,Barut1980a,Perelomov1986a,WMZhang1990a,Knapp2002a}.
The decomposition of $\mathrm{IWH}$ similar to the Harish-Chandra~\cite{Harish-Chandra1956a,Knapp1986a,Knapp2002a} (a.k.a.\ ``Gauss''~\cite{Zelobenko1973a,Barut1980a,Perelomov1986a}) decomposition is
\begin{equation}\label{xHC}
	x= e^{a^\dag \nu} e^{-\Ho r +\Omega z}e^{a \mu^*}\,,
\end{equation}
with purity co\"ordinate $r \in \R$, which we will call the ruler, central co\"ordinate $z=-s+i\psi \in \C$, phase-space co\"ordinates $\nu, \mu \in \C$, and $\Omega=\hbar1_\Hb$.
The decomposition of $\mathrm{IWH}$ similar to the Cartan decomposition~\cite{Helgason1978a,Barut1980a,Knapp2002a,Knapp1986a} is
\begin{equation}\label{xCartan}
	x= D_\beta e^{i\Omega\phi}e^{-\Ho r -\Omega\ell}D_\alpha^\inv\,,
\end{equation}
with different central co\"ordinates $\phi, \ell \in \R$, the same ruler~$r$, and phase-space co\"ordinates $\beta,\,\alpha \in \C$ appearing in the conventional displacement operator $D_\alpha\equiv e^{a^\dagger\alpha-a\alpha^*}$.
Introducing these decompositions and using them to transform equation~\ref{introSPQM} into standard It\^o-form stochastic differential equations (SDEs)~\cite{Ito1950a,Ito1996a,Chirikjian2009a,Gardiner2009a,Gardiner2021a} and to transform equation~\ref{diffuse} into a co\"ordinate Fokker-Planck-Kolmogorov (FPK) diffusion equation~\cite{Gardiner2009a,Gardiner2021a} is the subject of section~\ref{Coord}, which also solves those SDEs and (mostly) solves the FPK diffusion equation.
As a function of the registers of SPQM, we find that the ruler satisfies
\begin{equation}
		r_T=2\kappa T.
\end{equation}
For the remaining Harish-Chandra co\"ordinates, we find the phase points to follow Ornstein-Uhlenbeck~\cite{Uhlenbeck1930a,Gardiner2009a,Gardiner2021a} and GGW processes,
\begin{equation}
\nu[dw_{[0,T)}] =\int_0^{T_-}\!\!\sqrt\kappa\,dw_t\,e^{-2\kappa (T-t)}
	\hspace{15pt}
	\text{and}
	\hspace{15pt}
\mu[dw_{[0,T)}] =\int_0^{T_-}\!\!\sqrt\kappa\,dw_t\,e^{-2\kappa t}\,,
\end{equation}
where $T_-\equiv T-dt$, and the center to follow a quadratic functional process,
\begin{equation}
z[dw_{[0,T)}] =\frac12\int_0^{T_-}\!\!\int_0^{T_-}\!\!\kappa\,dw_t^*\,dw_s\,e^{-2\kappa |t-s|}\big(1+\textrm{sgn}(t-s)\big)\,,
\end{equation}
where used is the sign function, $\textrm{sgn}(u)=u/|u|$ for $u\ne0$ and $\textrm{sgn}(0)=0$.  The Cartan phase-space co\"ordinates follow linear functionals,
\begin{equation}\label{CartanbetaTalphaT}
\beta[dw_{[0,T)}]  = \int_0^{T_-}\sqrt\kappa\,dw_t\,\frac{\cosh2\kappa t}{\sinh 2\kappa T}
\hspace{15pt}
\text{and}
\hspace{15pt}
\alpha[dw_{[0,T)}] = \int_0^{T_-}\sqrt\kappa\,dw_t\,\frac{\cosh2\kappa (T-t)}{\sinh 2\kappa T}
\,.
\end{equation}
The Cartan central co\"ordinates $\ell$ and $\phi$ follow from the co\"ordinate transformation between Harish-Chandra and Cartan co\"ordinates (equations
\ref{eq:postrans}, \ref{eq:unitrans}, \ref{eq:posgauge}, and \ref{eq:unigauge}), but we do not write those solutions explicitly here.

As for the KOD~$D_t(x)$, we will not be able to solve analytically for the distribution over all seven dimensions.
Summing over the center, however,
\begin{equation}
	Z \equiv \left\{e^{1 z}\,:\,z\in\C\right\}\vartriangleleft\mathrm{IWH}\,,
\end{equation}
gives a \emph{reduced SPQM unraveling\/} of the total operation,
\begin{align}
	\Z^{(\text{SPQM})}_T =\int_{\lowerintsub{\mathrm{IWH}/Z}}\hspace{-1.5em}d^5\!\mu(Zx) \,C_T(Zx) \,\Odot\!\left(D_\beta\,e^{-\Ho r}D_\alpha^\dag\right)\,,
\end{align}
where the integral is over the quotient group $\textrm{IWH}/Z$, which we call the \emph{Reduced Instrumental Weyl-Heisenberg Group\/} (RIWH).  The reduced Kraus-operator distribution (RKOD),
\begin{equation}\label{CTZxintro}
\boxed{
	\vphantom{\Bigg(}
	\hspace{15pt}
    C_T (Zx)
    \equiv\int_Z\!d\phi\,d\ell\, D_T(x)e^{-2\ell}
    = \int \D\mu\!\left[dw_{[0,T)}\right]e^{-2\ell[dw_{[0,T)}]}\,\delta\Big(Zx,Z\gamma\!\left[dw_{[0,T)}\right]\Big)\,,
	\hspace{10pt}
	\vphantom{\Bigg)}
}
\end{equation}
is a marginal over the center that includes the Cartan center factor $e^{-2\ell}$.  We call $C_T(Zx)$ the \emph{Cartan-section reduced distribution function}, and we are able to solve for it from its FPK diffusion equation.  The solution is a Gaussian with ill-defined normalization,
\begin{equation}
	d^5\!\mu(Zx)\,C_T(Zx) =2\sinh r\,dr\,\delta(r-2\kappa T)\frac{d^2\beta}{\pi} \frac{1}{\Sigma_T}e^{-|\beta-\alpha|^2/\Sigma_T}\frac{d^2\alpha}{\pi}\,,
\end{equation}
where the mean-square distance between the two phase points is given by
\begin{equation}
	\Sigma_T = \kappa T - \tanh\kappa T\,.
\end{equation}
The normalization factor $2\sinh2\kappa T$ is particularly interesting, as the POVM~completeness relation for the SPQM instrument boils down to (assuming $\hbar=1$)
\begin{equation}
	2\sinh 2\kappa T \int \!\frac{d^2\alpha}{\pi}\, D_\alpha\,e^{-\Ho 4\kappa T}D_\alpha^\dag =1_\Hb\,,
\end{equation}
and this can be recognized as equivalent to the result of energy quantization,
\begin{equation}
	\tr e^{-\Ho 4\kappa T} = \sum_{n=0}^\infty e^{-(n+\frac12) 4\kappa T} = \frac{1}{2\sinh 2\kappa T}\,.
\end{equation}
This demonstrates that the KOD can be considered an alternative to energy quantization.
Indeed, the operator $\Ho$ does not appear in SPQM as an energy observable, but rather as a term required by the positivity of sampling measurement records.  The late-time limit of the completeness relation is of particular interest: when $T\gg1/\kappa$, $D_\alpha\,e^{-\Ho4\kappa T} D_\alpha^\dagger=e^{-2\kappa T}\proj\alpha$, showing that the SPQM POVM elements approach Glauber coherent states; the completeness relation shows that it does so uniformly,
\begin{align}
1_\Hb=\int\frac{d^2\alpha}{\pi}\proj\alpha\,,
\end{align}
thus giving the coherent-state resolution of the identity.

Everything in our analysis, the path integrals, the KODs, the FPK diffusion equations, the SDEs---everything!---follows from the path integral~\ref{overallZT} for the overall quantum operation $\Z^{(\text{SPQM})}_T$, which integrates over the sample-paths as they appear in the time-ordered Kraus operator~$	L^{(\text{SPQM})}\left[dw_{[0,T)}\right]$.  Yet as the analysis develops in sections~\ref{Moving} and~\ref{Coord}, there emerges a disconnect between the reduced distribution $C_T(Zx)$, which expresses the completeness relation, and the SDEs and stochastic integrals for the phase-point co\"ordinates: $C_t(Zx)$ has ill-defined normalization, and it is not the weight function whose moments are those of the Cartan phase-point variables, $\beta$ and $\alpha$, as they are expressed in the stochastic-integral solutions~\ref{CartanbetaTalphaT}.  The disconnect is all about the real center term, $e^{-\Omega\ell}$, which scales the Kraus operators when they are written in Cartan co\"ordinates.

The three faces of the KOD stochastic trinity---path integrals, diffusion equations, and SDEs---having been sundered in section~\ref{Coord}, we re-unite them in section~\ref{PathIntegral}.   The vehicle for the reunion is the \emph{Harish-Chandra-section reduced distribution function},
\begin{align}
\boxed{
	\vphantom{\Bigg(}
	\hspace{15pt}
    B_T(Zx)\equiv \int_Z\!d\psi\,ds\, D_T(x)e^{-2s}
    =\int \D\mu\!\left[dw_{[0,T)}\right]e^{-2s[dw_{[0,T)}]}\,\delta\Big(Zx,Z\gamma\!\left[dw_{[0,T)}\right]\Big)\,,\label{BTZxPathIntegralintro}
	\hspace{10pt}
	\vphantom{\Bigg)}
}
\end{align}
where $-s$ is the real part of the Harish-Chandra center co\"ordinate $z$.  The distribution $B_T(Zx)$ can be determined in two equivalent ways: first, from the FPK diffusion equation for $B_T(Zx)$ and, second, from applying the stochastic integrals for the Harish-Chandra phase-plane variables, $\nu$ and $\mu$, to the above path-integral expression for $B_T(Zx)$.  The stochastic trinity thus restored, one returns to completeness through the relation
\begin{equation}
	C_T(Zx)=e^{f(Zx)}B_T(Zx)\,,
\end{equation}
where $f(Zx)$ is a quadratic function of the phase-plane variables  that comes directly out of the co\"ordinate transformation between Cartan and Harish-Chandra co\"ordinates (equations \ref{quadCartan} and \ref{eq:posgauge}); $e^{2f(Zx)}$ can be regarded as a positive gauge transformation~\cite{Chaichian2001a}.

Section~\ref{remarks} concludes with musings on the stochastic trinity and the Lie-group manifolds that house instrument evolution.

\vfill\pagebreak

\section{IWH and Co\"ordinate-Free Right-Invariant Motion}\label{Moving}

This section defines the \emph{Simultaneous P\&Q Measurement} (SPQM) process and presents the Instrumental Weyl-Heisenberg Group, $G=\mathrm{IWH}$, as the universal covering group of \hbox{SPQM}.

Section \ref{PosObs} introduces Kraus operators and the concept of observables generating positive transformations instead of unitary transformations.
Section \ref{IWH} introduces the SPQM process and the group $\mathrm{IWH}$.
Section \ref{KDist} explains how the SPQM instrument is universal and defines a \emph{Kraus-operator distribution (or density)\/} (KOD) over $\mathrm{IWH}$.
Section \ref{RightDers} explains how the KOD diffuses over time with the introduction of \emph{right-invariant derivatives}, a differential-geometric technology that will be unfamiliar to most physicists and  quantum scientists.
Section \ref{RightOnes} explains how the sample-paths of the Kraus-operator diffusion are described with the introduction of the \emph{right-invariant one-forms}, which are dual to right-invariant derivatives.

The diffusion equation and stochastic differential equations in sections \ref{RightDers} and \ref{RightOnes} will be solved in section \ref{Coord}.

\subsection{Observables and Infinitesimal Positive Transformation}\label{PosObs}

While observables are often considered to be infinitesimal generators of unitary transformations, they can also generate \emph{positive transformations}.
Let $X$ be a Hermitian observable, $\kappa$ be a real number with units of inverse time, and $dW$ be a standard Wiener increment~\cite{Wiener1921b,Wiener1921a,Wiener1924a,Kac1947a,Kac1959a,Chaichian2001a,Gardiner2009a,Gardiner2021a}, which has measure
\begin{equation}
	\boxed{
		\vphantom{\Bigg(}
		\hspace{15pt}
		d\mu(dW)=\frac{d(dW)}{\sqrt{2\pi dt}} e^{-dW^2/2dt}\,.
		\hspace{15pt}
		\vphantom{\Bigg)}
	}
\end{equation}
Unitary transformations can be infinitesimally generated either deterministically or stochastically, such as
\begin{equation}\label{infinitesimalunitary}
	e^{-iX\kappa\,dt}
	\hspace{20pt}
	\text{or}
	\hspace{20pt}
	\sqrt{d\mu(dW)}e^{-iX\sqrt{\kappa}\,dW}\,.
\end{equation}
Positive transformations on the other hand are fundamentally stochastic.
In addition, the infinitesimal generators of positive transformations are not of the canonical form, with a single parameter conjugate to the infinitesimal generator.
The positive transformations we will be interested in do not involve jump operators, as in photodetection~\cite{MDSrinivas1981a,CSJackson2022a}, but rather are differential, with infinitesimal generators of the form~\cite{KJacobs1998a,KJacobs2006a,LMartin2015a,CSJackson2023a}
\begin{equation}\label{positive}
	\boxed{
		\vphantom{\Bigg(}
		\hspace{15pt}
		\sqrt{d\mu(dW)}L_X(dW)\equiv\sqrt{d\mu(dW)}e^{-X^2 \kappa dt + X\sqrt{\kappa}\,dW}\,.
		\hspace{15pt}
		\vphantom{\Bigg)}
	}
\end{equation}
As a set, these positive transformations define a \emph{measuring instrument}, complete over the Hilbert space, $\Hb$, according to the relation,
\begin{equation}
	\int_\R d\mu(dW) L_X(dW)^\dag L_X(dW)=1_\Hb\,;
\end{equation}
the operators $L_X(dW)$ are known as \emph{Kraus operators}~\cite{Kraus1983a,Nielsen2000a},
and the set of elements, $\{d\mu(dW) L_X(dW)^\dag L_X(dW)\}$, is known as a \emph{positive-operator-valued measure\/} or \hbox{POVM}.
We will call the positive transformations of equation~\ref{positive} \emph{differential Kraus operators\/} and their set a \emph{differential instrument}, both in anticipation of the (multi-dimensional) differential geometry coming up and to contrast these Kraus operators with jump operators.

The form of these infinitesimally generated positive transformations comes from the requirement that the total operation be completely positive and trace preserving.
In particular, define the superoperator $A \odot B $ by
\begin{equation}
	A \odot B(\rho) \equiv A \rho B
\end{equation}
and the (selective, Kraus-rank-one) \emph{operations},
\begin{equation}
	\Odot(L) \equiv L \!\odot\! L^\dag\,.
\end{equation}
Then the aforementioned Kraus operators define a \emph{total operation},
\begin{equation}\label{totalopXdW}
	\int d\mu(dW)\,\Odot\!\left(e^{-X^2 \kappa dt + X\sqrt{\kappa}\,dW}\right)
	= e^{-\frac12\kappa\,dt\,\ad_X^2}\,,
\end{equation}
where defined is the \emph{adjoint\/} superoperator
\begin{equation}
	\ad_X \equiv X\odot 1 - 1\odot X\,.
\end{equation}
In this context, the infinitesimal generator or \emph{Lindbladian},
\begin{equation}
	-\frac12\ad_X^2 = X \odot X - \frac12\left({X^2\odot 1+1\odot X^2}\right)\,,
\end{equation}
defines $X$ as a \emph{Lindblad operator}~\cite{Lindblad1976a}.

Kraus operators can be interpreted as an \emph{indirect measurement\/}~\cite{vonNeumann1932a,Arthurs1965a,Kraus1983a,Nielsen2000a,CSJackson2023a}, where in the differential case a meter with initial meter wavefunction,
\begin{equation}
	\sqrt{dq}\,\langle q | 0\rangle
	=
	\sqrt{\frac{dq}{\sqrt{2\pi \sigma^2}}e^{-q^2/2\sigma^2}}\,,
\end{equation}
is displaced by the system according to the interaction
\begin{equation}\label{SMInteraction}
	-\frac{i}{\hbar}Hdt
	=
	\sqrt{\kappa\,dt}\,X \otimes\!\left({-}2\sigma\frac{\partial}{\partial q}\right)\,,
\end{equation}
which in turn registers some ``position'' $q$, fixing
\begin{equation}
	dW = \sqrt{dt}\,\frac{q}{\sigma}\,,
\end{equation}
so that
\begin{equation}
	\sqrt{d\mu(dW)}\,L_X(dW)
	=
	\sqrt{dq}\,\langle q|e^{-iH\,dt/\hbar}|0\rangle
\end{equation}
(for further details, see~\cite{CSJackson2023a}).

An irresistible sidenote, developed more generally and in more detail in~\cite{CSJackson2023a}, is that the stochastic unitary transformations of equation~\ref{infinitesimalunitary} follow from the same meter interaction~\ref{SMInteraction}, but with registration of the meter momentum~$p$, instead of its position~$q$.  As such, the stochastic unitary transformations have a total operation identical to equation~\ref{totalopXdW},
\begin{equation}
\int d\mu(dW)\, \Odot\!\left(e^{-iX\sqrt{\kappa}\,dW}\right)
= e^{-\frac12\kappa\,dt\,\ad_X^2}\,,
\end{equation}
This alternative unraveling of the total operation corresponds to a symmetry of the general Lindbladian,
\begin{equation}
	\L(A) = A \odot A^\dag - \frac12\Big({A^\dag A\odot 1+1\odot A^\dag A}\Big)\,,
\end{equation}
which is
\begin{equation}
	\L(-iX)=\L(X)\,,
\end{equation}
so that $-iX$ is an alternative Lindblad operator of the total operation.

\subsection{SPQM and the Group IWH}\label{IWH}

The subject of this paper is the continual (or continuous) simultaneous observation of the canonical observables $P$ and $Q$, defined by the canonical commutation relations,
\begin{equation}
	\boxed{
		\vphantom{\Bigg(}
		\hspace{15pt}
		[Q,P] = i\Omega
		\hspace{15pt}
		\text{and}
		\hspace{15pt}
		[\Omega,Q] = [\Omega,P] = 0\,.
		\hspace{10pt}
		\vphantom{\Bigg)}
	}
\end{equation}
As a Lie algebra of infinitesimal generators, these observables are usually considered to generate unitary displacement operators,
\begin{equation}
	D_\alpha = e^{iQ\alpha_2-iP\alpha_1}\,,
	\hspace{20pt}
	\text{where}
	\hspace{20pt}
	\alpha = \frac{\alpha_1+i\alpha_2}{\sqrt2}\,,
\end{equation}
which together define the 3-dimensional \emph{Unitary Weyl-Heisenberg Group},
\begin{equation}\label{UWHG}
	\boxed{
		\vphantom{\Bigg(}
		\hspace{15pt}
		K = \mathrm{UWH} \equiv \left\{D_\alpha\,e^{i\Omega \phi}\,:\,\alpha\in\C,\;\phi\in\R\right\}.
		\hspace{15pt}
		\vphantom{\Bigg)}
	}
\end{equation}
If the generators operate irreducibly on the Hilbert space $\Hb$, then by Shur's lemma $\Omega=\hbar1_\Hb$ for some $\hbar \in \R$ (because $g\Omega g^\inv = \Omega$ for all $g \in \mathrm{UWH}$.)
Assuming that $\hbar$ is finite, that is $\hbar \neq 0$, all such representations are essentially equivalent because the observables can always be rescaled so as to make the choice $\hbar=1$.  Therefore, it is usually assumed that
\begin{equation}
	\boxed{
		\vphantom{\Bigg(}
		\hspace{15pt}
		\Omega = 1_\Hb\,.
		\hspace{15pt}
		\vphantom{\Bigg)}
	}
\end{equation}
We shall now assume $\hbar=1$, but we will continue nonetheless to use $\Omega$ for the infinitesimal generator to emphasize its existence is not defined by the Hilbert space, but instead by the canonical commutation relations.
In particular, what is defined by the Hilbert space is the relation $\Omega^2 = \Omega$, which is associative algebraic and not Lie algebraic.

For quantization, the Unitary Weyl-Heisenberg Group is supplemented with the unitary group generated by
\begin{equation}
	\Ho \equiv \frac{P^2+Q^2}{2}\,,
\end{equation}
defining the 4-dimensional \emph{Dynamical Weyl-Heisenberg Group},
\begin{equation}
	\mathrm{DWH} \equiv \left\{e^{-i \Ho s}D_\alpha\,e^{i\Omega \phi}\,:\,s\in\R,\;\alpha\in\C,\;\phi\in\R\right\}\,.
\end{equation}
Here, the use of ``dynamical'' refers to the analogies between $\Ho$ and the classical Hamiltonian of a simple harmonic oscillator, which quantum mechanics was originally founded upon.\\

The observables $P$ and $Q$ can be measured simultaneously in the sense that the positive transformations they generate commute infinitesimally (to order $dt$),
\begin{align}\label{Ldw}
	L(dw)&\equiv L_Q\left(dW^q\right)L_P\left(dW^p\right)\\
	&= e^{-Q^2\kappa dt + Q \sqrt\kappa\,dW^q}e^{-P^2\kappa dt + P \sqrt\kappa\,dW^p}\\
	&= e^{-(Q^2+P^2)\kappa dt + Q \sqrt\kappa\,dW^q + P \sqrt\kappa\,dW^p + \frac12[Q,P] \kappa\,dW^q dW^p}\\
	&= e^{-(Q^2+P^2)\kappa dt + Q \sqrt\kappa\,dW^q + P \sqrt\kappa\,dW^p}\\
	&=L_P\!\left(dW^p\right)L_Q\!\left(dW^q\right)\,,
\end{align}
so long as the Wiener outcome increments, $dW^q$ and $dW^p$, are independent, their joint Wiener measure being
\begin{equation}
	d\mu(dw) \equiv d\mu(dW^q)d\mu(dW^p) = \frac{d^2(dw)}{\pi dt}e^{-dw^*dw/dt} \,.
\end{equation}
Here we switch to using a complex Wiener increment
\begin{equation}
dw \equiv \frac{dW^q+idW^p}{\sqrt2}\,.
\end{equation}
Continually repeating this simultaneous measurement for a finite amount of time, $T$, defines the overall Kraus operators,
\begin{equation}
		L\!\left[dw_{[0,T)}\right] = \mathcal{T}\!\!\prod_{k=0}^{T/dt-1}L(dw_{kdt})\,,
\end{equation}
where $\mathcal{T}\prod$ denotes a time-ordered product.
It is important to appreciate that while the Kraus operators commute infinitesimally, they do not commute over finite amounts of time.
Finally, these Kraus operators are accompanied by the Wiener path measure,
\begin{equation}\label{complexWienermeasure}
	\D\mu\!\left[dw_{[0,T)}\right]
    = \prod_{k=0}^{T/dt-1}\!d\mu(dw_{kdt})
    =\left(\prod_{k=0}^{T/dt-1}\!d^{\,2}\!\big(dw_{k dt}\big)\right)\!\!
        \left(\frac{1}{\pi dt}\right)^{T/dt}\exp\!\left(-\int_0^{T_-}\frac{|dw_t|^2}{dt}\,\right)\,,
\end{equation}
which is written here in terms of the complex Wiener increments.
In summary, we have defined a time-dependent instrument with Kraus operators
\begin{equation}\label{SPQM}
	\boxed{
		\vphantom{\Bigg(}
		\hspace{15pt}
		\sqrt{\D\mu\!\left[dw_{[0,T)}\right]}\,L\!\left[dw_{[0,T)}\right] = \sqrt{\D\mu\!\left[dw_{[0,T)}\right]}\mathcal{T}
        \exp\!\left(\int_0^T \!\!\!-\Ho 2\kappa dt + Q\sqrt{\kappa}\,dW_t^q + P\sqrt{\kappa}\,dW_t^p\right)\,,
		\hspace{15pt}
		\vphantom{\Bigg)}
	}
\end{equation}
where $\mathcal{T}\exp\int$ is the time-ordered exponential.
This is the \emph{Simultaneous P\&Q Measurement\/} (SPQM.)
It is worth repeating that $\Ho$ appears here due to the form of the differential positive transformations of equation~\ref{positive} and in this context is not a Hamiltonian because it is not generating unitary transformations.

While SPQM has been considered as far back as~\cite{Barchielli1982a}, it has not been fully solved before.
There are two other ways of measuring $P$ and $Q$ simultaneously that are important to distinguish from SPQM: the Arthurs-Kelly measurement~\cite{Arthurs1965a,Braunstein1991a} and the Goetsch-Graham-Wiseman (GGW) model of heterodyne detection~\cite{Barchielli1991a,Goetsch1994a,Wiseman1994b,Wiseman1996a,CSJackson2022a}.
The Arthurs-Kelly measurement has the same system-meter interaction as the SPQM process, but is different in that Arthurs and Kelly imagine continually interacting the same two meters with the system until the measurement is terminated, whereas in the SPQM process the system interacts with many pairs of meters successively, registering the complex Wiener path $dw_{[0,T)}$.
The GGW model of heterodyne detection has the same many-meter model as the SPQM process, but each system-meter interaction is energy conserving (the so-called leaky cavity), whereas the system-meter interaction of the SPQM process is the meter displacement of equation~\ref{SMInteraction}.

The total operation of the SPQM process is a Wiener-like path integral,
\begin{equation}\label{totalZT}
	\Z_T\equiv \int \D\mu\!\left[dw_{[0,T)}\right]\Odot\Big(L\!\left[dw_{[0,T)}\right]\!\Big)\,,
\end{equation}
which is absolutely trivial to solve,
\begin{equation}
    \Z_T
	= \left(\int d\mu(dw)\,\Odot\Big(e^{-\Ho 2\kappa dt + Q\sqrt{\kappa}dW_t^q + P\sqrt{\kappa}dW_t^p}\Big)\right)^{\circ T/dt}
	= e^{-\frac{1}{2}\kappa T(\ad_Q^2+\ad_P^2)}\,.
\end{equation}
The interest of this article, however is entirely in the manifold diffusion process defined by SPQM, where equation~\ref{SPQM} is understood to define sample-paths in a finite-dimensional manifold.
The infinitesimal generators of these sample-paths are $Q$, $P$, and $\Ho$.
By simply considering their Lie brackets to first order,
\begin{align}
	[Q,P]&=i\Omega\,,\\
	[\Ho,Q]&=-iP\,,\\
	[\Ho,P]&=iQ\,,
\end{align}
and second order,
\begin{align}
	\big[[\Ho,Q],Q\big]&=-\Omega\,,\\
	\big[[\Ho,P],P\big]&=-\Omega\,,\\
	\big[\Ho,[\Ho,Q]\big]&=Q\,,\\
	\big[\Ho,[\Ho,P]\big]&=P\,,
\end{align}
we can see that SPQM defines a 7-dimensional manifold, a representation of what we will call the \emph{Instrumental Weyl-Heisenberg Group},
\begin{equation}
	\boxed{
		\vphantom{\Bigg(}
		\hspace{15pt}
		G = \mathrm{IWH} \equiv \left\{e^{-\Ho r}e^{Qq}e^{Pp}D_\alpha\,e^{i\Omega \phi}e^{-\Omega \ell}\,:\,r\in\R,\;q,p\in\R,\;\alpha\in\C,\;\phi,\ell\in\R\right\}.
		\hspace{15pt}
		\vphantom{\Bigg)}
	}
\end{equation}
There is a fourth and final Weyl-Heisenberg group worth defining, the 6-dimensional \emph{Complex Weyl-Heisenberg Group},
\begin{equation}
\C\mathrm{WH} \equiv \left\{e^{Qq}e^{Pp}e^{-\Omega \ell}D_\alpha\,e^{i\Omega \phi}\,:\,q,p\in\R,\alpha\in\C,\phi,\ell\in\R\right\},
\end{equation}
which is a maximal normal subgroup of $\mathrm{IWH}$, called the derived subgroup.
In Lie-group terminology, $\mathrm{IWH}$ is said to be solvable while $\C\mathrm{WH}$ is said to be nilpotent or unipotent~\cite{Barut1980a,Knapp2002a}.
In particular, the derived series of $G$ is
\begin{equation}
	G = \mathrm{IWH} \vartriangleright \C\mathrm{WH} \vartriangleright Z \vartriangleright 1\,,
\end{equation}
where defined is the \emph{center} of $G$,
\begin{equation}\label{centerdef}
	\boxed{
		\vphantom{\Bigg(}
		\hspace{15pt}
		Z \equiv \left\{e^{i\Omega \phi}e^{-\Omega \ell}\,:\,\phi,\ell\in\R\right\}.
		\hspace{15pt}
		\vphantom{\Bigg)}
	}
\end{equation}

Before proceeding, it is useful to introduce the complex-amplitude (a.k.a.\ annihilation) operator
\begin{align}
a\equiv\frac{1}{\sqrt2}(Q+iP)\,,
\end{align}
which has Lie bracket
\begin{align}
[a,a^\dagger]=\Omega\,.
\end{align}
We have
\begin{align}
\Ho=\smallfrac12(aa^\dagger+a^\dagger a)=a^\dagger a+\smallfrac12\Omega\,.
\end{align}
Note also that the displacement operator,
\begin{align}
D_\alpha=e^{iQ\alpha_2-iP\alpha_1}=e^{a^\dagger\alpha-a\alpha^*}\,,
\end{align}
has the ordered forms,
\begin{align}
D_\alpha
&=e^{-\frac12i\Omega\alpha_1\alpha_2}e^{iQ\alpha_2}e^{-iP\alpha_1}
=e^{\frac12i\Omega\alpha_1\alpha_2}e^{-iP\alpha_1}e^{iQ\alpha_2}\label{DorderedPQ}\\
&=e^{-\frac12|\alpha|^2\Omega}e^{a^\dagger\alpha}e^{-a\alpha^*}=e^{\frac12|\alpha|^2\Omega}e^{-a\alpha^*}e^{a^\dagger\alpha}
\label{Dorderedaadag}
\,,
\end{align}
which will prove useful in relating co\"ordinate systems and in evaluating right-invariant derivatives.  The first form in equation~\ref{Dorderedaadag} is usually called normal ordering, and the second form is called antinormal ordering.

\subsection{Haar Measure, Dirac Delta, and Kraus-Operator Distribution Function}\label{KDist}

Many readers will interpret the groups defined in the previous section as matrix groups, where the observables are quantized in the usual way.
The exponentials can be understood more abstractly, however, as generating path-connections, and this way of thinking gives rise to what is called the \emph{universal covering group}~\cite{Poincare2010a,Weyl1913a,Bourbaki1989a,Pontrjagin1946a,Knapp2002a,CSJackson2023a}.
We will now start to consider more seriously the points of $\mathrm{IWH}$ in this universal fashion and think of the instrument of SPQM as a representation of $G=\mathrm{IWH}$.
The map $L:\C^{T/dt}\longrightarrow\GL(\Hb)$ from the set of paths, $\C^{T/dt}$, to the operator space, $\GL(\Hb)$, can be factored into two maps,
\begin{equation}
	\boxed{
		\vphantom{\Bigg(}
		\hspace{15pt}
		L = R \circ \gamma\,,
		\hspace{15pt}
		\vphantom{\Bigg)}
	}
\end{equation}
where $\gamma:\C^{T/dt}\longrightarrow G$ maps Wiener paths to the universal covering group and  $R:G\longrightarrow\GL(\Hb)$ is the representation, mapping the universal cover to the space of linear operators on $\Hb$~\cite{Knapp2002a}.  We have denoted a sample-path by $dw_{[0,T)}$, and we now start denoting elements of the instrumental group by $x$.  To drive the notation home, we note that the instrumental group element and Kraus operator associated with a sample path are denoted by
\begin{align}
x_T =\gamma[dw_{[0,T)}]
\qquad\quad\textrm{and}\qquad\quad
L[dw_{[0,T)}]=R\big(x_T\big)\,.
\end{align}
The distinction between $L$ and $\gamma$ emphasizes that the time-ordered exponential of equation~\ref{SPQM} actually defines a diffusion problem on the instrumental group that is independent of the spectral information inherent in the definition of a linear operator.
In particular, this means the entire analysis of this article is independent of whether or how $\Ho$ is quantized.
(Remember, $\Ho$ is quantized in the usual way the moment $\Hb$ is assumed to be irreducible and to have a ground state.)

As does every finite-dimensional Lie group, $G=\mathrm{IWH}$ has a right-invariant \emph{Haar measure\/}~\cite{Haar1933a,vonNeumann1999a,Nachbin1965a,Bourbaki2004a},
\begin{equation}
	\boxed{
		\vphantom{\Bigg(}
		\hspace{15pt}
		d\mu(xg) =  d\mu(x)\,.
		\hspace{15pt}
		\vphantom{\Bigg)}
	}
\end{equation}
As is \emph{not} always the case for (solvable) Lie groups~\cite{Barut1980a}, it turns out that this right-invariant measure is also equal to the left-invariant measure,
\begin{equation}
	\boxed{
		\vphantom{\Bigg(}
		\hspace{15pt}
		d\mu(x) = d\mu(gx)\,,
		\hspace{15pt}
		\vphantom{\Bigg)}
	}
\end{equation}
and this left invariance will turn out to be important for the interpretation of the SPQM process as a diffusion, as will be seen in the very next subsection.
Comments about the existence and uniqueness of the Haar measure will be given in appendices~\ref{FrameTrans} and~\ref{FrameTransHC}, but first it is worth taking it for granted and appreciating what can be done with it.

With the Haar measure, we can introduce the accompanying singular distributions or ``\emph{Dirac deltas},'' defined by the property (sometimes called reproduction~\cite{Brif1999a,STAli1999a}) that for any function $f$ of $G=\mathrm{IWH}$,
\begin{equation}
	\boxed{
		\vphantom{\Bigg(}
		\hspace{15pt}
		\int_G d\mu(x)\,\delta(y,x)f(x)=f(y)\,.
		\hspace{15pt}
		\vphantom{\Bigg)}
	}
\end{equation}
From the invariance properties of the Haar distribution, the Dirac delta distributions inherit the corresponding covariance properties,
\begin{equation}
	\boxed{
		\vphantom{\Bigg(}
		\hspace{15pt}
		\delta(xg,yg)=\delta(x,y)=\delta(gx,gy)\,.
		\hspace{15pt}
		\vphantom{\Bigg)}
	}
\end{equation}
With the Haar measure and the Dirac delta, we can define a universal instrument by adding up all of the Wiener paths that end at the same Kraus operator, starting from the origin.
This becomes visible by considering the total operation,
\begin{align}
	\Z_T &= \int \D\mu\!\left[dw_{[0,T)}\right]\Odot\Big(L\!\left[dw_{[0,T)}\right]\!\Big)\\
	 &= \int \D\mu\!\left[dw_{[0,T)}\right]\Odot\Big(R\big(\gamma\big[dw_{[0,T)}\big]\big)\Big)\\
	 &= \int \D\mu\!\left[dw_{[0,T)}\right]\Odot\Big(R\big(\gamma\big[dw_{[0,T)}\big]\big)\Big)\int_G d\mu(x)\,\delta\Big(x,\gamma\big[dw_{[0,T)}\big]\Big)\\
	 &= \int_G d\mu(x)\,\Odot\big(R(x)\big)\int \D\mu\!\left[dw_{[0,T)}\right]\delta\Big(x,\gamma\!\left[dw_{[0,T)}\right]\Big)\,.
\end{align}
In summary, the SPQM process defines a \emph{universal instrument}, which unravels the total operation,
\begin{equation}\label{ZT1}
	\boxed{
		\vphantom{\Bigg(}
		\hspace{15pt}
		\Z_T = \int_G d\mu(x)\,D_T(x)\;\Odot\big(R(x)\big)\,,
		\hspace{10pt}
		\vphantom{\Bigg)}
	}
\end{equation}
according to a Haar-based \emph{Kraus-operator distribution function\/} (KOD),
\begin{equation}\label{KODF}
\boxed{
	\vphantom{\Bigg(}
	\hspace{15pt}
	D_T(x)\equiv\int \D\mu\!\left[dw_{[0,T)}\right]\delta\Big(x,\gamma\!\left[dw_{[0,T)}\right]\Big)\,,
	\hspace{10pt}
	\vphantom{\Bigg)}
}
\end{equation}
which is defined by a Wiener path integral~\cite{Wiener1921b,Wiener1921a,Wiener1924a,Kac1947a,Kac1959a,Chaichian2001a}.  The total operation is a completely positive, trace-preserving superoperator.  The trace-preserving property is equivalent to saying that the POVM elements, $d\mu(x)\,D_T(x)\,R(x)^\dagger R(x)$, satisfy a completeness relation,
\begin{align}\label{completenessDT}
1_\sH=\int_G\!d\mu(x)\,D_T(x)\,R(x)^\dagger R(x)\,.
\end{align}

The term ``universal'' refers to the fact that this description of the instrument is common to every representation and comes from the concepts of universal covering group and universal enveloping algebra~\cite{Bourbaki1989a,Dixmier1996a,CSJackson2023a}.
It is important to understand that the universal covering group, $\mathrm{IWH}$, is defined purely by the local structure (that is, the Lie algebra) of the observables and the quadratic term, here $\Ho$, which accompanies the observables due to the nature of differential positive transformations.
In particular, this means $G=\mathrm{IWH}$ is not defined by the Hilbert space of states.
This ability to describe the measuring instrument without reference to a Hilbert space is so striking that we give it a name: \emph{universal instrument autonomy\/}~\cite{CSJackson2022a,CSJackson2023a}. SPQM is in a very special class of universal instruments for which the universal covering group is finite dimensional; we dub such instruments \emph{principal instruments}~\cite{CSJackson2023a}.

\subsection{Diffusion Equation in Terms of Right-Invariant Derivatives}\label{RightDers}

The definition of the Kraus-operator distribution function~\ref{KODF} can be thought of as a Feynman-Kac formula~\cite{Kac1947a,Kac1959a,Chaichian2001a} for the solution of a Fokker-Planck-Kolmogorov (FPK) diffusion equation~\cite{Gardiner2009a,Gardiner2021a}.
This FPK diffusion equation can be obtained easily with the help of the so-called right-invariant derivatives~\cite{Knapp1988a,Frankel2012a,Kitaev2018a,CSJackson2021a,CSJackson2023a},
\begin{equation}
	\boxed{
		\vphantom{\Bigg(}
		\hspace{15pt}
		\Rinv{X}[f](x) \equiv \lim_{h\to0}\frac{f(e^{Xh}x)-f(x)}{h}\,,
		\hspace{15pt}
		\vphantom{\Bigg)}
	}
\end{equation}
which can be seen to have commutators~\cite{CSJackson2023a}
\begin{equation}
	\Rinv{X}\Rinv{Y}-\Rinv{Y}\Rinv{X}= -\Rinv{[X,Y]}\,.
\end{equation}
This negative sign is why left-invariant derivatives are usually considered instead of right-invariant ones.
Nevertheless, because the convention is to consider operators as acting to the right we are more-or-less stuck with having to consider a right-invariant basis of local transformations.

With the right-invariant derivatives, $D_{t+dt}$ can then be expanded about $t$ in the standard way.  We start with
\begin{align}
	D_{t+dt}(x)&=\int \D\mu\!\left[dw_{[0,t+dt)}\right]\delta\Big(x,\gamma\!\left[dw_{[0,t+dt)}\right]\Big)\\
	&=\int d\mu(dw_t)\int\D\mu\!\left[dw_{[0,t)}\right]\delta\Big(x,\gamma(dw_t)\gamma\!\left[dw_{[0,t)}\right]\Big)\,,\label{eq:Dtplusdt}
\end{align}
where
\begin{align}\label{eupdelta}
\gamma(dw_t)=e^{\updelta_t}\,,\hspace{15pt}
\updelta_t\equiv-\Ho 2\kappa\,dt + Q \sqrt\kappa\,dW^q + P \sqrt\kappa\,dW^p
=-\Ho 2\kappa\,dt + a\sqrt\kappa\,dw_t^* + a^\dagger\sqrt\kappa\,dw_t
\end{align}
is the purely group-theoretic version of $L(dw_t)$ of equation~\ref{Ldw}, that is, $L(dw_t)=R\big(\gamma[dw_t]\big)$.  Here we also define the \emph{forward generator\/} $\updelta_t$: \emph{$\gamma(dw_t)=e^{\updelta_t}$ is the fundamental differential positive operator for SPQM, and the forward generator $\updelta_t$ is thus the core mathematical object for the theory of the SPQM instrument.}  Continuing with equation~\ref{eq:Dtplusdt}, we have
\begin{align}\label{Linv}
	D_{t+dt}(x)
    &=\int d\mu(dw_t)\int\D\mu\!\left[dw_{[0,t)}\right]\delta\Big(\gamma(dw_t)^\inv x,\gamma\!\left[dw_{[0,t)}\right]\Big)\\\label{CK}
	&=\int d\mu(dw_t)\,D_t\big(\gamma(dw_t)^\inv x\big)\\
	&=\int d\mu(dw_t)\,D_t\big(e^{-\updelta_t}x\big)\\\label{ri}
    &=\int d\mu(dw_t)\,e^{-\Rinv{\updelta_t}}[D_t]\,\big(x\big)\\\label{Ito}
	&=\int d\mu(dw_t)\,\left(D_t(x)-\Rinv{\updelta_t}[D_t](x)+\frac12\Rinv{\updelta_t}\big[\Rinv{\updelta_t}[D_t]\big](x)\right)\\
    &=D_t(x)+\kappa\,dt\,\Delta\big[D_t](x)\,,
\end{align}
where
\begin{equation}
\boxed{
	\vphantom{\Bigg(}
	\hspace{15pt}
	\Delta
    \equiv 2\Rinv{\Ho}+\frac12\big(\Rinv{Q}\,\Rinv{Q}+\Rinv{P}\,\Rinv{P}\big)
	\hspace{15pt}
	\vphantom{\Bigg)}
}\label{ForGen}
\end{equation}
is the \emph{FPK forward generator}.
Equation~\ref{Linv} is where the left invariance of the Haar measure is used.
Equation~\ref{CK} is the analog of a Chapman-Kolmogorov equation for the distribution function~\cite{Gardiner2009a,Gardiner2021a}.
Equation~\ref{ri} moves $e^{\updelta_t}$ outside the argument of the KOD to become an exponential of the right-invariant derivative
\begin{equation}\label{eq:CSMP}
	\boxed{
		\vphantom{\Bigg(}
		\hspace{15pt}
		\Rinv{\updelta_t}=-\Rinv{\Ho}2\kappa\,dt + \Rinv{Q}\sqrt{\kappa}\,dW_t^q + \Rinv{P}\sqrt{\kappa}\,dW_t^p\,,
		\hspace{15pt}
		\vphantom{\Bigg)}
	}
\end{equation}
which we call the \emph{vector-valued SPQM increment}.
Equation \ref{Ito} Taylor expands the distribution function to second order, that is, to order $dt$, as required by the It\^o rule for the Wiener outcome increments.  The remaining step to the FPK forward generator $\Delta$ involves averaging over the Wiener distribution $d\mu(dw_t)$; in this averaging the deterministic term $-2\kappa\,dt\Rinv{\Ho}$ in $\Rinv{\updelta_t}$ contributes a first-derivative term to the FPK forward generator, whereas the Wiener outcome increment terms in $\Rinv{\updelta_t}$ contribute second-derivative diffusion terms.

In summary, the KOD of SPQM evolves according to the FPK diffusion equation,
\begin{equation}\label{FPKE}
	\boxed{
		\vphantom{\Bigg(}
		\hspace{15pt}
		\frac{1}{\kappa}\frac{\partial}{\partial t} D_{t}(x)=\Delta [D_t](x)\,,
		\hspace{15pt}
		\vphantom{\Bigg)}
	}
\end{equation}
with initial condition
\begin{equation}\label{D0deltax1}
	D_0(x) = \delta(x,1)\,,
\end{equation}
where defined is the group identity
\begin{equation}
	1 \equiv e^0\,,
\end{equation}
which is the origin of $\mathrm{IWH}$.
Equation \ref{FPKE} will be (mostly) solved in section \ref{solveFPK} after having established two co\"ordinate systems.  The subtlest thing about equation~\ref{FPKE} is remembering that the Lie algebra of the three directions apparent in the equation means that the motion beyond first order is actually 7-dimensional.

Before proceeding to SDEs, we draw attention to an important property of right-invariant derivatives.  The reader might already be thinking about this property by wondering why we didn't write $\Rinv{\updelta_t}$ and $\Delta$ in terms of the right-invariant derivatives associated with $a$ and $a^\dagger$.  The reason is that the map from operators to right-invariant derivatives, $X \longmapsto \Rinv{X}$, is only $\R$-linear and not $\C$-linear~\cite{CSJackson2023a}.  This means that $\Rinv{Q}$, $\Rinv{P}$, $\Rinv{iQ}$, and $-\Rinv{iP}$ are $\R$-linearly independent, with $\Rinv{Q}$ and $\Rinv{P}$ displacing Kraus operators in positive directions on IWH and $\Rinv{iQ}$ and $-\Rinv{iP}$ displacing in unitary directions.  The right-invariant derivatives $\Rinv{a}$, $\Rinv{a^\dagger}$, $-\Rinv{ia}$, and $\Rinv{ia^\dagger}$ each represent a different way of combining equally displacements in the positive and unitary directions.  In view of this, it is instructive to note that the vector-valued SPQM increment~(\ref{eq:CSMP}) has the form,
\begin{align}
\Rinv{\updelta_t}
=-\Rinv{\Ho}2\kappa\,dt+\frac12\sqrt\kappa\,dw_t\big(\Rinv{a}+\Rinv{a^\dagger}+i\,\Rinv{ia}-i\,\Rinv{ia^\dagger}\big)
+\frac12\sqrt\kappa\,dw_t^*\big(\Rinv{a}+\Rinv{a^\dagger}-i\,\Rinv{ia}+i\,\Rinv{ia^\dagger}\big)
\end{align}
This means, in particular, that
\begin{align}
\Delta\neq 2\Rinv{\Ho}+\frac12\big(\Rinv{a}\,\Rinv{a^\dagger}+\Rinv{a^\dagger}\Rinv{a}\big)\,.
\end{align}

\subsection{Sample-Path SDEs in Terms of Right-Invariant One-Forms}\label{RightOnes}

As has been mentioned, the time-ordered exponential of the SPQM process (equation~\ref{SPQM}) can be interpreted as defining sample-paths in the 7-dimensional manifold $G=\mathrm{IWH}$.
Sample-paths are usually described by stochastic-differential equations (SDEs).  We finish this section by explaining how such SDEs can be expressed in terms of the right-invariant structure.\\

The basis of right-invariant derivatives,
\begin{equation}
	\Big\{\Rinv{e_\nu}\Big\}\equiv\Big\{-\Rinv{\Ho},\Rinv{Q},\Rinv{P},-\Rinv{\,iP},\Rinv{\,iQ},-\Rinv{\Omega},\Rinv{\,i\Omega}\Big\}\,,
\end{equation}
defines a dual basis of right-invariant one-forms,
\begin{equation}
	\boxed{
		\vphantom{\Bigg(}
		\hspace{15pt}
		\theta^\mu\!\Big(\Rinv{e_\nu}\Big)\equiv\delta^\mu_\nu\,,
		\hspace{15pt}
		\vphantom{\Bigg)}
	}
\end{equation}
In terms of the right-invariant one-forms, the Haar measure has a simple expression in terms of wedge products,
\begin{equation}\label{HaarFormula}
d\mu(x) = \theta^{-\Ho}\wedge\theta^{Q}\wedge\theta^{P}\wedge\theta^{iQ}\wedge\theta^{-iP}\wedge\theta^{-\Omega}\wedge\theta^{i\Omega}.
\end{equation}
Also in terms of the right-invariant one-forms, the SDEs equivalent to equation~\ref{SPQM} are given by reading off the coefficient conjugate to the corresponding generator in the vector-valued SPQM increment~\ref{eq:CSMP},
\begin{align}\label{SDE1}
	\theta^{i\Omega}\big(\Rinv{\updelta_t}\big)	&=	0\,,\\
	\theta^{-\Omega}\big(\Rinv{\updelta_t}\big)	&=	0\,,\label{SDE2}\\
	\theta^{-iP}\big(\Rinv{\updelta_t}\big)	&=	0\,,\\
	\theta^{iQ}\big(\Rinv{\updelta_t}\big)	&=	0\,,\\
	\theta^{Q}\big(\Rinv{\updelta_t}\big)	&=	\sqrt\kappa\,dW_t^q\,,\\
	\theta^{P}\big(\Rinv{\updelta_t}\big)	&=	\sqrt\kappa\,dW_t^p\,,\\
    \theta^{-\Ho}\big(\Rinv{\updelta_t}\big)	&=	2\kappa\,dt\,.\label{SDE7}
\end{align}

These SDEs can be broken into two types:
the first-order SDEs,
\begin{equation}\label{firstorderSDEs}
	\boxed{
		\vphantom{\Bigg(}
		\hspace{15pt}
		\theta^{-\Ho}\big(\Rinv{\updelta_t}\big)	=	2\kappa\,dt\,,
		\hspace{15pt}
		\theta^{Q}\big(\Rinv{\updelta_t}\big)	=	\sqrt\kappa\,dW_t^q\,,
		\hspace{15pt}
		\theta^{P}\big(\Rinv{\updelta_t}\big)	=	\sqrt\kappa\,dW_t^p\,,
		\hspace{15pt}
		\vphantom{\Bigg)}
	}
\end{equation}
and the Pfaffians,
\begin{equation}\label{Pfaffians}
	\boxed{
		\vphantom{\Bigg(}
		\hspace{15pt}
		\theta^{-iP}\big(\Rinv{\updelta_t}\big)	=
		\theta^{iQ}\big(\Rinv{\updelta_t}\big)	=
		\theta^{-\Omega}\big(\Rinv{\updelta_t}\big)	=
		\theta^{i\Omega}\big(\Rinv{\updelta_t}\big)	=	0\,.
		\hspace{15pt}
		\vphantom{\Bigg)}
	}
\end{equation}
These equations will be solved in section~\ref{solveSDEs} after having established a co\"ordinate system.

The stochastic equations~\ref{firstorderSDEs} and~\ref{Pfaffians} are almost obvious by definition, but there is a subtlety that requires attention.  The right-invariant derivatives and one-forms live in the tangent and cotangent spaces to the group manifold~$G$ and thus are based on linear transformations that use the chain rule of ordinary calculus.  Hence the stochastic equations that come from the right-invariant one-forms are Stratonovich-form SDEs~\cite{Gardiner2009a,Gardiner2021a}---this means mid-point evaluation of coefficients of stochastic increments---and should be converted to the It\^o-form SDEs in which coefficients are evaluated at the beginning of the increment.  In the context of IWH, the only place this subtlety makes a difference is in the SDEs that come from $\theta^{i\Omega}$ and $\theta^{-\Omega}$.  Jackson and Caves~\cite{CSJackson2021a,CSJackson2023a} introduced the \emph{modified Maurer-Cartan stochastic differential\/} (MMCSD) as a way to get to the It\^o-form equations directly.  The MMCSD is an example of the It\^o correction in SDEs~\cite{Gardiner2009a,Gardiner2021a}, specifically the It\^o correction that arises when one transforms between a stochastic variable and the exponential function of that variable, as occurs in the forward generator~\ref{eupdelta}.  In this paper we get directly to It\^o-form SDEs in a different way, when we consider the Harish-Chandra decomposition in section~\ref{trans}.

A further subtlety about the right-invariant one-forms is that they have ``curl'' in the same sense as Gibbs would have defined.
In the standard language of forms, this is because the exterior algebra of forms is equivalent to the Lie algebra of derivatives~\cite{Knapp1988a},
\begin{equation}
	[\Rinv{e_\mu},\Rinv{e_\nu}]=-{c_{\mu\nu}}^\lambda \Rinv{e_\lambda}
	\hspace{20pt}
	\Longleftrightarrow
	\hspace{20pt}
	d\theta^\lambda = \frac12 {c_{\mu\nu}}^\lambda \,\theta^\mu\! \wedge \theta^\nu\,.
\end{equation}
This equivalence is standard in modern differential topology, but an introduction is included in appendix~\ref{app:ext}; though no use will be made of the ``curls'' $d\theta^\lambda$ in this article, they are given for completeness in equations~\ref{curl1}--\ref{curl7}.

\vfill\pagebreak

\section{IWH and Two Co\"ordinate Systems}\label{Coord}

Having introduced the Instrumental Weyl-Heisenberg Group, $G=\mathrm{IWH}$, a co\"ordinate system needs to be established so that we can locate the sample-paths of the SPQM process and follow their propagation.
If the concept of a universal covering group introduced in the previous section is unclear, seeing how equation~\ref{SPQM} is equivalent to a set of co\"ordinate SDEs should help to appreciate that $G=\mathrm{IWH}$ and the SPQM process are independent of matrix representation.\\

We will use two co\"ordinate systems, analogous to what are called Harish-Chandra~\cite{Harish-Chandra1956a,Knapp1986a,Knapp2002a} (a.k.a.\ ``Gauss''~\cite{Zelobenko1973a,Barut1980a,Perelomov1986a}) decompositions and Cartan decompositions~\cite{Helgason1978a,Barut1980a,Knapp2002a,Knapp1986a}.
These decompositions were originally designed in the context of semisimple Lie groups~\cite{Gilmore2002a,Barut1980a,Knapp2002a,Chern1952a,Howe2011a}, of which $\mathrm{IWH}$ is quite the opposite (in the sense of the Levi-Malcev decomposition).
In spite of this distinction, it is very useful to think of $\mathrm{IWH}$ in many ways \emph{as if\/} it were semisimple.
The Harish-Chandra decomposition is easier to prove first and will allow us to make connections between the SPQM process and two processes more familiar to physicists, the Ornstein-Uhlenbeck process and the Goetsch-Graham-Wiseman (GGW) heterodyne measuring process.
The Cartan decomposition is better suited for considering the \hbox{POVM}.

Section \ref{semisemi} identifies the analogs of the various elements of semisimple group theory.
Section \ref{trans} proves the Harish-Chandra decomposition of $\mathrm{IWH}$ in a way that also produces the corresponding It\^o-form co\"ordinate SDEs, which are immediately recognized and solved.
Section \ref{Cartan} introduces the Cartan decomposition and the transformations to Harish-Chandra co\"ordinates and presents the right-invariant derivatives and one-forms in both co\"ordinate systems.
Section \ref{solveSDEs} solves the SDEs of the SPQM process in both Cartan and Harish-Chandra co\"ordinates.
Section \ref{solveFPK} solves most of the FPK diffusion equation of the SPQM process in Cartan co\"ordinates, by which we mean introducing the Cartan-section reduced distribution function and solving for it.
Section \ref{Alt} explains how the solution of the FPK diffusion equation means that the POVM of the SPQM process offers an alternative perspective on the meaning of energy quantization.

\subsection{Usual Elements of Semisimple Lie Group Theory}\label{semisemi}

As introduced in the previous section, the Lie group of interest is the so-called Instrumental Weyl-Heisenberg Group
\begin{equation}
		G = \mathrm{IWH} \equiv \left\{e^{-\Ho r}e^{Qq}e^{Pp}D_\alpha\,e^{i\Omega \phi}e^{-\Omega \ell}\,:\,r\in\R,\;q,p\in\R,\;\alpha\in\C,\;\phi,\ell\in\R\right\}\,.
\end{equation}
Although $G$ is literally solvable, with derived series,
\begin{equation}
	G = \mathrm{IWH} \vartriangleright \C\mathrm{WH} \vartriangleright Z \vartriangleright 1\,,
\end{equation}
and center,
\begin{equation}
		Z \equiv \left\{e^{i\Omega \phi}e^{-\Omega \ell}\,:\,\phi,\ell\in\R\right\},
\end{equation}
it can be navigated in much the same way as a semisimple group.
While a bit of the terminology~\cite{Knapp2002a} will be used here, the theory of semisimple groups will be more-or-less glossed over.
The purpose of this subsection is basically to label the various subgroups that will prove to be both meaningful and useful for navigating $\mathrm{IWH}$ and therefore understanding \hbox{SPQM}.
The significance of these subgroups should become apparent as they are applied.

The map from the SPQM instrument to the SPQM POVM,
\begin{equation}
\boxed{
	\vphantom{\Bigg(}
	\hspace{15pt}
	\pi(x) = x^\dag x\,,
	\hspace{15pt}
	\vphantom{\Bigg)}
}
\end{equation}
defines the SPQM POVM as similar to a symmetric space, albeit a nonRiemannian one, with \emph{Cartan group involution}
\begin{equation}
		x^\iota \equiv x^{-\dag} = (x^{\inv})^\dag = (x^{\dag})^\inv\,.
\end{equation}
The subgroup of transformations that are even under the Cartan involution is the usual Unitary Weyl-Heisenberg Group of equation~\ref{UWHG}:
\begin{equation}
\boxed{
	\vphantom{\Bigg(}
	\hspace{15pt}
	K \equiv  \Big\{x \in G\,:\,x^\iota = x\Big\} = \pi^\inv(1) =\mathrm{UWH}\,.
	\hspace{15pt}
	\vphantom{\Bigg)}
}
\end{equation}
On the other hand, the remainder of $G$ displaces from the origin of the \emph{symmetric space},
\begin{equation}
\boxed{
	\vphantom{\Bigg(}
	\hspace{15pt}
	\mathcal{E} \equiv \pi(G) \cong K\backslash G\,.
	\hspace{15pt}
	\vphantom{\Bigg)}
}
\end{equation}
Considering the conjugation action of $K$ on $\E$, almost all of the $K$-conjugacy classes can be parameterized by the \emph{Cartan subgroup},
\begin{equation}
	\boxed{
		\vphantom{\Bigg(}
		\hspace{15pt}
		A \equiv \left\{e^{-\Ho r}e^{-\Omega \ell}\,:\,r,\ell\in\R\right\}\,,
		\hspace{15pt}
		\vphantom{\Bigg)}
	}
\end{equation}
and regular POVM elements are invariant under the \emph{commutant},
\begin{equation}
	\boxed{
		\vphantom{\Bigg(}
		\hspace{15pt}
		M \equiv \left\{k\in K\,:\,\forall a\in A,kak^\inv = a\right\}=\left\{e^{i\Omega \phi}\,:\,\phi\in\R\right\}.
		\hspace{15pt}
		\vphantom{\Bigg)}
	}
\end{equation}
Thus the $K$-conjugacy classes have the topology of the familiar phase space,
\begin{equation}
	K/M \cong \C\,.
\end{equation}
Indeed, it is the ``almost all'' feature where $G$ and $\E$ depart from semisimple groups and Riemannian symmetric spaces, since positive transformations of the form $e^{Qq+Pp}$ are characteristically not in the conjugacy classes of $A$ (see appendix~\ref{app:CoordTrans} for additional perspective).
Finally, important also are the \emph{maximal nilpotent (or unipotent) subgroup},
\begin{equation}
\boxed{
	\vphantom{\Bigg(}
	\hspace{15pt}
	N \equiv \left\{e^{a\mu^*}\,:\,\mu\in\C\right\}\,,
	\hspace{15pt}
	\vphantom{\Bigg)}
}
\end{equation}
and perhaps the most important, the \emph{Borel subgroup},
\begin{equation}
\boxed{
	\vphantom{\Bigg(}
	\hspace{15pt}
	B \equiv A\ltimes N = \left\{e^{-\Ho r}e^{-\Omega \ell}e^{a\mu^*}\,:\,r,\ell\in\R,\;\mu\in\C\right\}.
	\hspace{15pt}
	\vphantom{\Bigg)}
}
\end{equation}

This group-theoretic context now in hand, we stress that the most important groups for what follows are $G=\textrm{IWH}$ itself, the center $Z$ of $G$, and the quotient group $G/Z=\mathrm{IWH}/Z$.  The center $Z$ contains a phase, which is of no importance, and a normalization, which is the main source of difficulty in analyzing SPQM.  The cosets $Zx\in G/Z$ are parametrized by what we call the ruler, $r$, and by two complex phase-space parameters.  One of these complex phase-space parameters is associated with the POVM, and the other parametrizes a post-measurement displacement operator.  We call $G/Z$ the \emph{Reduced Instrumental Weyl-Heisenberg Group\/} (RIWH).  It is isomorphic to the adjoint group of $G$, but given the way we will use multipliers on the cosets $Zx$, we prefer to think of $G$ as a central extension of $G/Z$.

\subsection{Harish-Chandra Decomposition and SDEs as a Proof by Transfinite Induction}\label{trans}

The 7-dimensional Instrumental Weyl-Heisenberg Group $G$ affords a \emph{Harish-Chandra decomposition}, $G = N^\dag\! M A N$, where every element can be decomposed into the form
 \begin{equation}\label{HC}
 	\boxed{
 		\vphantom{\Bigg(}
 		\hspace{15pt}
 		x = e^{a^\dag \nu}e^{-\Ho r +\Omega z}e^{a\mu^*}\,,
 		\hspace{15pt}
 		\vphantom{\Bigg)}
 	}
 \end{equation}
defining seven \emph{Harish-Chandra co\"ordinates} $(\nu,r,z,\mu)\in\C\!\times\!\R\!\times\!\C\!\times\!\C$.
We often break the complex co\"ordinates into real and imaginary parts,
\begin{equation}\label{HCrealimag}
	\nu = \frac{\nu_1+i\nu_2}{\sqrt2}\,,
	\hspace{20pt}
	\mu = \frac{\mu_1+i\mu_2}{\sqrt2}\,,
	\hspace{20pt}
	z = -s+i\psi\,.
\end{equation}
The co\"ordinate $r$ we call the ruler, $\nu$ and $\mu$ are the postmeasurement and POVM phase-plane co\"ordinates, and $z$ is the IWH center co\"ordinate.

A proof that this decomposition exists for every element of the SPQM process (equation~\ref{SPQM}) is not difficult if we allow ourselves to apply transfinite induction:
At time $t=dt$ (the first infinitesimal increment), it is easy to see that the decomposition exists because (see equation~\ref{eupdelta})
\begin{align}
    x_{dt}=e^{\updelta}
	&=e^{-\Ho 2\kappa\,dt+a^\dag\sqrt\kappa\,dw_0+a\sqrt\kappa\,dw_0^*}\\
	&=e^{-\Ho 2\kappa dt} e^{a^\dag\sqrt\kappa\,dw_0+a\sqrt\kappa\,dw_0^*}\\
	&=e^{a^\dag\!\sqrt{\kappa}\,dw_0} e^{-\Ho 2\kappa dt+\Omega\frac12\kappa|dw_0|^2} e^{a\sqrt{\kappa}\,dw_0^*}.
\end{align}
Trivially, this also means that the decomposition exists for any finite integer $n$ and infinitesimal time $t=ndt$ simply because the one-parameter subgroups commute to infinitesimal order, so long as the Wiener increments are independent.  Now for the transfinite step.
If we assume that the decomposition holds for a finite time $t$,
\begin{equation}
		x_t = e^{a^\dag \nu_t}e^{-\Ho r_t +\Omega z_t}e^{a\mu_t^*}\,,
\end{equation}
then an increment later in the SPQM process we have
\begin{align}
	x_{t+dt}
	&=e^{\updelta_t}x_t\\
	&=e^{a^\dag\!\sqrt{\kappa}\,dw_t} e^{-\Ho 2\kappa dt+\Omega\frac12\kappa|dw_t|^2}e^{a\sqrt{\kappa}\,dw_t^*}\;
	e^{a^\dag \nu_t}e^{-\Ho r_t +\Omega z_t}e^{a\mu_t^*}\\
	&=e^{a^\dag\!\sqrt{\kappa}\,dw_t} e^{-\Ho 2\kappa dt}\;
	e^{a^\dag\nu_t}e^{a\sqrt{\kappa}\,dw_t^*}\;
	e^{-\Ho r_t+\Omega(z_t+\frac12\kappa|dw_t|^2+\nu_t\sqrt\kappa\,dw_t^*)}e^{a\mu_t^*}\\
	&=e^{a^\dag(e^{-2\kappa dt}\nu_t+\!\sqrt{\kappa}\,dw_t)}e^{-\Ho(r+2\kappa dt)
        +\Omega(z+\frac12\kappa |dw_t|^2+\nu_t\sqrt\kappa dw_t^*)}e^{a(\mu_t+e^{-r_t}\sqrt\kappa\,dw_t)^*}\,.
\end{align}
This concludes the proof of the Harish-Chandra decomposition for the SPQM process and $G=\mathrm{IWH}$.\\

A consequence of the proof is that the SPQM process in Harish-Chandra co\"ordinates is equivalent to the It\^o-form SDEs~\cite{Ito1950a,Ito1996a,Chirikjian2009a,Gardiner2009a,Gardiner2021a},
\begin{align}
	dr_t &= 2\kappa\,dt\,,\label{timeInc}\\
	d\nu_t&=-2\kappa\,dt\,\nu_t +\sqrt\kappa\,dw_t\label{OUP}\,,\\
	d\mu_t&=e^{-r_t}\sqrt\kappa\,dw_t\,,\label{GGW}\\
	-ds_t+id\psi_t=dz_t&=\frac12\kappa |dw_t|^2+\nu_t \sqrt\kappa\,dw_t^*\,.\label{latterCarl}
\end{align}
Although these are It\^o-form SDEs, notice that we did not use the It\^o rule in deriving them; in particular, we do not set $|dw_t|^2=dt$ in the SDE for the center co\"ordinate $z$.  We solve these equations for the initial condition $r_0=\nu_0=\mu_0=z_0=0$.  These initial values are chosen so that $x_0=1$, in agreement with the $\delta$-function initial condition~\ref{D0deltax1} for the {KOD}.  It is straightforward to see that the first three SDEs have as solution,
\begin{equation}
	\boxed{
		\vphantom{\Big(}
		\hspace{15pt}
		r_T = 2\kappa T\,,\label{timeIncsolution}
		\hspace{15pt}
		\vphantom{\Big)}
	}
\end{equation}
\vspace{-16pt}
\begin{equation}
	\boxed{
		\vphantom{\bigg(}
		\hspace{15pt}
		\nu_T=\int_0^{T_-} \!\!\sqrt{\kappa}\,dw_t\,e^{-2\kappa (T-t)}\,,\label{OUPsolution}
		\hspace{15pt}
		\vphantom{\bigg)}
	}
\end{equation}
\vspace{-16pt}
\begin{equation}
	\boxed{
		\vphantom{\bigg(}
		\hspace{15pt}
		\mu_T=\int_0^{T_-} \!\!\sqrt\kappa\,dw_t\,e^{-2\kappa t}\,,\label{GGWsolution}
		\hspace{15pt}
		\vphantom{\bigg)}
	}
\end{equation}
where $T_-\equiv T-dt$.  The fourth equation is solved by plugging the solution for $\nu_t$ into the equation for $z_t$ and integrating, with the result that
\begin{equation}
	\boxed{
		\vphantom{\Bigg(}
		\hspace{15pt}
        -s_T+i\psi_T=z_T=\frac12\int_0^{T_-}\!\!\int_0^{T_-}\!\!\kappa\,dw_t^*\,dw_s\,e^{-2\kappa |t-s|}\big(1+\textrm{sgn}(t-s)\big)\,,
		\hspace{15pt}
		\vphantom{\Bigg)}
	}\label{latterCarlsolution}
\end{equation}
where
\begin{align}
\textrm{sgn}(u)=
\begin{cases}
1,&u>0,\\
0,&u=0,\\
-1,&u<0,
\end{cases}
\end{align}
is the sign function.  There being subtleties in deriving and interpreting this solution, it is worked out carefully in appendix~\ref{zsolution}.  The solutions for the real and imaginary parts of $z$ are
\begin{align}
-s_T&=\frac12\int_0^{T_-}\!\!\int_0^{T_-}\!\!\kappa\,dw_t^*\,dw_s\,e^{-2\kappa |t-s|}\,,\label{sT}\\
i\psi_T&=\frac12\int_0^{T_-}\!\!\int_0^{T_-}\!\!\kappa\,dw_t^*\,dw_s\,e^{-2\kappa |t-s|}\textrm{sgn}(t-s)\,.\label{psiT}
\end{align}

The SPQM posterior phase-point variable, $\nu_T$ of equation~\ref{OUPsolution}, is a linear functional in the measurement record and can be recognized as an Ornstein-Uhlenbeck (OU) process~\cite{Uhlenbeck1930a,Gardiner2009a,Gardiner2021a}.
The SPQM prior phase-point variable, $\mu_T$ of equation~\ref{GGWsolution}, is a linear functional in the measurement record and can be recognized as the same as the Goetsch-Graham-Wiseman (GGW) heterodyne process.
(To be clear, the equations for the remaining variables of the GGW heterodyne process would instead be $r_T^{\text{GGW}}=2\kappa T$, $\nu_T^{\text{GGW}}=0$, and $z_T^{\text{GGW}}=\kappa T$~\cite{CSJackson2022a}; in particular, the GGW heterodyne process corresponds to the Borel subgroup $B$.)
The center SPQM variable, $z_T$ of equation~\ref{latterCarlsolution}, is a quadratic functional in the measurement record; the solution is identical to that obtained by deriving a quantum fluctuation-dissipation theorem for the correlation of $\nu$ and $\mu$.\\

It is worth taking a moment to reflect on equations \ref{timeIncsolution}--\ref{latterCarlsolution}.
Equation \ref{timeIncsolution}~tells us that if $\Ho$ is quantized in the standard way, then the Kraus operators collapse, with an $e$-folding time $\tau = 1/2\kappa$, to a scaled outer product of coherent states of the form
\begin{equation}\label{LTlarge}
	L_{T\gg1/\kappa}
    \sim e^{-\kappa T+z_T}e^{a^\dagger\nu_T}\proj0 e^{a\mu_T^*}
    =e^{i\psi_T}e^{\frac12(|\nu_T|^2+|\mu_T|^2)-s_T-\kappa T}\ket{\nu_T}\bra{\mu_T}\,.
\end{equation}
The complementarity in time of the OU and GGW processes (equations~\ref{OUPsolution} and~\ref{GGWsolution}) is also interesting: whereas the post-measurement variable $\nu_T$ of equation~\ref{OUPsolution} depends only on the end of the outcome register, the POVM variable $\nu_T$ of equation~\ref{GGWsolution} depends only on the beginning of the register.
It is thus reasonably clear that the POVM of the SPQM process culminates in the usual ``measurement in the coherent-state basis,''
\begin{equation}\label{csPOVM}
	1_\Hb = \int \frac{d^2\mu}{\pi}\proj{\mu}\,,
\end{equation}
where $d^2\mu=\frac12 d\mu_1 d\mu_2$.  This is just like GGW heterodyne detection, except that the post-measurement state is not vacuum but instead scrambled over phase space.  That the POVM variable $\mu_T$ ceases to evolve after a few $e$-foldings seems to be an important feature in the interpretation of SPQM as a measuring process~\cite{vonNeumann1932a}.  All this said, there remains the elephant in the room that prompts us to label these conclusions as ``reasonably clear'': the elephant is the factor that scales the long-time Kraus operator in equation~\ref{LTlarge}; although equation~\ref{LTlarge} makes it absolutely clear that the Kraus operator approaches an outer product of coherent states, not clear at all is how the POVM approaches a uniform distribution of coherent states, as required by the completeness relation~\ref{csPOVM}.  Figuring this out is, in some sense, what the rest of the paper is about.

\subsection{Cartan Decomposition and Various Transformations}\label{Cartan}

As far as what the SPQM process is ultimately doing in time, section~\ref{trans} in many ways says it all---except for that elephant in the room.  We now turn our attention to a detailed understanding of the measuring process at finite times, which requires addressing the elephant.
Key to this is the realization that the Harish-Chandra decomposition is not well suited to telling us the behavior of the \hbox{POVM}.
Rather, a Cartan decomposition of the Instrumental Weyl-Heisenberg Group, $G=KAK$, works best for this purpose.
A straightforward calculation (left to appendix~\ref{app:CoordTrans}) shows that almost every group element---and every Kraus operator---with a Harish-Chandra decomposition also affords a \emph{Cartan decomposition},
\begin{equation}\label{Cartandecomp}
	\boxed{
		\vphantom{\Bigg(}
		\hspace{15pt}
		x = \left(D_\beta\, e^{i\Omega \phi}\right)e^{-\Ho r-\Omega\ell }D_\alpha^\inv\,,
		\hspace{15pt}
		\vphantom{\Bigg)}
	}
\end{equation}
and therefore the POVM elements decompose as
\begin{equation}
\boxed{
	\vphantom{\Bigg(}
	\hspace{15pt}
	x^\dag x = D_\alpha\,e^{-\Ho 2r-\Omega2\ell }D_\alpha^\inv\,,
	\hspace{15pt}
	\vphantom{\Bigg)}
}
\end{equation}
where the \emph{Cartan co\"ordinates}, $(\beta,\phi,r,\ell,\alpha)\in\C\!\times\!\R\!\times\!\R\!\times\!\R\!\times\!\C$, can be obtained by the co\"ordinate transformations (the ruler $r$ is shared by the two co\"ordinate systems),
\begin{align}
\beta&=\frac{e^r\nu + \mu}{2\sinh r}\,,\\
\alpha&=\frac{e^r\mu + \nu}{2\sinh r}\,,\\
\ell&=s - f\,,\\
\phi &=\psi -\xi\,.
\end{align}
Here $f$ and $\xi$ are functions of the ruler and the phase-plane co\"ordinates, that is, functions on the RIWH~$G/Z$,
\begin{align}
\begin{split}\label{f}\
f&=\frac{(|\nu|^2+|\mu|^2)e^r+\nu^*\mu+\nu\mu^*}{4\sinh r}=\frac{|\nu+\mu|^2}{4(1-e^{-r})}+\frac{|\nu-\mu|^2}{4(1+e^{-r})}\\
&=\frac12\Big(|\beta|^2+|\alpha|^2-\beta^*\alpha e^{-r}-\beta\alpha^*e^{-r}\Big)=\frac{1-e^{-r}}{4}|\beta+\alpha|^2+\frac{1+e^{-r}}{4}|\beta-\alpha|^2\,,
\end{split}\\
\begin{split}
\xi&=\frac{\nu^*\mu-\nu\mu^*}{4i\sinh r}=\frac{(\nu-\mu)^*(\nu+\mu)-(\nu-\mu)(\nu+\mu)^*}{8i\sinh r}\\
&=e^{-r}\frac{\beta^*\alpha-\beta\,\alpha^*}{2i}=e^{-r}\frac{(\beta-\alpha)^*(\beta+\alpha)-(\beta-\alpha)(\beta+\alpha)^*}{4i}\,.\label{xi}
\end{split}
\end{align}
Notice the singularity in Cartan co\"ordinates at $r=0$, about which there is further discussion throughout the remainder of the paper and, in particular, in appendix~\ref{app:CoordTrans}.\\

As we wish to solve equation~\ref{FPKE} and equations~\ref{SDE1}--\ref{SDE7}, more to our purpose are the transformations from the right-invariant moving frame to the Cartan
co\"ordinate frame.
A calculation of these frame transformations can be found in appendix~\ref{FrameTrans}, but we include them here for continuity; we also include the transformations
to the Harish-Chandra co\"ordinate frame, which are worked out in appendix~\ref{FrameTransHC}.
For the FPK diffusion equation~\ref{FPKE}, we require the transformations of the derivatives,
\begin{align}\label{DerTrans}
\begin{array}{rclcl}
	\vphantom{\Big(}\Rinv{i\Omega}&=&\partial_\phi&=&\partial_\psi\\
	\vphantom{\Big(}-\Rinv{\Omega}&=&\partial_\ell&=&\partial_s \\
	\vphantom{\Big(}-\Rinv{iP}&=&\partial_{\beta_1}-\frac12\beta_2\partial_\phi&=&\partial_{\nu_1}-e^{-r}\partial_{\mu_1}+\frac12\nu_1\partial_s-\frac12\nu_2\partial_\psi\\
	\vphantom{\Big(}\Rinv{iQ}&=&\partial_{\beta_2}+\frac12\beta_1\partial_\phi&=&\partial_{\nu_2}-e^{-r}\partial_{\mu_2}+\frac12\nu_2\partial_s+\frac12\nu_1\partial_\psi \\
	\vphantom{\bigg(}\Rinv{Q}&=&\nabla_1 -\beta_1\partial_\ell+\displaystyle{\frac{\beta_2\cosh r-\alpha_2}{2\sinh r}}\partial_\phi
        &=&\nabla_1-\frac12\nu_1\partial_s+\frac12\nu_2\partial_\psi\\
	\vphantom{\bigg(}\Rinv{P}&=&\nabla_2 -\beta_2\partial_\ell-\displaystyle{\frac{\beta_1\cosh r-\alpha_1}{2\sinh r}}\partial_\phi
        &=&\nabla_2-\frac12\nu_2\partial_s-\frac12\nu_1\partial_\psi\\
	\vphantom{\bigg(}-\Rinv{\Ho}&=&\partial^{\textrm{C}}_{r}-\beta_1\nabla_1-\beta_2\nabla_2+\displaystyle{\frac{\beta_1^2+\beta_2^2}{2}\partial_\ell
        +\frac{\beta_1\alpha_2-\beta_2\alpha_1}{2\sinh r}\partial_\phi}&=&\partial^{\textrm{HC}}_{r}-\nu_1\partial_{\nu_1}-\nu_2\partial_{\nu_2}
\end{array}\,,
\end{align}
where
\begin{equation}
	\nabla_j \equiv \frac{1}{\sinh r}\left(\partial_{\alpha_j}+\cosh r\,\partial_{\beta_j}\right)=\partial_{\nu_j}+e^{-r}\partial_{\mu_j}\,.
\end{equation}
The two co\"ordinate systems share the co\"ordinate $r$, but the partial derivative with respect to $r$ is, of course, different in the two systems.  In the above equation we distinguish $\partial_r$ in the two systems, but we do this nowhere else because it is always clear in which co\"ordinate system we are operating.  It is worth recording for working in terms of the complex phase-space variables that
\begin{align}\label{nablac}
\nabla&\equiv\frac{1}{\sqrt2}(\nabla_1-i\nabla_2)
=\frac{1}{\sinh r}\left(\partial_{\alpha}+\cosh r\,\partial_{\beta}\right)=\partial_{\nu}+e^{-r}\partial_{\mu}\,,\\
\nabla^*&=\frac{1}{\sqrt2}(\nabla_1+i\nabla_2)
=\frac{1}{\sinh r}\left(\partial_{\alpha^*}+\cosh r\,\partial_{\beta^*}\right)=\partial_{\nu^*}+e^{-r}\partial_{\mu^*}\,.
\label{nablacstar}
\end{align}

For the SDEs~\ref{SDE1}--\ref{SDE7}, we require the transformations of the one-forms,
\begin{align}\label{FormTrans}
\begin{array}{rclcl}
	\vphantom{\Big(}\theta^{i\Omega}&=&d\phi+\smallfrac12(\beta_2 d\beta_1-\beta_1d\beta_2)
        +\smallfrac12(\alpha_2 d\alpha_1-\alpha_1d\alpha_2)+\cosh r\,(\beta_1d\alpha_2-\beta_2d\alpha_1)
        &=&d\psi+\frac12 e^r(\nu_1d\mu_2-\nu_2d\mu_1)\\
	\vphantom{\Big(}\theta^{-\Omega}&=&	d\ell + \frac12(\beta_1^2+\beta_2^2)dr+\sinh r \,(\beta_1 d\alpha_1 +\beta_2 d\alpha_2)
        &=&ds+\frac12 e^r(\nu_1d\mu_1+\nu_2d\mu_2)\\
	\vphantom{\Big(}\theta^{-iP}&=&d\beta_1 -\cosh r\, d\alpha_1&=&\frac12\big(d\nu_1-e^r d\mu_1+\nu_1 dr\big)\\
	\vphantom{\Big(}\theta^{iQ}&=&d\beta_2 -\cosh r\, d\alpha_2&=&\frac12\big(d\nu_2-e^r d\mu_2+\nu_2 dr\big)\\
	\vphantom{\Big(}\theta^{Q}&=&\beta_1 dr+\sinh r\, d\alpha_1&=&\frac12\big(d\nu_1+e^r d\mu_1+\nu_1 dr\big)\\
	\vphantom{\Big(}\theta^{P}&=&\beta_2 dr+\sinh r\, d\alpha_2&=&\frac12\big(d\nu_2+e^r d\mu_2+\nu_2 dr\big)\\
	\vphantom{\Big(}\theta^{-\Ho}&=&dr&=&dr
\end{array}\,.
\end{align}
With the one-form transformations in hand, the Haar measure~\ref{HaarFormula} in the two co\"ordinate systems is
\begin{equation}\label{d7x}
	\boxed{
		\vphantom{\Bigg(}
		\hspace{15pt}
		d^7\!\mu(x)=d\phi\, d\ell\, \frac{d^2\beta}{\pi}\,dr\,\sinh^2\!r\,\frac{d^2\alpha}{\pi}
        =d\psi\,ds\,\frac{d^2\nu}{2\pi}\,dr\,e^{2r}\frac{d^2\mu}{2\pi}\,.
		\hspace{15pt}
		\vphantom{\Bigg)}
	}
\end{equation}
Here and elsewhere complex phase-plane measures are denoted by $d^2\beta=\frac12d\beta_1 d\beta_2$. The factors of $1/\pi$ in the Cartan measure are conventional in quantum optics and ultimately come from the coherent-state completeness relation~\ref{csPOVM}.  The factors of $1/2\pi$ in Harish-Chandra co\"ordinates then follow from the transformation from Cartan to Harish-Chandra co\"ordinates.  The left invariance of these measures is discussed in appendices~\ref{FrameTrans} and~\ref{FrameTransHC}.  While we are considering measures, we should record the reduced measure on the 5-dimensional group $\textrm{RIWH}=\textrm{IWH}/Z$:
\begin{equation}\label{d5Zx}
	\boxed{
		\vphantom{\Bigg(}
		\hspace{15pt}
		d^5\!\mu(Zx)=\frac{d^7\!\mu(x)}{d\mu(Z)}=\frac{d^2\beta}{\pi}\,dr\,\sinh^2\!r\,\frac{d^2\alpha}{\pi}
        =\frac{d^2\nu}{2\pi}\,dr\,e^{2r}\frac{d^2\mu}{2\pi}\,.
		\hspace{15pt}
		\vphantom{\Bigg)}
	}
\end{equation}
Here $d\mu(Z)=d\phi\,d\ell=d\psi\,ds$ is the measure on the center~$Z$.

Accompanying the co\"ordinate Haar measure $d^7\!\mu(x)$ are the co\"ordinate forms of the conjugate $\delta$-function,
\begin{align}
\delta(x,x')
&=\delta(\phi-\phi')\,\delta(\ell-\ell')\,\frac{1}{\sinh^2\!r}\delta(r-r')\,\pi\delta^2(\beta-\beta')\,\pi\delta^2(\alpha-\alpha')\label{delta7Cartan}\\
&=\delta(\psi-\psi')\,\delta(s-s')\,e^{-2r}\delta(r-r')\,2\pi\delta^2(\nu-\nu')\,2\pi\delta^2(\mu-\mu')\label{delta7HC}\,.
\end{align}
There are obvious co\"ordinate forms for the $\delta$-function $\delta(Zx,Zx')$ that is conjugate to $d^5\!\mu(Zx)$:
\begin{align}
\delta(Zx,Zx')
&=\frac{1}{\sinh^2\!r}\delta(r-r')\,\pi\delta^2(\beta-\beta')\,\pi\delta^2(\alpha-\alpha')\label{delta5Cartan}\\
&=e^{-2r}\delta(r-r')\,2\pi\delta^2(\nu-\nu')\,2\pi\delta^2(\mu-\mu')\label{delta5HC}\,.
\end{align}
We are especially interested in $\delta(x,1)$.  Because the identity $1$ has Harish-Chandra co\"ordinates $\phi=\ell=r=0$ and $\nu=\mu=0$, we have
\begin{align}\label{deltax1HC}
\delta(x,1)=\delta(\phi)\,\delta(s)\,\delta(r)\,2\pi\delta^2(\nu)\,2\pi\delta^2(\mu)\,.
\end{align}
The Cartan form of $\delta(x,1)$ requires more attention because of the co\"ordinate singularity at $r=0$ (see appendix~\ref{app:CoordTrans} for discussion). The singularity is about more than just the $1/\sinh^2\!r$ in the Cartan form of the $\delta$-function, although that singularity is the root of the difficulties that require attention.  We provide the necessary attention in appendix~\ref{deltax1}.

\subsection{Solving the SDEs}\label{solveSDEs}

Section~\ref{RightOnes} left off by showing that SPQM corresponds to the right-invariant Stratonovich-form SDEs~\ref{SDE1}--\ref{SDE7}.
With the frame transformations~\ref{FormTrans} of the previous section at hand, the three first-order SDEs of equation~\ref{firstorderSDEs} find the following expressions,
\begin{align}
2\kappa dt=\theta^{-\Ho}(\Rinv{\updelta_t})&=dr\,,\label{CSDE1}\\
\begin{split}
	\sqrt\kappa\,dw_t=\frac{1}{\sqrt2}\Big(\theta^Q(\Rinv{\updelta_t})+i\theta^P(\Rinv{\updelta_t})\Big)
	&=\beta\,dr+\sinh r\, d\alpha\\
	&=\frac12\big(d\nu+e^r d\mu+\nu\,dr\big)\,,\label{CSDE23}
\end{split}
\end{align}
and the four Pfaffians~\ref{Pfaffians} give
\begin{align}
\begin{split}
0=\frac{1}{\sqrt2}\Big(\theta^{-iP}(\Rinv{\updelta_t})+i\theta^{iQ}(\Rinv{\updelta_t})\Big)&=d\beta-\cosh r\,d\alpha\\
&=\frac12\big(d\nu-e^r d\mu+\nu\,dr\big)\,,\label{CSDE45}
\end{split}\\
\begin{split}
0=\theta^{-\Omega}(\Rinv{\updelta_t})&=d\ell + \frac12(\beta_1^2+\beta_2^2)dr+\sinh r \,(\beta_1 d\alpha_1 +\beta_2 d\alpha_2)\\
&=ds+\frac12 e^r(\nu_1d\mu_1+\nu_2d\mu_2)\,,\label{CSDE6}
\end{split}\\
\begin{split}
0=\theta^{i\Omega}(\Rinv{\updelta_t})
&=d\phi+\frac12(\beta_2 d\beta_1-\beta_1d\beta_2)+\frac12(\alpha_2 d\alpha_1-\alpha_1d\alpha_2)+\cosh r\,(\beta_1d\alpha_2-\beta_2d\alpha_1)\\
&=d\psi+\frac12 e^r(\nu_1d\mu_2-\nu_2d\mu_1)\,,\label{CSDE7}
\end{split}
\end{align}
where the evaluation of the co\"ordinate one-forms on the vector-valued SPQM increment, $\Rinv{\updelta_t}$, is no longer denoted.

Equation~\ref{CSDE1} is the ruler equation~\ref{timeInc}, with solution $r_t=2\kappa t$ for initial condition $r_0=0$.  Summing and differencing equations~\ref{CSDE23} and~\ref{CSDE45}, one finds that
\begin{align}
	\begin{split}
		\sqrt\kappa\,dw_t&=d\beta+\beta\,dr-e^{-r}d\alpha\\
		&=d\nu+\nu\,dr\,,
	\end{split}\\
\begin{split}
	\sqrt\kappa\,dw_t&=-d\beta+\beta\,dr+e^r d\alpha\\
	&=e^r d\mu\,.
\end{split}
\end{align}
As was discussed in section~\ref{RightOnes}, these are Stratonovich-form SDEs, which means that the coefficients are evaluated at the midpoint, $t+\frac12\,dt$, of the increment---technically the midpoint has no status in the stochastic calculus, so one can regard midpoint evaluation as $a_{t+dt/2}=\frac12(a_t+a_{t+dt})=a_t+\frac12 da_t$---but for these SDEs, that evaluation does not produce an It\^o correction (because $dr_t = 2\kappa dt$ has no stochastic term), and the equations can be read as It\^o-form SDEs, with the coefficients evaluated at the beginning of the increment.  When $r$ and $dr$ are substituted into these SDEs, the equations for the Harish-Chandra complex phase-point co\"ordinates, $\nu$ and $\mu$, become those of equations~\ref{OUP} and~\ref{GGW}, with solutions~\ref{OUPsolution} and~\ref{GGWsolution} for initial conditions $\nu_0=0$ and $\mu_0=0$.

The SDEs for Cartan phase-space points are
\begin{align}\label{betaSDE}
d\beta&=\cosh2\kappa t\,d\alpha\,,\\
d\alpha&=\csch2\kappa t\,(-2\kappa\beta\,dt+\sqrt\kappa\,dw_t)\,.
\label{alphaSDE}
\end{align}
For integrating, these SDEs are more profitably written as
\begin{align}\label{betaalphaSDE1}
d(\nu e^{2\kappa t})=d(\beta e^{2\kappa t}-\alpha)&=e^{2\kappa t}\sqrt\kappa\,dw_t\,,\\
d\mu=d(-\beta e^{-2\kappa t}+\alpha)&=e^{-2\kappa t}\sqrt\kappa\,dw_t\,,
\label{betaalphaSDE2}
\end{align}
and the reason is that these are the same as integrating the SDEs for the Harish-Chandra phase-plane co\"ordinates.  The solutions, satisfying initial conditions $\nu_0=\mu_0=0$,
\begin{align}
\nu_Te^{2\kappa T}=\beta_Te^{2\kappa T}-\alpha_T&=\int_0^{T_-} \!\!\sqrt{\kappa}\,dw_t\,e^{2\kappa t}\,,\\
\mu_T=-\beta_Te^{-2\kappa T}+\alpha_T&=\int_0^{T_-} \!\!\sqrt{\kappa}\,dw_t\,e^{-2\kappa t}\,,
\end{align}
are  those obtained in equations~\ref{OUPsolution} and~\ref{GGWsolution} for the Harish-Chandra phase-space points.  Summarizing, we have that
the solutions for the ruler and the Cartan phase-space co\"ordinates are
\begin{equation}
	\boxed{
		\vphantom{\Bigg(}
		\hspace{15pt}
		r_T=2\kappa T\,,
		\hspace{25pt}
        \beta_T = \int_0^{T_-}\sqrt\kappa\,dw_t\,\frac{\cosh2\kappa t}{\sinh 2\kappa T}\,,
        \hspace{25pt}
		\alpha_T = \int_0^{T_-}\sqrt\kappa\,dw_t\,\frac{\cosh2\kappa (T-t)}{\sinh 2\kappa T}\,.
		\hspace{15pt}
		\vphantom{\Bigg)}
	}\label{solvedSDEs}
\end{equation}

The Stratonovich-form SDE for the Harish-Chandra center co\"ordinate $z$ follows from the SDEs~\ref{CSDE6} and~\ref{CSDE7},
\begin{align}\label{dzStrat}
dz_t=-ds_t+id\psi_t=e^{r_{t+dt/2}}\nu_{t+dt/2}d\mu_t^*\,.
\end{align}
Converted to It\^o form, this equation is
\begin{align}\label{dzIto}
	\begin{split}
		dz_t&=\big(\nu_t+\smallfrac12 d\nu_t)\,e^{r_t}\,d\mu_t^*\\
		&=(\nu_t+\smallfrac12\sqrt\kappa\,dw_t)\sqrt\kappa dw_t^*\\
		&=\smallfrac12\kappa|dw_t|^2+\nu_t\,dw_t^*\,,
	\end{split}
\end{align}
in agreement with equation~\ref{latterCarl}.  The solution for initial condition~$z_0=0$ is carefully worked out in appendix~\ref{zsolution} and given in equation~\ref{latterCarlsolution}.

The Cartan normalization and phase-space co\"ordinates have Stratonovich-form SDEs,
\begin{align}
-d\ell &= \smallfrac12(\beta_1^2+\beta_2^2)dr+\sinh r(\beta_1 d\alpha_1 +\beta_2 d\alpha_2)\,,\\
d\phi&=\smallfrac12(\beta_1 d\beta_2-\beta_2d\beta_1)+\smallfrac12(\alpha_1 d\alpha_2-\alpha_2d\alpha_1)+\cosh r\,(\beta_2d\alpha_1-\beta_1d\alpha_2)\,,
\end{align}
where all the coefficients are evaluated at the midpoint $t+\frac12 dt$.  Converted to equivalent It\^o-form SDEs, these equations become
\begin{align}\label{dl1}
-d\ell_t
&=|\beta|^2\,dr+\sinh r(\beta\,d\alpha^*+\beta^*\,d\alpha)+\smallfrac12\sinh r(d\beta\,d\alpha^*+d\beta^*\,d\alpha)\\
&=\coth2\kappa t\,\kappa|dw_t|^2-2|\beta_t|^2\kappa\,dt+\beta_t\sqrt\kappa\,dw_t^*+\beta_t^*\sqrt\kappa\,dw_t\,,\label{dl2}\\
id\phi
&=\smallfrac12(\beta^*d\beta-\beta\,d\beta^*+\alpha^*d\alpha-\alpha\,d\alpha^*)
+\cosh r(\beta\,d\alpha^*-\beta^*d\alpha)+\smallfrac12\cosh r(d\beta\,d\alpha^*-d\beta^*\,d\alpha)\label{dphi1}\\
&=\csch 2\kappa t\,(\alpha_t\beta_t^*-\alpha_t^*\beta_t)\kappa\,dt
+\smallfrac12(\beta_t\coth2\kappa t-\alpha_t\,\csch2\kappa t)\sqrt\kappa\,dw_t^*
-\smallfrac12(\beta_t^*\coth2\kappa t-\alpha_t^*\csch2\kappa t)\sqrt\kappa\,dw_t\,.\label{dphi2}
\end{align}
Though more complicated, these equations have a similar character to the SDEs for the Harish-Chandra center co\"ordinate $z=-s+i\psi$.  The It\^o correction for $d\ell$ is the last term in equation~\ref{dl1}, and it becomes the $\coth$ term at the beginning of equation~\ref{dl2}, whereas the It\^o correction for $d\phi$ at the end of equation~\ref{dphi1} vanishes.  We could solve these equations directly, but it is both easier and more productive to combine the solution for $z$ with the co\"ordinate transformation to Cartan co\"ordinates, thus finding
\begin{equation}\label{solvedellphi}
	\boxed{
		\vphantom{\Bigg(}
		\hspace{15pt}
		\ell_T = s_T-f_T
		\hspace{15pt}
		\text{and}
		\hspace{15pt}
		\phi_T=\psi_T-\xi_T\,,
		\hspace{15pt}
		\vphantom{\Bigg)}
	}
\end{equation}
where $s_T$ and $\psi_T$ are the Harish-Chandra solutions~\ref{sT} and~\ref{psiT} and $f_T$ and $\xi_T$ are the functions~\ref{f} and~\ref{xi} with all the co\"ordinates evaluated at time~$T$.

\subsection{Solving Most of the FPK Diffusion Equation}\label{solveFPK}

Section~\ref{RightDers} left off showing that the sample-paths of SPQM diffuse according to the KOD $D_t(x)$, which satisfies the FPK equation~\ref{FPKE} with initial condition $D_0(x)=\delta(x,1)$.  The crucial mathematical object in the diffusion equation is the FPK forward generator $\Delta$, which is written in terms of right-invariant derivatives in equation~\ref{ForGen}.

With the frame transformations of equation~\ref{DerTrans} at hand, it is easy to express the three pieces of the forward generator in Cartan co\"ordinates,
\begin{align}
	2\Rinv{\Ho}=-2\partial_{r}
    +2\bigg(\beta_1\nabla_1+\beta_2\nabla_2-\frac{\beta_1^2+\beta_2^2}{2}\partial_\ell\bigg)+ \partial_\phi B_{\Ho}\,,
\end{align}
\begin{align}
\frac12\Rinv{Q}\Rinv{Q}
&=\frac12\!\left(\nabla_1 -\beta_1\partial_\ell+\frac{\beta_2\cosh r-\alpha_2}{2\sinh r}\partial_\phi\right)^2\\
&=\frac12\big(\nabla_1-\beta_1\partial_\ell\big)^2+\frac12\partial_\phi B_Q\\
&=\frac12\Big(\nabla_1^2 -\big(\nabla_1\beta_1+\beta_1\nabla_1\big)\partial_\ell+\beta_1^2\partial_\ell^2\Big)+\frac12\partial_\phi B_Q\\
&=\frac12\Big(\nabla_1^2 -\big(\coth r+2\beta_1\nabla_1\big)\partial_\ell+\beta_1^2\partial_\ell^2\Big)+\frac12\partial_\phi B_Q\\
&=-\frac12\coth r\,\partial_\ell+\frac12\nabla_1^2-\partial_\ell\bigg(\beta_1\nabla_1 -\frac12\beta_1^2\partial_\ell\bigg)+\frac12\partial_\phi B_Q\,,
\end{align}
\begin{align}
\frac12\Rinv{P}\Rinv{P}
&=\frac12\!\left(\nabla_2 -\beta_2\partial_\ell -\frac{\beta_1\cosh r-\alpha_1}{2\sinh r}\partial_\phi\right)^2\\
&=\frac12\big(\nabla_2 -\beta_2\partial_\ell\big)^2+\frac12\partial_\phi B_P\\
&=-\frac12\coth r\,\partial_\ell+\frac12\nabla_2^2-\partial_\ell\bigg(\beta_2\nabla_2 -\frac12\beta_2^2\partial_\ell\bigg)+\frac12\partial_\phi B_P\,,
\end{align}
where we introduce the quantities
\begin{align}
B_{\Ho}&=\frac{\beta_2\alpha_1-\beta_1\alpha_2}{\sinh r}\,,\\
B_Q&=\frac{\beta_2\cosh r-\alpha_2}{\sinh r}\big(\nabla_1 -\beta_1\partial_\ell\big)
+\left(\frac{\beta_2\cosh r-\alpha_2}{2\sinh r}\right)^2\partial_\phi\,,\\
B_P&=-\frac{\beta_1\cosh r-\alpha_1}{\sinh r}\big(\nabla_2 -\beta_2\partial_\ell\big)
+\left(\frac{\beta_1\cosh r-\alpha_1}{2\sinh r}\right)^2\partial_\phi\,,
\end{align}
which are independent of $\phi$ and commute with $\partial_\phi$.  The $\phi$-derivative terms quickly disappear from the analysis, ultimately because $\phi$ is irrelevant to the instrument elements as a consequence of the symmetry $\Odot(e^{i\Omega\phi}L)=\Odot(L)$.  Putting these expressions together, the FPK forward generator~\ref{ForGen} in Cartan co\"ordinates is
\begin{equation}
	\boxed{
		\vphantom{\Bigg(}
		\hspace{15pt}
		\Delta= -2\partial_r -\coth r\,\partial_\ell + \frac12\Big(\nabla_1^2+\nabla_2^2\Big)
		+(2-\partial_\ell)\bigg(\beta_1\nabla_1+\beta_2\nabla_2-\frac{\beta_1^2+\beta_2^2}{2}\partial_\ell\bigg)+ \partial_\phi B\,,
		\hspace{15pt}
		\vphantom{\Bigg)}
	}\label{Cartanforward}
\end{equation}
where $B=B_{\Ho}+\frac12 B_Q+\frac12 B_P$.  It is worth noting that
\begin{align}
\frac12\Big(\nabla_1^2+\nabla_2^2\Big)=\nabla^*\nabla=\nabla\nabla^*\,,
\end{align}
where $\nabla$ and $\nabla^*$, defined in equations~\ref{nablac} and~\ref{nablacstar}, are derivatives with respect to complex phase-space co\"ordinates.

Because of the cubic and quartic nature of the last few terms in equation~\ref{Cartanforward}, we do not hope to find a complete analytic solution to equation~\ref{FPKE}.
However, ``$5/7$-ths" of the distribution can be analyzed quite easily.
Remember that we are interested in the instrument, and observe that the instrument elements can be partitioned by reconsidering the total operation~\ref{ZT1} as
\begin{align}\label{ZTunravelD}
	\Z_T &= \int_G \!d^7\!\mu(x)\,D_T(x) \Odot(x)\\
	&= \int_{\lowerintsub{G/Z}}\!\!\!\!d^5\!\mu(Zx)\left(\int_Z\!d\phi\,d\ell\,D_T(x) e^{-2\ell}\right)\Odot\!\left(D_\beta\,e^{-\Ho r}D_\alpha^\dag\right)\\
	&=\int_{\lowerintsub{G/Z}}\!\!\!\!d^5\!\mu(Zx)\,C_T(Zx)\,\Odot\!\left(D_\beta\,e^{-\Ho r}D_\alpha^\dag\right)\,,\label{ZTunravelC}
\end{align}
where we use the coset measure $d^5\!\mu(Zx)$ of equation~\ref{d5Zx} and define the \emph{Cartan-section reduced distribution function},
\begin{equation}\label{CTZx}
	\boxed{
		\vphantom{\Bigg(}
		\hspace{15pt}
		C_T(Zx)\equiv\int_Z\! d\phi\,d\ell \, D_T(x) e^{-2\ell}\,.
		\hspace{15pt}
		\vphantom{\Bigg)}
	}	
\end{equation}
Readers uncomfortable with the coset notation can think that in this equation, $x=e^{i\Omega\phi}e^{-\Omega\ell}D_\beta\,e^{-\Ho r}D_\alpha^\dagger$ and $Zx=D_\beta\,e^{-\Ho r}D_\alpha^\dagger$.  Even more prosaically, one can regard $D_T$ as being a function of all seven Cartan co\"ordinates and $C_T$ as being a function of five of them, the ruler~$r$ and the complex phase-space co\"ordinates $\beta$ and $\alpha$.  Our excuse---quite a good excuse, really---for using the co\"ordinate-independent coset notation is that we will elaborate on this distribution function and another one in section~\ref{PathIntegral}, but working there mainly in Harish-Chandra co\"ordinates.

Equation~\ref{ZTunravelC} is a new unraveling of the SPQM instrument, which we call the \emph{reduced SPQM instrument}, with instrument elements
\begin{align}
d^5\!\mu(Zx)\,C_T(Zx)\,\Odot(D_\beta\,e^{-\Ho r}D_\alpha^\dag)\,,
\end{align}
in which the Cartan reduced distribution $C_T(Zx)$ is conjugate to the operation~$\Odot(D_\beta\,e^{-\Ho r}D_\alpha^\dag)$.  If $\Ho$ is quantized in the standard way, we have at late times,
\begin{align}
D_\beta\,e^{-\Ho r}D_\alpha^\dagger\big|_{T\gg1/\kappa}\sim e^{-\kappa T}D_{\beta_T}\proj0 D_{\alpha_T}^\dagger=e^{-\kappa T}\ket{\beta_T}\bra{\alpha_T}\,.
\end{align}
This makes clear that $C_T(Zx)$ is the elephant in the room hinted at in section~\ref{trans}: it is the phase-space weighting function that is crucial for the POVM completeness relation, which takes the form
\begin{align}\label{completenessCT}
1_\sH=\int_G\!d^7\!\mu(x)\,D_T(x)\,x^\dagger x=\int_{\lowerintsub{G/Z}}\!\!\!\!\!d^5\!\mu(Zx)\,C_T(Zx)\,D_\alpha^\dagger e^{-\Ho 2r}D_\alpha\,.
\end{align}
As an elephant, however, $C_t(Zx)$ is not normalized to unity---indeed, its normalization is ill-defined.  Moreover, $C_t(Zx)$ is not the weight function whose moments are those of the Cartan phase-point variables $\beta$ and $\alpha$ according to the SDE solutions~\ref{solvedSDEs}.  The distribution function that does give these moments is the straight marginal of $D_T(x)$ over the center~$Z$,
\begin{equation}
	D_T(Zx) \equiv \int_Z d\phi\,d\ell \, D_T(x)\,.
\end{equation}
This distribution is normalized and has finite moments, those coming from the SDE solutions~\ref{solvedSDEs}.  For those very reasons, however, $D_T(Zx)$ cannot possibly give rise to a POVM completeness relation; it will not be seen again in this paper.

To derive an evolution equation for $C_t(Zx)$ of equation~\ref{CTZx}, one takes its time derivative, substitutes the FPK equation~\ref{FPKE} into the integral, and pushes the FPK forward generator $\Delta$ of equation~\ref{Cartanforward} through the center integrals by integrating by parts.  Integration by parts on $\phi$ gets rid of the derivatives with respect to $\phi$, and integration by parts on $\ell$ translates to substituting $\partial_\ell\rightarrow2$, resulting in the partial differential equation (PDE),
\begin{equation}
	\boxed{
		\vphantom{\Bigg(}
		\hspace{15pt}
		\frac{1}{\kappa}\frac{\partial}{\partial t}C_t(Zx) = \Big({-}2\partial_r-2\coth r+ \nabla^*\nabla\Big)[C_t](Zx)\,.\label{Bloch}
		\hspace{15pt}
		\vphantom{\Bigg)}
	}	
\end{equation}
where recall that $\nabla^*\nabla=\frac12(\nabla_1^2+\nabla_2^2)$.  This PDE is ballistic in the ruler~$r$---solution proportional to $\delta(r-2\kappa t)$---and Gaussian preserving in the phase-space variables, but as a consequence of the $-2\coth r$ term, the PDE does not preserve normalization~\cite{Kac1947a,Kac1959a,Chaichian2001a}.

To solve for $C_t(Zx)$ requires knowing $D_{dt}(x)$, which is, when $dt\rightarrow0$, the $\delta$-function $\delta(x,1)$.  This can be done fairly easily by evaluating the Cartan-co\"ordinate solutions~\ref{solvedSDEs} and~\ref{solvedellphi} at $T=dt$.  We perform that task in appendix~\ref{deltax1}, where we also identify all the $\delta$-function forms for initial conditions.  The result for $D_{dt}(x)$ is
\begin{align}
D_{dt}(x)=
\delta(\phi)\,\delta\Big(\ell+\smallfrac12\kappa\,dt|\beta+\alpha|^2\Big)\,
\frac{1}{\sinh^2\!r}\delta(r-2\kappa\,dt)\,
2\pi\frac{\kappa\,dt}{\pi}e^{-\kappa\,dt|\beta+\alpha|^2}\,2\pi\delta^2(\beta-\alpha)\,.
\end{align}
The distinctive feature of this distribution is the wide, normalized Gaussian in $\beta+\alpha$, which limits to a uniform distribution in $\beta+\alpha$ as $dt\rightarrow0$.  What the wide Gaussian is about is that the identity is represented by Cartan co\"ordinates $\phi=\ell=r=0$ and $\beta=\alpha$, with $\beta+\alpha$ free to take on any complex value.  With this expression, it is easy to see that
\begin{equation}
	\boxed{
		\vphantom{\Bigg(}
		\hspace{15pt}
		C_{dt}(Zx)=\int_Z\! d\phi\,d\ell \, D_{dt}(x) e^{-2\ell}
		=\frac{2}{r}\delta(r-2\kappa dt)\,\pi\delta^2(\beta-\alpha)\,.
		\hspace{15pt}
		\vphantom{\Bigg)}
	}	
\end{equation}
The integral of this distribution has a zero from the $1/r$ behavior multiplying the $\sinh^2r$ in the measure $d^5\!\mu(Zx)$ and an infinity from the uniformity in $\beta+\alpha$; therefore, the normalization is ill defined.  This ill-defined normalization is, however, exactly what is needed to give a well-defined \hbox{POVM}.  For these reasons, $C_{dt}(Zx)$ does not limit to $\delta(Zx,Z1)$ as $dt\rightarrow0$, for which see appendix~\ref{deltax1}.

The initial condition $C_{dt}(Zx)$ is independent of $\beta+\alpha$, and the distribution $C_t(Zx)$ remains so under the PDE~\ref{Bloch}.  To see the consequences most clearly, it is useful to transform to sum and difference variables,
\begin{align}
\beta_{\pm}=\beta\pm\alpha\,,
\end{align}
in which the covariant derivative $\nabla$ of equation~\ref{nablac} becomes
\begin{align}
\nabla=\coth(r/2)\partial_{\beta_+}+\tanh(r/2)\partial_{\beta_-}\,.
\end{align}
Therefore the weight function evolves according to the PDE,
\begin{equation}
		\frac{\partial}{\partial t}C_t(Zx)
        = \kappa\Big({-}2\coth r-2\partial_r+\tanh^2(r/2)\partial_{\beta_-^*}\partial_{\beta_-}\Big)C_t(Zx)\,,
\end{equation}
with solution
\begin{equation}\label{CTZxsolution}
\boxed{
	\vphantom{\Bigg(}
	\hspace{15pt}
		C_T(Zx) =\frac{1}{\sinh r}\delta(r-2\kappa T)\,2\pi\frac{1}{\pi\Sigma_T}e^{-|\beta-\alpha|^2/\Sigma_T}\,.
        \hspace{15pt}
	\vphantom{\Bigg)}
}
\end{equation}
The width of the difference in phase points, $\Sigma_T$, satisfies the differential equation, $d\Sigma_t/dt=\kappa\tanh^2\!\kappa t$, with solution, given initial condition $\Sigma_0=0$,
\begin{equation}\label{SigmaT}
\boxed{
	\vphantom{\Bigg(}
	\hspace{15pt}
	\Sigma_T = \kappa T - \tanh\kappa T\,.
	\hspace{15pt}
	\vphantom{\Bigg)}
}
\end{equation}

In summary, the SPQM instrument can be considered as the \emph{reduced-SPQM-instrument unraveling},
\begin{equation}
	\boxed{
		\vphantom{\Bigg(}
		\hspace{15pt}
		\Z_T =\int d^5\!\mu(Zx) \,C_T(Zx) \,\Odot\!\left(D_\beta\,e^{-\Ho r}D_\alpha^\dag\right)\,,
		\hspace{15pt}
		\vphantom{\Bigg)}
	}
\end{equation}
with instrument elements
\begin{equation}
	\boxed{
		\vphantom{\Bigg(}
		\hspace{15pt}
		d^5\!\mu(Zx)\,C_T(Zx)\,\Odot(D_\beta\,e^{-\Ho r}D_\alpha^\dagger)=2\sinh2\kappa T\,dr\,\delta(r-2\kappa T)\,\frac{d^2\alpha}{\pi}\,
            \frac{d^2\beta}{\pi\Sigma_T}\,e^{-|\beta-\alpha|^2/\Sigma_T}\,\Odot(D_\beta\,e^{-\Ho r}D_\alpha^\dagger)\,,
		\hspace{15pt}
		\vphantom{\Bigg)}
	}
\end{equation}
where the width $\Sigma_T$ of the difference in phase points is given by equation~\ref{SigmaT}.
There are four notable features in the temporal behavior of the reduced SPQM instrument:
\begin{enumerate}
\item
The ruler $r$ (or purity parameter) is ballistic, which means that $e^{-\Ho r}$ collapses exponentially to $e^{-\kappa T}\proj0$ in the standard quantization. More generally, $D_\beta\,e^{-r\Ho}D_\alpha^\dagger$ collapses exponentially at late times to an outer product of coherent states, $e^{-\kappa T}\oprod{\beta}{\alpha}$.
\item
The dependence on the future and past phase-space parameters, $\beta$ and $\alpha$, is only in their difference.
\item
The distribution of the difference spreads out very slowly for small times, as $\Sigma_T \propto T^3$, and then for long times becomes normal diffusion, with $\Sigma_T \propto T$.
\item
There is a center normalization, $2\sinh 2\kappa T$, that increases over time.
\end{enumerate}
This center normalization is the focus of the next section, which finds that the elephant provides an alternative perspective on the quantum.

\subsection{POVM as an Alternative Perspective on the Quantum}\label{Alt}

The center normalization just mentioned is remarkable in that it is equivalent to traditional energy quantization.
Specifically, the completeness relation~\ref{completenessCT} for the SPQM process is
\begin{align}\label{POVMcompleteness}
	1_\Hb
    = 2\sinh2 \kappa T \int \frac{d^2\alpha}{\pi}D_\alpha\,e^{-\Ho 4\kappa T}D_\alpha^\dagger\int\frac{d^2\beta}{\pi\Sigma_T}e^{-|\beta-\alpha|^2/\Sigma_T}
    = 2\sinh2 \kappa T \int\frac{d^2\alpha}{\pi}D_\alpha\,e^{-\Ho 4\kappa T}D_\alpha^\dagger\,.
\end{align}
It is important to appreciate that for late times $T\gg1/\kappa$, when $e^{-\Ho 4\kappa T}$ collapses to $e^{-2\kappa T}\proj0$ in the standard quantization, this completeness relation becomes the coherent-state resolution of the identity of equation~\ref{csPOVM}:
\begin{align}
1_\Hb=\int\frac{d^2\alpha}{\pi}\proj\alpha\,.
\end{align}
The completeness relation says much more, however, when considered for arbitrary times $T$.  If $\Hb$ is an irreducible representation, Schur's lemma says that
\begin{align}
\int\frac{d^2\alpha}{\pi}D_\alpha\,e^{-\Ho 4\kappa T}D_\alpha^\dagger
=1_\Hb\tr\!\big(e^{-\Ho 4\kappa T}\big)\,.
\end{align}
The trace, which one recognizes as a partition function, is defined within the representation and is evaluated using traditional energy quantization as
\begin{align}
\tr\big(e^{-\Ho 4\kappa T}\big)
=\sum_{n=0}^{\infty} e^{-(n+\frac12)4\kappa T}
=\frac{1}{2\sinh2\kappa T}\,.
\end{align}

The center normalization in the POVM completeness relation thus evaluates the partition function of $e^{-\Ho 4\kappa T}$ without using traditional energy quantization.  This is to be expected in view of the Stone-von-Neumann theorem, but expected though it is, please appreciate how different the setting of this paper is from the original ideas of energy quantization and thermal equilibrium.  Remember that here the operator $\Ho$ comes from the trace-preserving character of the instrument.  It is not the energy of the system; indeed we have explicitly eschewed any notion of system energy or Hamiltonian.  The operator $\Ho$ plays the role of a ``dissipator,''  specifically a dissipator that damps the POVM to the coherent states, but there is no notion of energy associated with this dissipation.  The coherent states and $\Ho$ arise within a group structure constructed solely from the measured observables, $Q$ and $P$.  Moreover, the parameter conjugate to $\Ho$, the ruler $r$, is quite literally time, rather than an inverse temperature. It seems remarkable that this result holds, from the completeness of the SPQM POVM, without any assumption of a Hamiltonian, a ground state, or thermal equilibrium.

\vfill\pagebreak

\section{Reduced Distribution Functions and Feynman-Kac Path Integrals}\label{PathIntegral}

This section further considers reduced distribution functions, their path-integral expressions and diffusion equations, and their relation to SDEs.
The path-integral expression,
\begin{equation}\label{KOD}
	D_T(x) = \int \D\mu\!\left[dw_{[0,T)}\right] \delta(x,\gamma[dw_{[0,T)}])\,,
\end{equation}
is generally considered to be a Feynman-Kac formula~\cite{Kac1947a,Kac1959a,Chaichian2001a} for the associated diffusion equation.  Our analysis is rooted in the path integral~\ref{totalZT} for the overall SPQM quantum operation,
\begin{align}\label{totalop}
	\Z_T = \int \D\mu\!\left[dw_{[0,T)}\right]\Odot\Big(R\big(\gamma\big[dw_{[0,T)}\big]\big)\Big)\,.
\end{align}
Equation \ref{KOD} expresses the relation between a complex Wiener path $dw_{[0,T)}$ and a point in the group manifold~IWH, $\gamma\big[dw_{[0,T)}\big]$ and in turn the overall Kraus operator $R(\gamma\big[dw_{[0,T)}\big])$ written as a time-ordered product of incremental Kraus operators.
Equation \ref{KOD} defines the Kraus-operator distribution function (KOD) as the amalgamation of all paths that lead to the same Kraus operator.
The KOD inherits the path-integral expression from equation \ref{totalop}, and from this path integral one can derive a diffusion equation for the \hbox{KOD}.
The reason the path integral is called a Feynman-Kac formula is that everybody after Kac thinks about going in the opposite direction, starting with the diffusion equation and formulating an equivalent path integral.

Section~\ref{FKformulas} reviews the Cartan-section reduced distribution $C_T(Zx)$, introduces the Harish-Chandra-section reduced distribution $B_T(Zx)$ of Eq.~(\ref{BTZxPathIntegralintro}), and shows that these two are related by a positive gauge transformation.
Section~\ref{HCRDF} defines a normalized version of the Harish-Chandra reduced distribution, denoted by $\tilde B_T(Zx)$, and finds its path-integral expression in terms of a modified path-integration measure in which the outcome increments are correlated.
Section~\ref{PDEHCRDF} formulates the diffusion equations for $B_T(Zx)$ and $\tilde B_T(Zx)$, and section~\ref{SDEHCRDF} solves the path integral for $\tilde B_T(Zx)$ using the stochastic integrals for the Harish-Chandra phase-space co\"ordinates.

\subsection{Feynman-Kac Formulas}\label{FKformulas}

The \emph{ur} KOD of equation~\ref{KOD},
\begin{equation}\label{KODF2}
	D_T(x)\equiv\int \D\mu\!\left[dw_{[0,T)}\right]\delta\Big(x,\gamma\!\left[dw_{[0,T)}\right]\Big)\,,
\end{equation}
unravels $\Z_T$ over the universal domain of $G=\textrm{IWH}$,
\begin{equation}\label{ZTunravelD2}
		\Z_T = \int_G d^7\!\mu(x)\,D_T(x)\;\Odot(x)\,,
\end{equation}

Reduced distributions are defined on $G/Z = \mathrm{RIWH}$.  The first of these reduced distributions, introduced in equation~\ref{CTZx} as the Cartan-section reduced distribution,
\begin{equation}\label{CTZx2}
	C_T (Zx)\equiv\int_Z\!d\phi\,d\ell\, D_T(x)e^{-2\ell}\,,
\end{equation}
can also be defined by the Feynman-Kac path integral,
\begin{equation}
C_T (Zx)
= \int \D\mu\!\left[dw_{[0,T)}\right]e^{-2\ell[dw_{[0,T)}]}\,\delta\Big(Zx,Z\gamma\!\left[dw_{[0,T)}\right]\Big)\,,\label{CTPathIntegral}
\end{equation}
where $\ell[dw_{[0,T)}]=\ell_T$ is the solution of the SDE~\ref{dl2} for the Cartan center co\"ordinate $\ell$,
\begin{align}
-\ell[dw_{[0,T)}]=\int_0^{T_-}\!\Big(\big(\coth2\kappa t-2|\beta_t|^2\,\big)\kappa\,dt+\beta_t\sqrt\kappa\,dw_t^*+\beta_t^*\sqrt\kappa\,dw_t\Big)\,,
\end{align}
with the notation here emphasizing that this solution is a functional of the sample-path of Wiener outcome increments.  As was noted in equation~\ref{solvedellphi}, one can use the transformation from Harish-Chandra co\"ordinates to write
\begin{align}
-\ell[dw_{[0,T)}]=f_T-s[dw_{[0,T)}]\,,
\end{align}
where $f_T$ is the function of $G/Z=\textrm{RIWH}$ given in equation~\ref{f},
\begin{align}
2f(Zx)=e^{-r/2}\sinh(r/2)\,|\beta+\alpha|^2+e^{-r/2}\cosh(r/2)\,|\beta-\alpha|^2\,,
\end{align}
with the ruler and the phase-plane co\"ordinates evaluated at time~$T$, and
\begin{align}
-2s_T=-2s[dw_{[0,T)}]=\int_0^{T_-}\!\!\int_0^{T_-}\!\!\kappa\,dw_t^*\,dw_s\,e^{-2\kappa |t-s|}
\end{align}
is the stochastic integral~\ref{sT} for the Harish-Chandra center co\"ordinate $s$ (derived in appendix~\ref{zsolution}).  The function $C_T(Zx)$ unravels $\Z_T$ onto $\textrm{RIWH}=G/Z$ as in equation~\ref{ZTunravelC},
\begin{align}
	\Z_T =\int_{\lowerintsub{G/Z}}\!\!\!\!d^5\!\mu(Zx)\,C_T(Zx)\,\Odot\!\left(D_\beta\,e^{-\Ho r}D_\alpha^\dag\right)\,,\label{ZTunravelC2}
\end{align}
and $C_T(Zx)$ satisfies the FPK equation~\ref{Bloch},
\begin{equation}
	\frac{1}{\kappa}\frac{\partial}{\partial t}C_t(Zx) = \Big({-}2\partial_r-2\coth r+\nabla^*\nabla\Big)C_t(Zx)\,.\label{Bloch2}
\end{equation}

There is another natural reduced distribution, conjugate to the Harish-Chandra section,
\begin{align}\label{BTZx}
B_T (Zx)
&\equiv \int_Z\!d\psi\,ds\, D_T(x)e^{-2s}\\
&=\int \D\mu\!\left[dw_{[0,T)}\right]e^{-2s[dw_{[0,T)}]}\,\delta\Big(Zx,Z\gamma\!\left[dw_{[0,T)}\right]\Big)\,.\label{BTZxPathIntegral}
\end{align}
This \emph{Harish-Chandra reduced distribution function}, $B_T(Zx)$, unravels $\Z_T$ as
\begin{align}
	\Z_T =\int_{\lowerintsub{G/Z}}\!\!\!\!d^5\!\mu(Zx)\,B_T(Zx)\,\Odot\!\left(e^{a^\dagger\nu}e^{-\Ho r}e^{a\mu^*}\right)\,.\label{ZTunravelB}
\end{align}
The two reduced distribution functions are equivalent to one another by a positive gauge transformation~\cite{Chaichian2001a},
\begin{align}
	C_T (Zx)
    &=\int_Z\!d\phi\,d\ell\,D_T(x)e^{-2\ell}\\
    &=\int_Z\!d\psi\,ds\, D_T(x)e^{-2[s-f(Zx)]}\\
    &=e^{2f(Zx)}B_T(Zx)\,.
\end{align}
In the lingo of Feynman-Kac formulas~\cite{Chaichian2001a}, $2f(Zx)$ is the ``convective pressure.''

Section~\ref{solveFPK} introduced the Cartan reduced distribution~$C_T(Zx)$ and showed that it is the distribution that expresses POVM completeness.  The price for relevance to POVM completeness is that $C_T(Zx)$ has ill-defined normalization and is disconnected from the stochastic-integral solutions for the phase-space variables.  The next three sections investigate the Harish-Chandra reduced distribution~$B_T(Zx)$.  It is clear that $B_T(Zx)$ is not the right distribution for addressing POVM completeness, because of the Gaussian gauge function~$e^{-2f(Zx)}$, but this Gaussian gauge transformation is just what is needed to get a Gaussian distribution function that can be normalized and whose normalized version can be evaluated from the moments of the phase-space variables, albeit as we shall see, moments defined relative to a modified path-integration measure.

\subsection{Normalized Harish-Chandra Reduced Distribution Function and Modified Path-Integration Measure}
\label{HCRDF}

The Feynman-Kac formula~\ref{BTZxPathIntegral} for $B_T(Zx)$ suggests that we combine $e^{-2s[dw_{[0,T)}]}$ with the Weiner measure~\ref{complexWienermeasure},
\begin{align}
\D\mu\!\left[dw_{[0,T)}\right]e^{-2s[dw_{[0,T)}]}
&=\left(\prod_{k=0}^{T/dt-1}\!d^{\,2}\!\big(dw_{k dt}\big)\right)\!\!
        \left(\frac{1}{\pi dt}\right)^{T/dt}
        \exp\!\left({-}\int_0^{T_-}\frac{|dw_t|^2}{dt}+\int_0^{T_-}\!\!\int_0^{T_-}\!\!\kappa\,dw_t^*\,dw_s\,e^{-2\kappa |t-s|}\right)\,.
\end{align}
The quadratic functional in the exponential can be written as
\begin{align}
-\int_0^{T_-}\frac{|dw_t|^2}{dt}+\int_0^{T_-}\!\!\int_0^{T_-}\!\!\kappa\,dw_t^*\,dw_s\,e^{-2\kappa |t-s|}
&=-\frac{1}{dt}\int_{t=0}^{T_-}\int_{s=0}^{T_-}dw_t^*\,dw_s\Big(\delta_{ts}-\kappa\,dt\,e^{-2\kappa|t-s|}\Big)\\
&=-\frac{1}{dt}\sum_{k=0}^{N-1}\sum_{l=0}^{N-1}dw_k^*\,dw_l\Big(\delta_{kl}-\kappa\,dt\,e^{-2\kappa dt\,|k-l|}\Big)\,.
\end{align}
When converting between stochastic integrals and sums, we use $t_k=k\,dt$ ($t_N=T=N\,dt$) and $dw_k=dw_{kdt}=dw_{t_k}$.  Now define the real, symmetric, and positive $N\times N$ matrix $M_T$, whose matrix elements are
\begin{align}
M_{kl}=\delta_{kl}-\kappa\,dt\,e^{-2\kappa\,dt\,|k-l|}\,,
\end{align}
It is elegant for various formal expressions to introduce a continuous version of $M_T$, but we do not bother with that here, since we work with the sums that the stochastic integrals represent.  Notice that $M_T$ is a Toeplitz matrix, that is, $M_{k+j,l+j}=M_{kl}$.   Putting this together, we have
\begin{align}
\D\mu\!\left[dw_{[0,T)}\right]e^{-2s[dw_{[0,T)}]}
&=\left(\prod_{k=0}^{N-1}\!d^{\,2}\!\big(dw_{k dt}\big)\right)\!\!
        \left(\frac{1}{\pi dt}\right)^N
        \exp\!\left(-\frac{1}{dt}\sum_{k=0}^{N-1}\sum_{l=0}^{N-1}dw_k^*\,M_{kl}\,dw_l\right)\,,
\end{align}
which integrates over the Wiener outcome paths to
\begin{align}
\int\D\mu\!\left[dw_{[0,T)}\right]e^{-2s[dw_{[0,T)}]}=\left(\frac{1}{\pi dt}\right)^N\frac{\pi^N}{\det(M_T/dt)}=\frac{1}{\det M_T}\,.
\end{align}

This prompts us to define a new (normalized, zero-mean) Gaussian measure on the Wiener outcome paths,
\begin{align}
\sD\mu_M\!\left[dw_{[0,T)}\right]
&\equiv\det M_T\,\sD\mu\!\left[dw_{[0,T)}\right]e^{-2s[dw_{[0,T)}]}\\
&=\left(\prod_{k=0}^{N-1} d^{\,2}\!\big(dw_k\big)\right)
\left(\frac{\det M_T}{(\pi dt)^N}\right)
\exp\!\left(-\frac{1}{dt}\sum_{k=0}^{N-1}\sum_{l=0}^{N-1}dw_k^*\,M_{kl}\,dw_l\right)\label{DmuM}\,.
\end{align}
Relative to this measure, the outcome increments are correlated,
\begin{align}\label{dwdwM}
\langle dw_k^*\,dw_l\rangle_M=dt\,(M_T^\inv)_{kl}\,.
\end{align}
Notice that the increment correlation $\langle dw_k^*\,dw_l\rangle_M$, with $k$ and $l$ held fixed, changes as the total time $T$ changes.  The inverse matrix $M_T^\inv$ matrix is real, symmetric, and positive, all properties inherited from $M_T$.  The inverse does not inherit the Toeplitz property of $M_T$.  That $M_T$ is Toeplitz implies that it is persymmetric, that is, symmetric across the anti-diagonal; $M_T^\inv$ does inherit the persymmetry, which turns out to have an important consequence in section~\ref{SDEHCRDF}.

Returning to the Harish-Chandra reduced distribution~\ref{BTZx}, we see that its normalization can be written in several ways:
\begin{align}\label{normBTZx1}
\sN_T
&\equiv\int d^5\!\mu(Zx)\,B_T (Zx)\\
&=\int d^5\!\mu(Zx)\,C_T (Zx)e^{-2f(Zx)}\label{normBTZx2}\\
&=\int_Z\!d^7\!\mu(x)\,D_T(x)e^{-2s}\\
&=\int \D\mu\!\left[dw_{[0,T)}\right]e^{-2s[dw_{[0,T)}]}=\frac{1}{\det M_T}\label{normBTZx3}\,.
\end{align}
The normalized version of the Harish-Chandra reduced distribution is
\begin{align}\label{nBTZx}
\tilde B_T(Zx)\equiv\frac{1}{\sN_T}B_T(Zx)=\frac{e^{-2f(Zx)}}{\sN_T}C_T(Zx)\,.
\end{align}
Section~\ref{PDEHCRDF} finds the diffusion equations satisfied by the reduced Harish-Chandra distributions, and section~\ref{SDEHCRDF} uses the path integral for the normalized distribution,
\begin{align}\label{nBTZxPathIntegral}
	\boxed{
		\vphantom{\Bigg(}
		\hspace{15pt}
        \tilde B_T(Zx)
        =\int\D\mu_M\!\left[dw_{[0,T)}\right]\delta\Big(Zx,Z\gamma\!\left[dw_{[0,T)}\right]\Big)\,,
	   \hspace{15pt}
	   \vphantom{\Bigg)}
	}
\end{align}
to evaluate $\tilde B_T(Zx)$ from the SDE solutions for the Harish-Chandra co\"ordinates.

\subsection{Diffusion Equation for Harish-Chandra Reduced Distribution Function}
\label{PDEHCRDF}

The easiest way to get to the diffusion equation for $B_t(Zx)$ is to return to the FPK equation~\ref{FPKE} for $D_t(x)$, to write the FPK forward generator $\Delta$ in Harish-Chandra co\"ordinates, and then to marginalize over the center to get a PDE for $B_t(Zx)$.

By using the frame transformations~\ref{DerTrans}, it is easy to express the terms of the forward generator in Harish-Chandra co\"ordinates,
\begin{align}
	2\Rinv{\Ho}
    &=-2\partial_{r}+2\big(\nu_1\partial_{\nu_1}+\nu_2\partial_{\nu_2}\big)\\
    &=-2\partial_{r}-4+2\big(\partial_{\nu_1}\nu_1+\partial_{\nu_2}\nu_2\big)\,,
\end{align}
\begin{align}
\frac12\Rinv{Q}\Rinv{Q}
&=\frac12\!\left(\nabla_1-\smallfrac12\nu_1\partial_s+\smallfrac12\nu_2\partial_\psi\right)^2\\
&=\frac12\big(\nabla_1-\smallfrac12\nu_1\partial_s\big)^2+\frac12\partial_\psi A_Q\\
&=\frac12\Big(\nabla_1^2-\smallfrac12\partial_s\big(\nabla_1\nu_1+\nu_1\nabla_1\big)+\smallfrac14\nu_1^2\partial_s^2\Big)+\frac12\partial_\psi A_Q\\
&=\frac12\Big(\nabla_1^2-\smallfrac12\partial_s\big(2\nabla_1\nu_1-1\big)+\smallfrac14\nu_1^2\partial_s^2\Big)+\frac12\partial_\psi A_Q\\
&=\frac14\partial_s\bigg(1+\frac12\nu_1^2\partial_s\bigg)+\frac12\nabla_1^2-\frac12\partial_s\nabla_1\nu_1+\frac12\partial_\psi A_Q\,,
\end{align}
\begin{align}
\frac12\Rinv{P}\Rinv{P}
&=\frac12\!\left(\nabla_2-\smallfrac12\nu_2\partial_s-\smallfrac12\nu_1\partial_\psi\right)^2\\
&=\frac12\big(\nabla_2-\smallfrac12\nu_2\partial_s\big)^2+\frac12\partial_\psi A_P\\
&=\frac14\partial_s\bigg(1+\frac12\nu_2^2\partial_s\bigg)+\frac12\nabla_2^2-\frac12\partial_s\nabla_2\nu_2+\frac12\partial_\psi A_P\,,
\end{align}
where the terms
\begin{align}
A_Q&=\nu_2\big(\nabla_1-\smallfrac12\nu_1\partial_s\big)+\smallfrac14\nu_2^2\partial_\psi\,,\\
A_P&=-\nu_1\big(\nabla_2-\smallfrac12\nu_2\partial_s\big)-\smallfrac14\nu_1^2\partial_\psi\,,
\end{align}
are independent of $\psi$ and commute with $\partial_\psi$.
Putting these expressions together, the FPK forward generator~\ref{ForGen} in Harish-Chandra co\"ordinates is
\begin{equation}
	\boxed{
		\vphantom{\Bigg(}
		\hspace{15pt}
		\Delta= -2\partial_r -4 + \frac12\partial_s\bigg(1+\frac{\nu_1^2+\nu_2^2}{4}\partial_s\bigg)+\frac12\Big(\nabla_1^2+\nabla_2^2\Big)
		+2\partial_{\nu_1}\nu_1+2\partial_{\nu_2}\nu_2-\frac12\partial_s\big(\nabla_1\nu_1+\nabla_2\nu_2\big)+ \partial_\psi A\,,
		\hspace{15pt}
		\vphantom{\Bigg)}
	}\label{HCforward}
\end{equation}
where $A=\frac12(A_Q+A_P)$.

To derive an evolution equation for $B_t(Zx)$ of equation~\ref{BTZx}, one follows the procedure outlined for $C_t(Zx)$ at equation~\ref{Bloch}, using here the rules that integration by parts on $s$ and $\psi$ make the substitutions $\partial_s\rightarrow2$ and $\partial_\psi\rightarrow0$.  The resulting PDE for $B_t(Zx)$ is
\begin{align}
		\frac{1}{\kappa}\frac{\partial}{\partial t}B_t(Zx)
        = \bigg({-}2\partial_r-3+\frac{\nu_1^2+\nu_2^2}{2}+\tilde\Delta\bigg)B_t(Zx)\,,\label{BlochHCreal}
\end{align}
where for brevity, we define a reduced generator for the phase-space-variable derivatives,
\begin{align}
\tilde\Delta
&\equiv\big(2\partial_{\nu_1}-\nabla_1\big)\nu_1+\big(2\partial_{\nu_2}-\nabla_2\big)\nu_2+\frac12\Big(\nabla_1^2+\nabla_2^2\Big)\\
&=\big(2\partial_\nu-\nabla\big)\nu+\big(2\partial_{\nu^*}-\nabla^*\big)\nu^*+\nabla^*\nabla\,.
\end{align}
Notice that
\begin{align}
2\partial_{\nu_j}-\nabla_j&=\partial_{\nu_j}-e^{-r}\partial_{\mu_j}\,,\\
2\partial_\nu-\nabla&=\partial_\nu-e^{-r}\partial_\mu\,.
\end{align}
Converting fully to complex phase-space co\"ordinates puts the PDE in the form
\begin{equation}
	\boxed{
		\vphantom{\Bigg(}
		\hspace{15pt}
		\frac{1}{\kappa}\frac{\partial}{\partial t}B_t(Zx)
        = \Big({-}2\partial_r-3+|\nu|^2+\tilde\Delta\Big)B_t(Zx)\,.\label{BlochHCcomplex}
		\hspace{15pt}
		\vphantom{\Bigg)}
	}	
\end{equation}
This PDE is ballistic in the ruler~$r$---solution proportional to $\delta(r-2\kappa t)$---and Gaussian preserving in the phase-space variables, but as a consequence of the term $-3+|\nu|^2$, it does not preserve normalization~\cite{Kac1947a,Kac1959a,Chaichian2001a}. The effect of the positive gauge transformation from $C_t(Zx)$ to $B_t(Zx)$ is two-fold: (i)~the norm-nonconserving ``potential'' term changes character, from a ruler-dependent $-2\coth r$ in the PDE~\ref{Bloch2} for $C_t(Zx)$ to a term $-3+|\nu|^2$ in the PDE~\ref{BlochHCcomplex} for $B_t(Zx)$, which depends on the posterior phase-space variable $\nu$; (ii)~there are first-derivative, ``vector-potential'' terms in the PDE for $B_t(Zx)$, corresponding to the Ornstein-Uhlenbeck behavior~\ref{OUP} of $\nu$, whereas there are no such terms in the PDE for $C_T(Zx)$.

We turn now to the task of converting the PDE~\ref{BlochHCcomplex} to the normalized distribution~$\tilde B_t(Zx)$.
The initial condition for the PDE~\ref{BlochHCcomplex} comes from inserting the Harish-Chandra $D_{dt}(x)$ of equation~\ref{DdtHC} into $B_{dt}(Zx)$ as it is expressed in the integral~\ref{BTZx} specialized to $T=dt$.  Taking the limit $dt\rightarrow0$, one finds the expected result that $B_0(Zx)$ is the $\delta$-function on $G/Z$ of equation~\ref{deltaZxZ1HC}:
\begin{align}
B_0(Zx)=\delta(Zx,Z1)=\delta(r)\,2\pi\delta^2(\nu)\,2\pi\delta^2(\mu)\,.
\end{align}
This is expected because the identity is represented uniquely in Harish-Chandra co\"ordinates by $\psi=s=r=0$, $\nu=\mu=0$.  The initial condition means that $B_t(Zx)$ is initially normalized to unity; thus the normalization factor $\sN_t$ has initial value $\sN_0=1$, and the normalized distribution has the same initial condition,
\begin{align}\label{nBTZxinitialdelta}
\tilde B_0(Zx)=B_0(Zx)=\delta(Zx,Z1)\,.
\end{align}

The normalization factor,
\begin{align}
\sN_T\equiv\int d^5\!\mu(Zx)\,B_T (Zx)=\int e^{2r}\,dr\,\frac{d^2\nu}{2\pi}\,\frac{d^2\mu}{2\pi}\,B_T(Zx)\,.
\end{align}
satisfies the differential equation,
\begin{align}
\frac{1}{\kappa}\frac{d\sN_t}{dt}
&=\int e^{2r}\,dr\,\frac{d^2\nu}{2\pi}\,\frac{d^2\mu}{2\pi}\,\Big({-}2\partial_r-3+|\nu|^2\Big)B_t(Zx)\\
&=\sN_t\int d^5\!\mu(Zx)\,\big(1+|\nu|^2\big)\tilde B_t(Zx)\,,
\end{align}
where the reader should notice that integration by parts on the ruler becomes the rule $\partial_r\rightarrow-2$.
This differential equation assumes the form
\begin{align}\label{dlnNt}
\frac{1}{\kappa}\frac{d\ln\sN_t}{dt}=1+n_t\,,
\end{align}
where
\begin{align}\label{nt}
n_t\equiv\langle|\nu_t|^2\rangle_M=\int d^5\!\mu(Zx)\,|\nu|^2\tilde B_t(Zx)
\end{align}
is the second moment of $\nu$ relative to the normalized distribution $\tilde B_t(Zx)$.  We place a subscript $M$ on this moment because we can use the path-integral expression~\ref{nBTZxPathIntegral} for $\tilde B_T(Zx)$ to re\"express it as
\begin{align}
n_T=\langle|\nu_T|^2\rangle_M=\int\D\mu_M\!\left[dw_{[0,T)}\right]\,|\nu[dw_{[0,T)}]|^2\,,
\end{align}
where $\nu[dw_{[0,T)}]=\nu_T$ is the stochastic-integral solution~\ref{OUPsolution} for $\nu$.  Once the stochastic integral is plugged into this equation, the correlations of the Wiener increments are evaluated according to the modified measure $\D\mu_M\!\left[dw_{[0,T)}\right]$, that is, as in equation~\ref{dwdwM}.

The PDE for the normalized reduced distribution~$\tilde B_t(Zx)$ of equation~\ref{nBTZx} now follows as
\begin{align}
	\frac{1}{\kappa}\frac{\partial}{\partial t}\tilde B_t(Zx)
    =\left(\!{-}\frac{1}{\kappa}\frac{d\ln\sN_t}{dt}-2\partial_r-3+|\nu|^2+\tilde\Delta\right)\!\tilde B_t(Zx)\,.
\end{align}
Inserting equation~\ref{dlnNt} gives
\begin{equation}
	\boxed{
		\vphantom{\Bigg(}
		\hspace{15pt}
    	\frac{1}{\kappa}\frac{\partial}{\partial t}\tilde B_t(Zx)
        =\Big({-}2\partial_r-4+|\nu|^2-n_t+\tilde\Delta\Big)\tilde B_t(Zx)\label{PDEnBTZx}\,.
		\hspace{15pt}
		\vphantom{\Bigg)}
	}	
\end{equation}
It is easy to see how this equation preserves normalization.  The presence of the moment~$n_t$, essential for normalization, makes the equation nonlinear, but it can still be solved easily.

To solve for $\tilde B_t(Zx)$, one notes that the PDE is ballistic in the ruler~$r$ and Gaussian preserving in the phase-space variables. Thus the solution has the form $\tilde B_t(Zx)=e^{-2r}\delta(r-2\kappa t)\Phi_t(Zx)$, where $\Phi_t(Zx)$ is a normalized, zero-mean Gaussian in the phase-space variables $\nu$ and $\mu$.  The derivatives in the PDE~\ref{PDEnBTZx} are invariant under complex conjugation and under simultaneous rephasing of the phase-space variables, that is, $\nu\rightarrow\nu e^{i\chi}$ and $\mu\rightarrow\mu^{i\chi}$; it is productive to think of the invariance under complex conjugation as invariance under the change of phase-space co\"ordinates $\nu\leftrightarrow\nu^*$ and $\mu\leftrightarrow\mu^*$.  It is useful to appreciate that all the diffusion equations in this paper share these invariance properties.  The invariance under simultaneous rephasing implies that $\Phi_t(Zx)$ is a zero-mean Gaussian, since it starts from a zero-mean $\delta$-function initial condition.  It further implies that $\Phi_t(Zx)$ is determined by the three nonzero second moments of the phase-space variables: $n_t$ of equation~\ref{nt} and
\begin{align}
m_t&\equiv\langle|\mu_t|^2\rangle_M=\int d^5\!\mu(Zx)\,|\mu|^2\tilde B_t(Zx)\,,\label{mt}\\
q_t&\equiv\langle\nu_t^*\mu_t\rangle_M=\langle\mu_t^*\nu_t\rangle_M=\int d^5\!\mu(Zx)\,\nu^*\mu\,\tilde B_t(Zx)\label{qt}\,,
\end{align}
with the invariance under complex conjugation implying that $q_t$ is real.  This form of the solution for $\tilde B_t(Zx)$ in hand, one derives from the PDE~\ref{PDEnBTZx} first-order temporal differential equations for the second moments $n_t$, $m_t$, and $q_t$.  Just as the ordinary differential equation (ODE) for the normalization factor involves a second moment, so the equations for the second moments involve fourth moments.  The Gaussian form of the solution relates the fourth moments to second moments, thus closing the system of differential equations.  The resulting three ODEs, for the derivatives of $n_t$, $m_t$, and $q_t$, have terms that are constant, linear, and quadratic in the moments; the presence of the quadratic terms makes these ODEs (coupled) Riccati equations.  The last step is to solve the three Riccati equations, with initial conditions $n_0=m_0=q_0=0$, which are implied by the $\delta$-function initial condition for $\tilde B_0(Zx)$.  This gives the solution for $\tilde B_t(Zx)$.  By integrating to find $\sN_T$, using the solution for $n_t$, one can backtrack to find $B_T(Zx)$ and $C_T(Zx)$ from equation~\ref{nBTZx}.

We have carried out this procedure of deriving the Riccati equations from the PDE~\ref{PDEnBTZx}, but we do not present that derivation in this paper, preferring instead to use a different method, which derives the Riccati equations from the path integral for $\tilde B_T(Zx)$.  Implementing that method is the final task of this paper, carried out in the next section.

\subsection{Normalized Harish-Chandra Reduced Distribution Function from its Path Integral}
\label{SDEHCRDF}

The path integral~\ref{nBTZxPathIntegral} for $\tilde B_T(Zx)$ can be written explicitly in terms of the $\delta$-function in Harish-Chandra co\"ordinates,
\begin{align}\label{nBTZxPathIntegral2}
\tilde B_T(Zx)
&=\int\D\mu_M\!\left[dw_{[0,T)}\right]\,e^{-2r}\delta(r-r_T)\,2\pi\delta^2(\nu-\nu_T)\,2\pi\delta^2(\mu-\mu_T)\\
&=e^{-4\kappa T}\delta(r-2\kappa T)\int\D\mu_M\!\left[dw_{[0,T)}\right]\,2\pi\delta^2\big(\nu-\nu[dw_{[0,T)}]\big)\,2\pi\delta^2\big(\mu-\mu[dw_{[0,T)}]\big)\,,
\end{align}
where $r_T=2\kappa T$, $\nu_T=\nu[dw_{[0,T)}]$, and $\mu_T=\mu[dw_{[0,T)}]$ are the solutions~\ref{timeIncsolution}--\ref{GGWsolution} to the SDEs for the ruler and the  Harish-Chandra phase-space co\"ordinates.  The measure for the path integral is a (normalized, zero-mean) Gaussian measure in the outcome increments $dw_t$; thus the path integral gives a normalized, zero-mean Gaussian in $\nu$ and $\mu$, which is determined by the three (real) moments introduced in the previous section:
\begin{align}
n_T=\langle|\nu_T|^2\rangle_M
&=\int d^5\!\mu(Zx)\,|\nu|^2\tilde B_T(Zx)=\int\D\mu_M\!\left[dw_{[0,T)}\right]\,|\nu[dw_{[0,T)}]|^2\,,\\
m_T=\langle|\mu_T|^2\rangle_M
&=\int d^5\!\mu(Zx)\,|\mu|^2\tilde B_T(Zx)=\int\D\mu_M\!\left[dw_{[0,T)}\right]\,|\mu[dw_{[0,T)}]|^2\,,\\
q_T=\langle\nu_T^*\mu_T\rangle_M=\langle\mu_T^*\nu_T\rangle
&=\int d^5\!\mu(Zx)\,\nu^*\mu\,\tilde B_T(Zx)=\int\D\mu_M\!\left[dw_{[0,T)}\right]\,\nu[dw_{[0,T)}]^*\mu[dw_{[0,T)}]\,.
\end{align}
In this context, that the first moments and all the other second moments of the phase-space variables are zero follows from the fact that the measure is invariant under simultaneous rephasing of all the outcome increments; the reality of $q_T$ follows from the fact that the measure is unchanged under the transformation $dw_{[0,T)}\rightarrow dw^*_{[0,T)}$.  These properties come from the fact that $M_T$ is real and symmetric.

Plugging in the stochastic-integral solutions for $\nu_T$ and $\mu_T$ puts these moments into the following form:
\begin{align}
n_T=\langle|\nu_T|^2\rangle_M
&=\kappa\sum_{k,l=0}^{N-1}\langle dw_k^*\,dw_l\rangle_M\,e^{-2\kappa(T-t_k)}e^{-2\kappa(T-t_l)}\\
&=\kappa\,dt\,\sum_{k,l=0}^{N-1} e^{-2\kappa\,dt\,(N-k)}(M_T^\inv)_{kl}e^{-2\kappa\,dt\,(N-l)}\,,\label{nuTsquaredM}\\
m_T=\langle|\mu_T|^2\rangle_M
&=\kappa\sum_{k,l=0}^{N-1}\langle dw_k^*\,dw_l\rangle_M e^{-2\kappa t_k}e^{-2\kappa t_l}\\
&=\kappa\,dt\,\sum_{k,l=0}^{N-1}e^{-2\kappa\,dt\,k}(M_T^\inv)_{kl}e^{-2\kappa\,dt\,l}\,,\label{muTsquaredM}\\
q_T=\langle\nu_T^*\mu_T\rangle_M=\langle\mu_T^*\nu_T\rangle_M
&=\kappa\sum_{k,l=0}^{N-1}\langle dw_k^*\,dw_l\rangle_M e^{-2\kappa(T-t_k)}e^{-2\kappa t_l}\\
&=\kappa\,dt\,\sum_{k,l=0}^{N-1}e^{-2\kappa\,dt\,(N-k)}(M_T^\inv)_{kl}e^{-2\kappa\,dt\,l}\label{nuTmuTM}\,.
\end{align}
In the final form of $q_T$, it is evident that $q_T$ is real.  Notice that these expressions satisfy the zero initial conditions.

That $M_T$ is a Toeplitz matrix introduces an additional, quite important symmetry.  The inverse matrix does not inherit the Toeplitz property of $M$, but it does inherit a less restrictive property.  That $M_T$ is Toeplitz implies that it is symmetric about the anti-diagonal, that is, $M_{kl}=M_{N-1-l,N-1-k}$.  A matrix that is symmetric about the anti-diagonal is called \emph{persymmetric}.  It is easy to show that the inverse of a persymmetric matrix is persymmetric, so $M_T^\inv$ satisfies $(M_T^\inv)_{kl}=(M_T^\inv)_{N-1-l,N-1-k}$.  Persymmetry has a major consequence for the three moments, which comes from manipulating $m_T$:
\begin{align}
m_T
&=\kappa\,dt\,\sum_{k,l=0}^{N-1}e^{-2\kappa\,dt\,(N-1-k)}(M_T^\inv)_{N-1-k,N-1-l}\,e^{-2\kappa\,dt\,(N-1-l)}\\
&=e^{4\kappa\,dt}\,\kappa\,dt\,\sum_{k,l=0}^{N-1}e^{-2\kappa\,dt\,(N-k)}(M_T^\inv)_{lk}e^{-2\kappa\,dt\,(N-l)}\,.
\end{align}
We can set $e^{4\kappa\,dt}=1$ and thus conclude that
\begin{align}
m_T=n_T\,.
\end{align}
The complementarity in time of $\nu_T$ and $\mu_T$ was discussed in section~\ref{trans}: the post-measurement variable $\nu_T$ of equation~\ref{OUPsolution} depends exponentially on the end of the outcome register, and the POVM variable $\mu_T$ of equation~\ref{GGWsolution} depends exponentially on the beginning of the register.  The persymmetry of $M_T$ and $M_T^\inv$ expresses that the beginning and end of the record look the same statistically, so it is not surprising that the persymmetry implies that $m_T=n_T$.

The next step, deriving Ricatti equations for the three moments, involves incrementing the moments from $T$ to $T+dT$.  The tedious part of this task is determining how $M_T^\inv$ increments, that is, finding  $M_{T+dT}^\inv$, and that can be done using the Schur complement.  We relegate this entire task to appendix~\ref{momentRiccatis} and here skip directly to the coupled Ricatti ODEs, taken from equations~\ref{nTRiccatiapp}--\ref{qTRiccatiapp}:
\begin{align}
\frac{1}{\kappa}\frac{dn_T}{dT}&=(1-n_T)^2\,,\label{nTRiccati}\\
\frac{1}{\kappa}\frac{dm_T}{dT}&=\big(q_T+e^{-2\kappa T}\big)^2\,,\label{mTRiccati}\\
\frac{1}{\kappa}\frac{dq_T}{dT}&=-q_T(1-n_T)+e^{-2\kappa T}(1+n_T)\,.\label{qTRiccati}
\end{align}
With the zero initial conditions, these have the solutions,
\begin{align}
n_T=m_T&=\frac{\kappa T}{1+\kappa T}\,,\\
q_T&=\frac{1}{1+\kappa T}-e^{-2\kappa T}=-\frac{\kappa T}{1+\kappa T}+2e^{-\kappa T}\sinh\kappa T\,.
\end{align}

It is quite instructive to notice that the equality $n_T=m_T$ and the reality of $q_T$ together imply that the sum and difference Harish-Chandra phase-space variables are uncorrelated,
\begin{align}
\langle(\nu_T\pm\mu_T)^*(\nu_T\mp\mu_T\rangle_M
=\langle|\nu_T|^2\rangle_M-\langle|\mu_T|^2\rangle_M\mp\langle\nu_T^*\mu_T\rangle_M\pm\langle\mu_T^*\nu_T\rangle_M=0\,,
\label{crossmomentsplusminus}
\end{align}
with second moments,
\begin{align}\label{momentsplusminus}
\langle|\nu_T\pm\mu_T|^2\rangle_M
=\langle|\nu_T|^2\rangle_M+\langle|\mu_T|^2\rangle_M\pm\langle\nu_T^*\mu_T\rangle_M\pm\langle\mu_T^*\nu_T\rangle_M
=2(n_T\pm q_T)\,,
\end{align}
where
\begin{align}
n_T+q_T&=2e^{-\kappa T}\sinh\kappa T\,,\\
n_T-q_T&=\frac{-1+\kappa T}{1+\kappa T}+e^{-2\kappa T}=2e^{-\kappa T}\cosh\kappa T\,\frac{\Sigma_T}{1+\kappa T}\,,
\end{align}
The width $\Sigma_T = \kappa T - \tanh\kappa T$ was introduced in equation~\ref{SigmaT}.
It is worth noting the early- and late-time behavior of the various moments:
\begin{align}
n_T=m_T&=
\begin{cases}
\kappa T\,,&\kappa T\ll1,\\
1\,,&\kappa T\gg1,
\end{cases}\\
q_T&=
\begin{cases}
\kappa T\,,&\kappa T\ll1,\\
1/\kappa T\,,&\kappa T\gg1,
\end{cases}\\
n_T+q_T&=
\begin{cases}
2\kappa T\,,&\kappa T\ll1,\\
1\,,&\kappa T\gg1,
\end{cases}\\
n_T-q_T&=
\begin{cases}
\frac23(\kappa T)^3\,,&\kappa T\ll1,\\
1\,,&\kappa T\gg1.
\end{cases}
\end{align}
At early times, $\nu$ and $\mu$ are tightly correlated, with their sum undergoing standard diffusion; at late times, they become uncorrelated, and their second moments saturate at 1.

The sum and difference variables being uncorrelated, the Gaussian path integral~\ref{nBTZxPathIntegral2} has the normalized solution,
\begin{align}
\tilde B_T(Zx)
&=e^{-4\kappa T}\delta(r-2\kappa T)\,
4\pi\frac{1}{2\pi(n_T+q_T)}\exp\!\bigg({-}\frac{|\nu+\mu|^2}{2(n_T+q_T)}\bigg)\,
4\pi\frac{1}{2\pi(n_T-q_T)}\exp\!\bigg({-}\frac{|\nu-\mu|^2}{2(n_T-q_T)}\bigg)\,.
\end{align}
Noting that
\begin{align}
(n_T+q_T)(n_T-q_T)=\sinh2\kappa T\,\frac{2\Sigma_T}{e^{2\kappa T}(1+\kappa T)}\,,
\end{align}
we can put the normalized solution in the form
\begin{align}
\tilde B_T(Zx)
&=\frac{1}{\sinh2\kappa T}\,\delta(r-2\kappa T)\,\frac{2 e^{-2\kappa T}(1+\kappa T)}{\Sigma_T}
\exp\!\bigg({-}|\nu+\mu|^2\frac{e^{\kappa T}}{4\sinh\kappa T}-|\nu-\mu|^2\frac{e^{\kappa T}}{4\cosh\kappa T}\frac{1+\kappa T}{\Sigma_T}\bigg)\,,
\label{nBTZx2}
\end{align}
which satisfies the $\delta$-function initial condition~\ref{nBTZxinitialdelta}.

To retrieve the unnormalized distribution $B_T(Zx)=\sN_T\tilde B_T(Zx)$, one needs $\det M_T=1/\sN_T$, which according to equation~\ref{detMTODE} or equation~\ref{dlnNt}, satisfies the equation
\begin{align}\label{detMTODEtext}
\frac{1}{\kappa}\frac{d\ln\det M_T}{dT}=-(1+n_T)=-2+\frac{1}{1+\kappa T}\,,
\end{align}
with solution, for initial condition $\det M_0=1$,
\begin{align}
\det M_T=e^{-2\kappa T}(1+\kappa T)=\frac{1}{\sN_T}\,.
\end{align}
The unnormalized distribution is therefore
\begin{flalign}
B_T(Zx)
&=\frac{1}{\sinh2\kappa T}\,\delta(r-2\kappa T)\,\frac{2}{\Sigma_T}
\exp\!\bigg({-}|\nu+\mu|^2\frac{e^{\kappa T}}{4\sinh\kappa T}-|\nu-\mu|^2\frac{e^{\kappa T}}{4\cosh\kappa T}\frac{1+\kappa T}{\Sigma_T}\bigg)\\
&=\frac{1}{\sinh2\kappa T}\,\delta(r-2\kappa T)\,\frac{2}{\Sigma_T}
\exp\!\bigg({-}|\beta+\alpha|^2 e^{-\kappa T}\sinh\kappa T-|\beta-\alpha|^2 e^{-\kappa T}\cosh\kappa T\frac{1+\kappa T}{\Sigma_T}\bigg)\,.
\label{BTZx2}
\end{flalign}
The second line transforms to Cartan phase-space co\"ordinates.  Unnormalizing changes the Gaussian's prefactor; transforming to Cartan co\"ordinates changes the Gaussian. The final step is to undo the gauge transformation to get back to the Cartan reduced distribution,
\begin{align}
C_T(Zx)=e^{f(Zx)}B_T(Zx)
=\frac{1}{\sinh2\kappa T}\,\delta(r-2\kappa T)\,\frac{2}{\Sigma_T}
\exp\!\bigg({-}\frac{|\beta-\alpha|^2}{\Sigma_T}\bigg)\,,
\end{align}
which matches the solution~\ref{CTZxsolution} obtained from the PDE~\ref{Bloch}.

It is fair to ask whether the point of this section is just to provide a different, more complicated route to the solution for $C_T(Zx)$?  We think it is more than that, and here's why.  It all comes back to the elephant in the room, that is, how one handles the normalization or scaling of the Kraus operators that comes from the center~$Z$.  The Cartan reduced distribution, defined in equation~\ref{CTZx} and determined from the diffusion equation~\ref{Bloch}, succeeds in representing POVM completeness by marginalizing the \emph{ur}-distribution~$D_T(x)$ over the Cartan-center normalization, $e^{-2l}=e^{2f(Zx)}e^{-2s}$; this gives a distribution uniform in the Cartan sum variable $\beta+\alpha$ and thus spread over all of phase space in a way that gives POVM completeness.  As a consequence, however, the moments of $C_T(Zx)$ are not those of the stochastic integrals for the phase-plane variables.  The Harish-Chandra reduced distribution~$B_T(Zx)$, defined in equation~\ref{BTZx}, marginalizes $D_T(x)$ over the Harish-Chandra-center normalization, $e^{-2s}$.  After normalization to unity by the factor $\sN_T$, the normalized distribution $\tilde B_T(Zx)$ is determined from the diffusion equation~\ref{PDEnBTZx} or by applying the path-integral expression~\ref{nBTZxPathIntegral}, with its modified path measure, to the stochastic integrals for the Harish-Chandra phase-space variables.   The route from $\tilde B_T(Zx)$ to POVM completeness runs backwards through the normalization factor $\sN_T$ and the anti-Gaussian gauge transformation $e^{2f(Zx)}$ and arrives at $C_T(Zx)$.  The point of this section, one might say, is to find and reveal these connections among path integrals, diffusion equations, and SDEs; discovering these connections, driven in this paper by a combination of necessity and opportunity, allows us to re-unite the three faces of the stochastic trinity.  Accessing the entire stochastic trinity by using a positive gauge transformation that is grounded in a problem's Lie group---this, we hope, might be generally applicable to Feynman-Kac formulas for non-normalization-preserving diffusion equations.

\vfill\pagebreak

\section{Concluding Remarks. The Stochastic Trinity}\label{remarks}

We set out on the project of analyzing simultaneous measurements of noncommuting observables~\cite{CSJackson2021a,CSJackson2023a} with the goal of showing that such measurements end up with a POVM in the overcomplete coherent-state basis.  We now think that we have uncovered something more ambitious, a distinctive new window into the space of quantum dynamics.  The formulation of the problem of continual, differential measurements invites one---compels one, really---to think in terms of the paths of Wiener outcome increments.  These outcome paths, as sample-paths drawn from the Wiener measure, know nothing about a space in which they are wandering.  When they are instantiated in the exponents of Kraus operators, however, the time-ordered products of the differential Kraus operators generate a (complex) Lie-group manifold, the instrumental Lie group, in which the Kraus operators are, in the way of groups, both the transformations and the moving points.  The motion in the instrumental Lie-group manifold is described by Kraus-operator SDEs or by the diffusion of the KOD, as embodied in a FPK diffusion equation.  The continuous, but not differentiable paths are handled effortlessly by the It\^o calculus of the outcome increments, with its terms of order $\sqrt{dt}$ and $dt$.  This requires getting just beyond the linear structure of vector fields (right-invariant derivatives) and one-forms (right-invariant one-forms) on the Lie-group manifold, as is evident from our discussion of Stratonovich vs.~It\^o SDEs and in the derivation of the FPK diffusion equation, where right-invariant derivatives end up as diffusive second derivatives.  We end up in a very comfortable place, working in all three corners of the stochastic trinity: path integrals, FPK diffusion equations, and SDEs, all three describing motion, equivalently, on the instrumental Lie-group manifold.

Why don't others find the same comfort in all three faces of the trinity?  Field theorists, interested in the propagator of closed-system dynamics, have a different way of handling the continuous, but not differentiable paths, coming from a Stratonovich-calculus way of dealing with the temporal derivatives in the kinetic terms in a Lagrangian. While they usually have an equivalent Schr\"odinger-like equation for the propagator, they do not have the analog of~SDEs, even though the problem often undergoes a Wick rotation into ``Euclidean spacetime.''

Open-systems theorists, both in condensed-matter physics and in quantum optics, generally work with master equations or stochastic master equations for the evolution of quantum states and sometimes with diffusion equations for a probability distribution associated with the states.  Those who start with diffusion equations can avail themselves of a Feynman-Kac formula for a path integral, but the connection to SDEs has generally not been made.  The reason for not using all three faces of the stochastic trinity is, we think, a failure to identify the appropriate Lie-group manifold; this failure is connected to the emphasis on the evolution of quantum states, which obscures nearly entirely the Lie-group manifold on which the open-system dynamics is occurring.

Suffice it to say that we think we have found something: the home of quantum dynamics, the Lie-group manifold that supports all three faces of the trinity.
The exhausted reader who has survived to read this concluding sentence of a very long paper might be pleased---or so we hope---to learn that our ambition is larger than was evident at the beginning.

\acknowledgments
CSJ thanks Mohan Sarovar for all of the helpful discussions and financial support through Sandia National Laboratories and thanks CMC for the incredibly fruitful collaboration.
CMC supported himself and is eternally grateful for the opportunity to work with \hbox{CSJ}.

This work was supported in part by the Center for Quantum Information and Control at the University of New Mexico and in part by the U.S. Department of Energy, Office of Science, Office of Advanced Scientific Computing Research, under the Quantum Computing Application Teams (QCAT) program. Sandia National Laboratories is a multimission laboratory managed and operated by NTESS, LLC., a wholly owned subsidiary of Honeywell International, Inc., for the U.S. DOE's NNSA under contract DE-NA-0003525. This paper describes objective technical results and analysis. Any subjective views or opinions that might be expressed in the paper do not necessarily represent the views of the U.S. Department of Energy or the United States Government.

\vfill\pagebreak

\appendix

\section{Exterior Derivatives of the Right-Invariant Basis}
\label{app:ext}

The Lie algebra of right-invariant derivatives is equivalent to the exterior algebra of right-invariant one-forms.  This can be seen by considering the second exterior differential of a scalar function,
\begin{align}
	0&=ddf \\
	&= d\Big(\Rinv{e_\mu}[f]\theta^\mu\Big)\\
	&= \Rinv{e_\mu}\Big[\Rinv{e_\nu}[f]\Big]\theta^\mu\!\wedge\theta^\nu+\Rinv{e_\lambda}[f]d\theta^\lambda\\
	&= \frac12\bigg(\Rinv{e_\mu}\Big[\Rinv{e_\nu}[f]\Big]-\Rinv{e_\nu}\Big[\Rinv{e_\mu}[f]\Big]\bigg)\theta^\mu\!\wedge\theta^\nu+\Rinv{e_\lambda}[f]d\theta^\lambda\\
	&= \frac12\big[\Rinv{e_\mu},\Rinv{e_\nu}\big][f]\,\theta^\mu\!\wedge\theta^\nu+\Rinv{e_\lambda}[f]d\theta^\lambda\\
	&= \Rinv{e_\lambda}[f]\Big({-}\frac12{c_{\mu\nu}}^\lambda\,\theta^\mu\!\wedge\theta^\nu+d\theta^\lambda\Big)\,,
\end{align}
which by linear independence means that
\begin{equation}
	[\Rinv{e_\mu},\Rinv{e_\nu}]=-{c_{\mu\nu}}^\lambda \Rinv{e_\lambda}
	\hspace{20pt}
	\Longleftrightarrow
	\hspace{20pt}
	d\theta^\lambda = \frac12 {c_{\mu\nu}}^\lambda \,\theta^\mu\! \wedge \theta^\nu\,.
\end{equation}
Specifically, if we write out all of the different Lie brackets,
\begin{align}
	\Ho	&=	\text{no such brackets}\,,\\
	Q &=	[\Ho,-iP]\,,\\
	P	&=	[\Ho,iQ]\,,\\
	-iP	&=	[\Ho,Q]\,,\\
	iQ	&=	[\Ho,P]\,,\\
	-\Omega	&=	[Q,iP]= [iQ,P]\,,\\
	i\Omega &= [Q,P]= [iQ,-iP],
\end{align}
we can read off all of the one-form differentials or ``curls'',
\begin{align}\label{curl1}
	d\theta^{i\Omega}	&=	\theta^{Q} \!\wedge \theta^{P}+\theta^{iQ} \!\wedge \theta^{-iP}\,,\\
	d\theta^{-\Omega}	&=	\theta^{-iP} \!\wedge \theta^{Q}+\theta^{iQ} \!\wedge \theta^{P}\,,\\
	d\theta^{-iP}	&=	\theta^{\Ho}\!\wedge\theta^{Q}\,,\\
	d\theta^{iQ}	&=	\theta^{\Ho}\!\wedge\theta^{P}\,,\\
	d\theta^{Q}	&=	 \theta^{\Ho}\!\wedge\theta^{-iP}\,,\\
	d\theta^{P}	&=	\theta^{\Ho}\!\wedge\theta^{iQ}\,,\\
	d\theta^{-\Ho}	&=	0\,,\label{curl7}
\end{align}

\vfill\pagebreak

\section{Global Co\"ordinate Transformation between Cartan and Harish-Chandra}
\label{app:CoordTrans}

The global transformation from Cartan to Harish-Chandra co\"ordinates is most easily accomplished by inserting the displacement-operator normal ordering of equation~\ref{Dorderedaadag} into the Cartan decomposition~\ref{Cartandecomp} and then pushing the factors around until they are in Harish-Chandra form:
\begin{align}
	\left(D_\beta\,e^{i\Omega \phi}\right)e^{-\Ho r-\Omega\ell}D_\alpha^\inv
	&=e^{-\Omega[\ell+\frac12(|\beta|^2+|\alpha|^2)]}e^{i\Omega\phi}e^{a^\dagger\beta}e^{-a\beta^*}e^{-\Ho r}e^{-a^\dagger\alpha}e^{a\alpha^*}\\
	&=e^{-\Omega[\ell+\frac12(|\beta|^2+|\alpha|^2)]}e^{i\Omega\phi}e^{a^\dagger\beta}e^{-a\beta^*}e^{-a^\dagger e^{-r}\alpha}e^{-\Ho r}e^{a\alpha^*}\\
	&=e^{-\Omega[\ell+\frac12(|\beta|^2+|\alpha|^2)]}e^{i\Omega\phi}e^{\Omega\beta^*\alpha e^{-r}}
        e^{a^\dagger\beta}e^{-a^\dagger e^{-r}\alpha}e^{-a\beta^*}e^{-\Ho r}e^{a\alpha^*}\\
    &=e^{-\Omega[\ell+\frac12(|\beta|^2+|\alpha|^2)]}e^{i\Omega\phi}e^{\Omega\beta^*\alpha e^{-r}}
        e^{a^\dagger\beta}e^{-a^\dagger e^{-r}\alpha}e^{-\Ho r}e^{-ae^{-r}\beta^*}e^{a\alpha^*}\\
	&=e^{a^\dag(\beta -e^{-r}\alpha)}e^{-\Ho r-\Omega[\ell+\frac12(|\beta|^2+\frac12|\alpha|^2)-\beta^*\alpha e^{-r}-i\phi]}e^{a(\alpha-e^{-r}\beta)^*}\,.
\end{align}
Identifying the parameters of the last line gives the co\"ordinate transformation,
\begin{align}
	\nu &= \beta - e^{-r}\alpha\,,\label{HCCnu}\\
	\mu &= \alpha - e^{-r}\beta\,,\label{HCCmu}\\
	-s+i\psi = z &= -\ell-f+i(\phi+\xi)\,,
\end{align}
where
\begin{align}
f&\equiv\frac12\Big(|\beta|^2+|\alpha|^2-\beta^*\alpha e^{-r}-\beta\alpha^*e^{-r}\Big)=\frac{1-e^{-r}}{4}|\beta+\alpha|^2+\frac{1+e^{-r}}{4}|\beta-\alpha|^2\,,\label{quadCartan}\\
\xi&\equiv e^{-r}\frac{\beta^*\alpha-\beta\,\alpha^*}{2i}=e^{-r}\frac{(\beta-\alpha)^*(\beta+\alpha)-(\beta-\alpha)(\beta+\alpha)^*}{4i}\,,
\end{align}
are functions of the RWIH~$G/Z$.

The inverse co\"ordinate transformation is
\begin{align}
	\beta&=\frac{e^r\nu + \mu}{2\sinh r}\,,\\
	\alpha&=\frac{e^r\mu + \nu}{2\sinh r}\,,\\
    \ell&=s-f\,,\label{eq:postrans}\\
    \phi&=\psi-\xi\,.\label{eq:unitrans}
\end{align}
It is quite useful to notice that the sum and difference phase-space co\"ordinates simply rescale under this transformation,
\begin{align}
\beta+\alpha&=\frac{\nu + \mu}{1- e^{-r}}=\frac{e^{r/2}(\nu+\mu)}{2\sinh(r/2)}\,,\label{eq:sumvariables}\\
\beta-\alpha&=\frac{\nu - \mu}{1+ e^{-r}}=\frac{e^{r/2}(\nu-\mu)}{2\cosh(r/2)}\,.\label{eq:diffvariables}
\end{align}
We also have
\begin{align}
f&=\frac{|\nu+\mu|^2}{4(1-e^{-r})}+\frac{|\nu-\mu|^2}{4(1+e^{-r})}=\frac{(|\nu|^2+|\mu|^2)e^r+\nu^*\mu+\nu\mu^*}{4\sinh r}\,,\label{eq:posgauge}\\
\xi&=\frac{(\nu-\mu)^*(\nu+\mu)-(\nu-\mu)(\nu+\mu)^*}{8i\sinh r}=\frac{\nu^*\mu-\nu\mu^*}{4i\sinh r}\,.\label{eq:unigauge}
\end{align}

It is crucial to appreciate that the transformation from Harish-Chandra to Cartan co\"ordinates is singular at $r=0$, with the consequence that positive transformations of the form $e^{Qq+Pp}$ are not represented in Cartan co\"ordinates.  At $r=0$ the Harish-Chandra decomposition becomes
\begin{align}
x_{r=0}
=e^{\Omega z}e^{a^\dag \nu}e^{a\mu^*}
=e^{\Omega(z-\frac12\nu\mu^*)}e^{a^\dag \nu+a\mu^*}
=e^{\Omega(z-\frac12\nu\mu^*)}D_{(\nu-\mu)/2}e^{a^\dag\frac12(\nu+\mu)+a\frac12(\nu+\mu)^*}\,.
\end{align}
Each displacement operator and each positive transformation of the form~$e^{Qq+Pp}$ is represented uniquely in the 4-plane of $\nu$ and $\mu$ co\"ordinates; in particular, the identity operator has co\"ordinates $r=0$ and $z=\nu=\mu=0$.
At $r=0$ the Cartan decomposition reduces to
\begin{align}
	x_{r=0}=e^{i\Omega\phi}e^{-\Omega\ell}D_\beta D_{-\alpha}=e^{i\Omega[\phi+(\beta^*\alpha-\beta\alpha^*)/2i]}e^{-\Omega\ell}D_{\beta-\alpha}\,.
\end{align}
The Cartan co\"ordinate singularity at $r=0$ is that $x_{r=0}$ does not depend on $\beta+\alpha$.  A displacement operator $D_{\beta-\alpha}=D_\tau$ is represented by a plane of $\beta+\alpha$ values, specifically, by all the co\"ordinates satisfying $\beta-\alpha=\tau$, $r=0$, $\ell=0$, and $\phi=\big(\beta+\alpha)^*\tau-(\beta+\alpha)\tau^*\big)/4i$.  Most importantly, the identity $1$ is represented by the plane of Cartan co\"ordinate values satisfying  $\beta=\alpha$ and $r=\phi=\ell=0$.  Positive transformations of the form~$e^{Qq+Pp}$ are not represented at all in Cartan co\"ordinates; this is not a problem because these positive operators lie on a boundary that is not accessible to the Kraus operators of SPQM.

\vfill\pagebreak

\section{Solution for Harish-Chandra Center Co\"ordinate}
\label{zsolution}

To solve the SDE~\ref{latterCarl} for the Harish-Chandra center co\"ordinate $z$, it is best to work with the sums over Wiener increments that underlie the It\^o stochastic integrals.  Thus we begin by writing the solution~\ref{OUPsolution} for the post-measurement Harish-Chandra co\"ordinate $\nu$ as
\begin{align}
	\nu_N=\sum_{k=0}^{T/dt-1}\sqrt{\kappa}\,dw_{kdt}\,e^{-2\kappa(T-t_k)}=\sum_{k=0}^{N-1}\sqrt{\kappa}\,dw_k\,e^{-2\kappa(N-k)dt}\,,
\end{align}
where $dw_k=dw_{kdt}$ and $t_k=k\,dt$ ($t_N=T=N\,dt$).  Notice that the initial condition $\nu_0=0$ is enforced by having no terms in the sum for $N=0$.
The solution for $z$ is
\begin{align}\label{zN}
z_N&=\frac12\sum_{k=0}^{N-1}\kappa|dw_k|^2+\sum_{k=1}^{N-1}\nu_k \sqrt\kappa\,dw_k^*\\
&=\frac12\sum_{k=0}^{N-1}\kappa|dw_k|^2+\sum_{k=1}^{N-1}\sum_{l=0}^{k-1}\kappa\,dw_k^*\,dw_l\,e^{-2\kappa|k-l|dt}\,,
\end{align}
where we omit the $k=0$ term on the second line since $\nu_0=0$ and where we can insert the absolute value because $k>l$.  Notice that, as for $\nu$, the way the initial condition $z_0=0$ is enforced is that there are no terms in the sums when $N=0$.  Now define
\begin{align}
y=\sum_{k=1}^{N-1}\sum_{l=0}^{k-1}\kappa\,dw_k^*\,dw_l\,e^{-2\kappa|k-l|dt}\,,
\end{align}
and manipulate $y^*$ by specifying the summing range in an equivalent way and then switching $k$ and $l$,
\begin{align}
y^*=\sum_{l=0}^{N-2}\sum_{k=l+1}^{N-1}\kappa\,dw_k\,dw_l^*\,e^{-2\kappa|k-l|dt}
=\sum_{k=0}^{N-2}\sum_{l=k+1}^{N-1}\kappa\,dw_k^*\,dw_l\,e^{-2\kappa|k-l|dt}\,.
\end{align}
With this result we an write our sums as including the entire range of values, $k,l=1\,\ldots,N-1$.  Converting to the real and imaginary parts of $z=-s+i\psi$, we have
\begin{align}
-s_N=\Re(z_N)&=\frac12\Bigg(\sum_{k=0}^{N-1}\kappa|dw_k|^2
+\sum_{k=1}^{N-1}\sum_{l=0}^{k-1}\kappa\,dw_k^*\,dw_l\,e^{-2\kappa|k-l|dt}+\sum_{k=0}^{N-2}\sum_{l=k+1}^{N-1}\kappa\,dw_k^*\,dw_l\,e^{-2\kappa|k-l|dt}\Bigg)\\
&=\frac12\sum_{k=0}^{N-1}\sum_{l=0}^{N-1}\kappa\,dw_k^*\,dw_l\,e^{-2\kappa|k-l|dt}\,,\\
i\psi_N=i\Im(z_N)
&=\frac12\Bigg(\sum_{k=1}^{N-1}\sum_{l=0}^{k-1}\kappa\,dw_k^*\,dw_l\,e^{-2\kappa|k-l|dt}-\sum_{k=0}^{N-2}\sum_{l=k+1}^{N-1}\kappa\,dw_k^*\,dw_l\,e^{-2\kappa|k-l|dt}\Bigg)\\
&=\frac12\sum_{k=0}^{N-1}\sum_{l=0}^{N-1}\kappa\,dw_k^*\,dw_l\,e^{-2\kappa|k-l|dt}\textrm{sgn}(k-l)\,,
\end{align}
where the sign function is defined as
\begin{align}
\textrm{sgn}(u)=
\begin{cases}
1,&u>0,\\
0,&u=0,\\
-1,&u<0.
\end{cases}
\end{align}
The center variable~\ref{zN} is therefore
\begin{align}
z_N=-s_N+i\psi_N=\sum_{k=0}^{N-1}\sum_{l=0}^{N-1}\kappa\,dw_k^*\,dw_l\,e^{-2\kappa|k-l|dt}\frac12\big(1+\textrm{sgn}(k-l)\big)\,,
\end{align}
where one recognizes
\begin{align}
\frac12\big(1+\textrm{sgn}(u)\big)=
H(u)=
\begin{cases}
1,&u>0,\\
\frac12,&u=0,\\
0&u<0.
\end{cases}
\end{align}
as the Heaviside step function with the choice $H(0)=\frac12$.  This appendix is really an exercise in relating the real and imaginary parts of $z$ to its symmetric and antisymmetric parts and, in the process, getting the weighting of the diagonal ($k=l$) term right---equivalently, making the right choice for the $u=0$ value of the sign and Heaviside functions.  The incremental It\^o calculus, with its quite explicit diagonal terms $\frac12\kappa|dw_k|^2$, leaves no doubt about the right choice.

Converting back to integrals gives
\begin{align}
-s_T&=\frac12\int_0^{T_-}\!\!\int_0^{T_-}\!\!\kappa\,dw_t^*\,dw_s\,e^{-2\kappa |t-s|}\,,\\
i\psi_T&=\frac12\int_0^{T_-}\!\!\int_0^{T_-}\!\!\kappa\,dw_t^*\,dw_s\,e^{-2\kappa |t-s|}\textrm{sgn}(t-s)\,,
\end{align}
which lead to the formula for $z_T=-s_T+i\psi_T$ in equation~\ref{latterCarlsolution}.  To use these integrals, one should return to the incremental sums.  If one started with the Stratonovich-form SDE~\ref{dzStrat} for $z$, one could convert it to an ordinary temporal differential equation with the Wiener increments replaced by $\delta$-functions.  Integrating that equation would give the same results as here, but only after careful attention to how to weight the diagonal ($t=s$) $\delta$-functions
in the integral.  Thinking in terms of the Stratonovich-form SDE for $z$, it is clear that solving the equation for $z$---what we have done in this appendix---is identical to deriving a fluctuation-dissipation theorem for the correlation of $\nu$ and~$\mu$.

\vfill\pagebreak

\section{Local Transformations Between Cartan-Co\"ordinate and Right-Invariant Frames}
\label{FrameTrans}

The Cartan decomposition of a position $x \in \mathrm{IWH}$ is
\begin{equation}
	x = \left(D_\beta\,e^{i\Omega \phi}\right)e^{-\Ho r-\Omega\ell}D_\alpha^\inv\,.
\end{equation}
Before differentiating $x$ with respect to the Cartan co\"ordinates, it is useful to differentiate the displacement operator with respect to its arguments using the ordered forms in equation~\ref{DorderedPQ}:
\begin{align}
\partial_{\alpha_1}D_\alpha&=\big({-}iP+\smallfrac12\alpha_2\,i\Omega\big)D_\alpha\,,\\
\partial_{\alpha_2}D_\alpha&=\big(iQ-\smallfrac12\alpha_1\,i\Omega\big)D_\alpha\,.
\end{align}
We will also need the conjugations
\begin{align}
\frac{1}{\sqrt2}D_\alpha(Q+iP)D_\alpha^\dagger=D_\alpha a D_\alpha^\dagger&=a-\alpha=\frac{1}{\sqrt2}\big(Q-\alpha_1+i(P-\alpha_2)\big)\,,
\end{align}
and
\begin{align}
e^{-r\Ho}a e^{r\Ho}&=ae^r\,,\\
e^{-r\Ho}a^\dagger e^{r\Ho}&=a^\dagger e^{-r}\,,\\
e^{-r\Ho}Q e^{r\Ho}&=Q\cosh r+iP\sinh r\,,\\
e^{-r\Ho}P e^{r\Ho}&=P\cosh r-iQ\sinh r\,.
\end{align}
Differentiating the position $x$ gives
\begin{align}
\begin{split}
	\partial_\phi x &= i\Omega\,x\,,
\end{split}\\
\begin{split}
	-\partial_\ell x&= \Omega\,x\,,
\end{split}\\
\begin{split}
	\partial_{\beta_1} x &= -iP\,x +\smallfrac12\beta_2\,i\Omega\,x\,,
\end{split}\\
\begin{split}
	\partial_{\beta_2} x &= iQ\,x-\smallfrac12\beta_1\,i\Omega\,x\,,
\end{split}\\
\begin{split}
	{-}\partial_{r}x
    &= D_\beta \Ho D_\beta^\dagger\,x\\
    &=\smallfrac12\big((Q-\beta_1)^2+(P-\beta_2)^2\big)x\\
	&= \Ho\,x-\beta_1\,Q\,x-\beta_2\,P\,x+\smallfrac12(\beta_1^2+\beta_2^2)\,\Omega\,x\,,
\end{split}\\
\begin{split}
	\partial_{\alpha_1}x
    &=D_\beta\,e^{-r\Ho}\big(iP+\smallfrac12\alpha_2\,i\Omega\big)e^{r\Ho}D_\beta^\dagger\,x\\
    &=D_\beta\big(iP\cosh r+Q\sinh r+\smallfrac12\alpha_2\,i\Omega\,\big)D_\beta^\dagger\,x\\
    &=\cosh r\,iP\,x+\sinh r\,Q\,x-\beta_1\sinh r\,\Omega\,x-(\beta_2\cosh r-\smallfrac12\alpha_2)\,i\Omega\,x\,,
\end{split}\\
\begin{split}
	\partial_{\alpha_2}x
    &=D_\beta\,e^{-r\Ho}\big({-}iQ-\smallfrac12\alpha_1\,i\Omega\big)e^{r\Ho}D_\beta^\dagger\,x\\
    &=D_\beta\big({-}iQ\cosh r+P\sinh r-\smallfrac12\alpha_1\,i\Omega\,\big)D_\beta^\dagger\,x\\
    &=-\cosh r\,iQ\,x+\sinh r\,P\,x-\beta_2\sinh r\,\Omega\,x+(\beta_1\cosh r-\smallfrac12\alpha_1)\,i\Omega\,x\,.
\end{split}
\end{align}

By the chain rule, we have the frame transformation,
\begin{align}
\begin{array}{rclllllll}
	\vphantom{\Big(}\partial_\phi &=&+\Rinv{i\Omega}\\
    \vphantom{\Big(}\partial_\ell &=&&{-}\Rinv{\Omega}\\
	\vphantom{\Big(}\partial_{\beta_1} &=&+\smallfrac12\beta_2\Rinv{i\Omega}&&-\Rinv{iP}\\
	\vphantom{\Big(}\partial_{\beta_2} &=&-\smallfrac12\beta_1\Rinv{i\Omega}&&&+\Rinv{iQ}\\
	\vphantom{\Big(}\partial_{\alpha_1} &=&-(\beta_2\cosh r-\smallfrac12\alpha_2)\Rinv{i\Omega}&-\beta_1\sinh r\,\Rinv{\Omega}&+\cosh r\,\Rinv{iP}&&+\sinh r\Rinv{Q}\\
	\vphantom{\Big(}\partial_{\alpha_2} &=&+(\beta_1\cosh r-\smallfrac12\alpha_1)\Rinv{i\Omega}&-\beta_2\sinh r\,\Rinv{\Omega}&&-\cosh r\,\Rinv{iQ}&&+\sinh r\Rinv{P}\\
	\vphantom{\Big(}\partial_{r}&=&&-\smallfrac12(\beta_1^2+\beta_2^2)\Rinv{\Omega}&&&+\beta_1\Rinv{Q}&+\beta_2\Rinv{P}&-\Rinv{\Ho}
\end{array}\,.
\end{align}
\vfill\pagebreak
Inverting the transformation gives
\begin{align}
	\Rinv{i\Omega}&=\partial_\phi\,,\\
	-\Rinv{\Omega}&=\partial_\ell\,,\\
	-\Rinv{iP}&=\partial_{\beta_1}-\smallfrac12\beta_2\partial_\phi\,,\\
	\Rinv{iQ}&=\partial_{\beta_2}+\smallfrac12\beta_1\partial_\phi\,,\\
	\Rinv{Q}&=\nabla_1 -\beta_1\partial_\ell +\frac{\beta_2\cosh r-\alpha_2}{2\sinh r}\partial_\phi\,,\\
	\Rinv{P}&=\nabla_2 -\beta_2\partial_\ell -\frac{\beta_1\cosh r-\alpha_1}{2\sinh r}\partial_\phi\,,\\
\begin{split}
	-\Rinv{\Ho} &= \partial_{r}-\beta_1\Rinv{Q}-\beta_2\Rinv{P}+\frac{\beta_1^2+\beta_2^2}{2}\Rinv{\Omega}\\
	&= \partial_{r}-\beta_1\nabla_1-\beta_2\nabla_2+\frac{\beta_1^2+\beta_2^2}{2}\partial_\ell
    +\frac{\beta_1\alpha_2-\beta_2\alpha_1}{2\sinh r}\partial_\phi\,,
\end{split}
\end{align}
where
\begin{equation}\label{nablaj}
	\nabla_j \equiv \frac{1}{\sinh r}\left(\partial_{\alpha_j}+\cosh r\,\partial_{\beta_j}\right)\,.
\end{equation}
Transposing the transformation gives
\begin{align}
    \begin{split}
	\theta^{i\Omega}	&=	d\phi+\smallfrac12\beta_2 d\beta_1-\smallfrac12\beta_1d\beta_2
        -(\beta_2\cosh r-\smallfrac12\alpha_2)\,d\alpha_1+(\beta_1\cosh r-\smallfrac12\alpha_1)d\alpha_2\\
	&=	d\phi+\smallfrac12(\beta_2 d\beta_1-\beta_1d\beta_2)+\smallfrac12(\alpha_2 d\alpha_1-\alpha_1d\alpha_2)
        +\cosh r\,(\beta_1d\alpha_2-\beta_2d\alpha_1)\,,
    \end{split}\\
	\theta^{-\Omega}	&=	d\ell + \smallfrac12(\beta_1^2+\beta_2^2)dr+\sinh r\,(\beta_1 d\alpha_1 +\beta_2 d\alpha_2)\,,\\
	\theta^{-iP}	&=	d\beta_1 -\cosh r\, d\alpha_1\,,\\
	\theta^{iQ}	&=	d\beta_2 -\cosh r\, d\alpha_2\,,\\
	\theta^{Q}	&=	\beta_1 dr+\sinh r\, d\alpha_1\,,\\
	\theta^{P}	&=	\beta_2 dr+\sinh r\, d\alpha_2\,,\\
	\theta^{-\Ho}	&=	dr\,.
\end{align}
With the one-form transformations in hand, the Haar measure~\ref{HaarFormula} in Cartan co\"ordinates is
\begin{equation}\label{d7xCartan}
		d^7\!\mu(x)=d\phi\, d\ell\, \frac{d^2\beta}{\pi}\,dr\,\sinh^2\!r\,\frac{d^2\alpha}{\pi}\,.
\end{equation}
The phase-plane measures are $d^2\beta=\frac12d\beta_1 d\beta_2$ and $d^2\beta=\frac12d\beta_1 d\beta_2$. The factors of $1/\pi$ are conventional in quantum optics and ultimately come from the coherent-state completeness relation~\ref{csPOVM}.

It is easy to show that the left-invariant derivatives and one-forms can be obtained from the right-invariant quantities by changing the sign of the quantities associated with anti-Hermitian operators and transforming the co\"ordinates according to $\phi\leftrightarrow-\phi$ and $\beta\leftrightarrow\alpha$.  The sign changes don't change the Haar measure, and the measure is unchanged by the co\"ordinate transformation, which shows that the Haar measure is both right- and left-invariant.

\vfill\pagebreak

\section{Local Transformations Between Harish-Chandra-Co\"ordinate and Right-Invariant Frames}
\label{FrameTransHC}

The Harish-Chandra decomposition of a position $x \in \mathrm{IWH}$ is
\begin{equation}\label{HC2}
		x = e^{a^\dag \nu}e^{-\Ho r +\Omega z}e^{a\mu^*}\,.
\end{equation}
Break the complex co\"ordinates into real and imaginary parts, as in equation~\ref{HCrealimag}.
Differentiating the position gives
\begin{align}
	\partial_{\nu_1} x&=\frac{1}{\sqrt2}\,a^\dag\,x\,,\\
	\partial_{\nu_2} x&=\frac{1}{\sqrt2}\,ia^\dag x\,,\\
	-\partial_{s} x&=\Omega\,x\,,\\
	\partial_{\psi} x&=\,i\Omega\,x\,,\\
    \begin{split}
	-\partial_{r} x&=e^{a^\dag \nu}\Ho e^{-a^\dag \nu}\,x\\
	&=\Ho\,x - \nu\,a^\dag x\\
	&=\Ho\,x - \nu_1\,\frac{1}{\sqrt2} a^\dag\,x - \nu_2\,\frac{1}{\sqrt2}ia^\dag x\,,
    \end{split}\\
    \begin{split}
	\partial_{\mu_1}x&=\frac{1}{\sqrt2}e^{a^\dag \nu}e^{-\Ho r}ae^{\Ho r}e^{-a^\dag \nu}x\\
	&=\frac{1}{\sqrt2}e^{a^\dag \nu}(e^r a)e^{-a^\dag \nu}x\\
	&=\frac{1}{\sqrt2}e^r (a-\nu\Omega)x\\
	&=e^r \left(\frac{1}{\sqrt2}a\,x-\frac12\nu_1\,\Omega\,x-\frac12\nu_2\,i\Omega\,x\right)\,,
    \end{split}\\
    \begin{split}
	\partial_{\mu_2}x&=-i\frac{1}{\sqrt2}e^{a^\dag \nu}e^{-\Ho r}ae^{\Ho r}e^{-a^\dag \nu}x\\
	&=e^r\!\left(\!{-}\frac{1}{\sqrt2}\,ia\,x-\frac12\nu_2\,\Omega\,x+\frac12\nu_1\,i\Omega\,x\right)\,.
    \end{split}
\end{align}
By the chain rule, we have the frame transformation,
\begin{align}
\begin{array}{rclllllll}
\vphantom{\Big(}\partial_{\psi}&=&\Rinv{i\Omega}\\
\vphantom{\Big(}-\partial_{s}&=&&\Rinv{\Omega}\\
\vphantom{\Big(}\partial_{\nu_1}&=&&&\frac{1}{\sqrt2}\Rinv{a^\dag}\\
\vphantom{\Big(}\partial_{\nu_2}&=&&&&\frac{1}{\sqrt2}\Rinv{ia^\dag}\\
\vphantom{\Big(}e^{-r}\partial_{\mu_1}&=&-\frac12\nu_2\Rinv{i\Omega}&-\frac12\nu_1\Rinv{\Omega}&&&\frac{1}{\sqrt2}\Rinv{a}\\
\vphantom{\Big(}e^{-r}\partial_{\mu_2}&=&\frac12\nu_1 \Rinv{i\Omega}&-\frac12\nu_2\Rinv{\Omega}&&&&-\frac{1}{\sqrt2}\Rinv{ia}\\
\vphantom{\Big(}-\partial_{r}&=&&&-\nu_1\frac{1}{\sqrt2}\Rinv{a^\dag}&-\nu_2\frac{1}{\sqrt2}\Rinv{ia^\dag}&&&\Rinv{\Ho}
\end{array}\,.
\end{align}
\vfill\pagebreak
Inverting the transformation gives
\begin{align}
\Rinv{i\Omega}&=\partial_{\psi}\,,\\
-\Rinv{\Omega}&=\partial_{s}\,,\\
\frac{1}{\sqrt2}\Rinv{a^\dag} &=\partial_{\nu_1}\,,\\
\frac{1}{\sqrt2}\Rinv{ia^\dag}&=\partial_{\nu_2}\,,\\
\frac{1}{\sqrt2}\Rinv{a}&=e^{-r}\partial_{\mu_1}-\frac12\nu_1\partial_s +\frac12\nu_2 \partial_\psi\,,\\
-\frac{1}{\sqrt2}\Rinv{ia}&=e^{-r}\partial_{\mu_2}-\frac12\nu_2\partial_s-\frac12\nu_1 \partial_\psi\,,\\
-\Rinv{\Ho}&=\partial_{r}-\nu_1\partial_{\nu_1}-\nu_2\partial_{\nu_2}\,.
\end{align}
Transposing the transformation gives
\begin{align}
\theta^{i\Omega}&=d\psi+\smallfrac12e^r(\nu_1d\mu_2-\nu_2d\mu_1)\,,\\
\theta^{-\Omega}&=ds+\smallfrac12e^r(\nu_1d\mu_1+\nu_2d\mu_2)\,,\\
{\sqrt2}\,\theta^{a^\dag} &= d\nu_1+\nu_1dr\,,\\
{\sqrt2}\,\theta^{ia^\dag}&=d\nu_2+\nu_2dr\,,\\
{\sqrt2}\,\theta^{a}&= e^r d\mu_1\,,\\
{\sqrt2}\,\theta^{-ia}&= e^r d\mu_2\,,\\
\theta^{-\Ho}&=dr\,.
\end{align}
It is instructive to convert the right-invariant derivatives and forms to position and momentum,
\begin{align}
-\Rinv{iP}&=\frac{1}{\sqrt2}(\Rinv{a^\dagger}-\Rinv{a})=\partial_{\nu_1}-e^{-r}\partial_{\mu_1}+\frac12\nu_1\partial_s-\frac12\nu_2\partial_\psi\,,\\
\Rinv{iQ}&=\frac{1}{\sqrt2}(\Rinv{ia^\dagger}+\Rinv{ia})=\partial_{\nu_2}-e^{-r}\partial_{\mu_2}+\frac12\nu_2\partial_s+\frac12\nu_1\partial_\psi\,,\\
\Rinv{Q}&=\frac{1}{\sqrt2}(\Rinv{a^\dagger}+\Rinv{a})=\nabla_1-\frac12\nu_1\partial_s+\frac12\nu_2\partial_\psi\,,\\
\Rinv{P}&=\frac{1}{\sqrt2}(\Rinv{ia^\dagger}-\Rinv{ia})=\nabla_2-\frac12\nu_2\partial_s-\frac12\nu_1\partial_\psi\,,
\end{align}
where the derivatives
\begin{equation}
	\nabla_j = \partial_{\nu_j}+e^{-r}\partial_{\mu_j}
\end{equation}
are equal to the derivatives of equation~\ref{nablaj} and
\begin{align}
	\theta^{-iP}	&= \frac{1}{\sqrt2}\big(\theta^{a^\dagger}-\theta^{a}\big)=\frac12\big(d\nu_1-e^r d\mu_1+\nu_1 dr\big)\,,\\
	\theta^{iQ}	&= \frac{1}{\sqrt2}\big(\theta^{ia^\dagger}+\theta^{ia}\big)=\frac12\big(d\nu_2-e^r d\mu_2+\nu_2 dr\big)\,,\\
	\theta^{Q}	&= \frac{1}{\sqrt2}\big(\theta^{a^\dagger}+\theta^{a}\big)=\frac12\big(d\nu_1+e^r d\mu_1+\nu_1 dr\big)\,,\\
	\theta^{P}	&= \frac{1}{\sqrt2}\big(\theta^{ia^\dagger}-\theta^{ia}\big)=\frac12\big(d\nu_2+e^r d\mu_2+\nu_2 dr\big)\,.
\end{align}

Comparing the expressions for right-invariant derivatives and one-forms in the Harish-Chandra co\"ordinates of this appendix with the expressions in Cartan co\"ordinates from appendix~\ref{FrameTrans}, one sees that these expressions relate the co\"ordinate partial derivatives and one-forms in the two co\"ordinate systems.  The relations between co\"ordinate partial derivatives and one-forms can, of course, be derived straighforwardly from the global co\"ordinate transformations in appendix~\ref{app:CoordTrans}.  We do not record these relations because the expressions in terms of the right-invariant quantities are more useful for our purposes.  We remind the reader that even though the ruler co\"ordinate $r$ is shared between Harish-Chandra and Cartan co\"ordinates, the partial derivative $\partial_r$ is different between the two systems because partial derivatives are defined by holding the other co\"ordinates constant.

The Haar measure~\ref{HaarFormula} in Harish-Chandra co\"ordinates is
\begin{equation}\label{d7xHC}
		d^7\!\mu(x)=d\psi\,ds\,\frac{d^2\nu}{2\pi}\,dr\,e^{2r}\frac{d^2\mu}{2\pi}\,.
\end{equation}
The factors of $1/2\pi$ in the Harish-Chandra phase-plane measures follow from transforming the Cartan Haar measure to Harish-Chandra co\"ordinates.  Just as for Cartan co\"ordinates, it is easy to show that the left-invariant one-forms yield the same measure.

\vfill\pagebreak

\section{Delta-Functions and the Singularity in Cartan Co\"ordinates}
\label{deltax1}

In this appendix we find the Cartan form of the $\delta$-function $\delta(x,1)$ in two ways.  The first uses that $D_0(x)=\delta(x,1)$, so we can step slightly away from $T=0$, to $T=dt$, and find $D_{dt}(x)$ as an expression that limits to $\delta(x,1)$ as $dt\rightarrow0$.  The second way is perhaps more straightforward: show that the expression for $\delta(x,1)$ in Cartan co\"ordinates transforms to the known form~\ref{deltax1HC} in Harish-Chandra co\"ordinates.  That said, since we need to know $D_{dt}(x)$ in section~\ref{solveFPK}, we start with the first way.

Starting from the general path-integral expression~\ref{KODF} for the KOD and setting $T=dt$,
\begin{align}\label{Ddtxappendix}
D_{dt}(x)=\int d\mu(dw_0)\,\delta\big(x,\gamma(dw_0)\big)\,,
\end{align}
and plugging in the Cartan-co\"ordinate expression~\ref{delta7Cartan} for the $\delta$-function, we have
\begin{align}
D_{dt}(x)
=\int d\mu(dw_0)\,\delta(\phi-\phi_{dt})\,\delta(\ell-\ell_{dt})\,\frac{1}{\sinh^2\!r}\delta(r-r_{dt})\,\pi\delta^2(\beta-\beta_{dt})\,\pi\delta^2(\alpha-\alpha_{dt})\,.
\end{align}
It is useful to work in terms of sum and difference co\"ordinates for the phase-space variables, so we note that
\begin{align}
\pi\delta^2(\beta-\beta_{dt})\,\pi\delta^2(\alpha-\alpha_{dt})
=2\pi\delta^2\big(\beta+\alpha-(\beta_{dt}+\alpha_{dt})\big)\,2\pi\delta^2\big(\beta-\alpha-(\beta_{dt}-\alpha_{dt})\big)\,.
\end{align}

We now specialize the stochastic-integral solutions for the co\"ordinates to the first increment.  At $T=dt$, $r_{dt}=2\kappa\,dt$ and the phase-space co\"ordinates are
\begin{align}
\nu_{dt}=\sqrt\kappa\,dw_0\,e^{-2\kappa dt}=\sqrt\kappa\,dw_0\,,\qquad
&\qquad\mu_{dt}=d\mu_0=\sqrt\kappa\,dw_0\,,\\
\nu_{dt}+\mu_{dt}=\sqrt\kappa\,dw_0(e^{-2\kappa dt}+1)=2\sqrt\kappa\,dw_0\,,\qquad
&\qquad\nu_{dt}-\mu_{dt}=\sqrt\kappa\,dw_0(e^{-2\kappa dt}-1)=-2\kappa dt\sqrt\kappa\,dw_0\,,\\
\beta_{dt}=\sqrt{\kappa}\,dw_0\,\csch 2\kappa dt=\frac{\sqrt\kappa\,dw_0}{2\kappa dt}\,,\qquad
&\qquad\alpha_{dt}=\sqrt\kappa\,dw_0\coth2\kappa dt=\frac{\sqrt\kappa\,dw_0}{2\kappa dt}\,,\\\
\beta_{dt}+\alpha_{dt}=\sqrt\kappa\,dw_0\coth\kappa dt=\frac{\sqrt\kappa\,dw_0}{\kappa dt}\,,\qquad
&\qquad \beta_{dt}-\alpha_{dt}=-\sqrt\kappa\,dw_0\tanh\kappa dt=-\kappa dt \sqrt\kappa\,dw_0\,.
\end{align}
In each case, the first expression is exact, and the second is the leading-order contribution.  What should give any reader pause are the divergences in the single-increment Cartan phase variables.  These divergences are an expression of the $r=0$ singularity in Cartan co\"ordinates, and they make the decisive contribution to our determination of $D_{dt}(x)$.  It is important to understand that there are subtleties in trying to extract the single-increment behavior directly from the SDEs, instead of from the stochastic integrals.  For example, the SDEs~\ref{OUP} and~\ref{GGW} say that $\nu_{dt}=d\nu_0=\sqrt\kappa\,dw_0$ and $\mu_{dt}=d\mu_0=\sqrt\kappa\,dw_0$, so one misses the leading-order contribution to $\nu_{dt}-\mu_{dt}$ because that leading-order contribution is zero from the perspective of the SDEs.  More tellingly, the $r=0$ singularity appears directly in equations~\ref{betaSDE} and~\ref{alphaSDE} as singularities in $d\beta_0$ and $d\alpha_0$, preventing one from determining the single-increment values $\beta_{dt}$ and $\alpha_{dt}$ from these equations.  The single-increment values can be read off the SDEs~\ref{betaalphaSDE1} and~\ref{betaalphaSDE2}, however, and this is because these are effectively equations for the Harish-Chandra co\"ordinates.

We also need the leading-order single-increment behavior of the several center co\"ordinates:
\begin{align}
z_{dt}&=-s_{dt}+i\psi_{dt}=\smallfrac12\kappa|dw_0|^2\,,\\
-\ell_{dt}&=f_{dt}-s_{dt}=f_{dt}=2\kappa\,dt\,\frac{|\beta_{dt}+\alpha_{dt}|^2}{4}=\frac{1}{2\kappa\,dt}\frac{|\nu_{dt}+\mu_{dt}|^2}{4}=\frac{\kappa|dw_0|^2}{2\kappa\,dt}\,,\\
\phi_{dt}&=\psi_{dt}-\xi_{dt}=0\,.
\end{align}
Please appreciate that, as indicated, the leading-order contribution to $\ell_{dt}$ comes entirely from $f_{dt}$.

Putting all this into equation~\ref{Ddtxappendix}, one finds
\begin{align}
\begin{split}
D_{dt}(x)
&=\delta(\phi)\int\frac{d^2(\sqrt\kappa\,dw_0)}{\pi\kappa\,dt}e^{-\kappa|dw_0|^2/\kappa\,dt}\,
\delta\bigg(\ell+\frac{\kappa|dw_0|^2}{2\kappa\,dt}\bigg)\,\frac{1}{\sinh^2\!r}\delta(r-2\kappa\,dt)\\
&\qquad\qquad\qquad\times2\pi\delta^2\bigg(\beta+\alpha-\frac{\sqrt\kappa\,dw_0}{\kappa\,dt}\bigg)\,2\pi\delta^2(\beta-\alpha)\,.
\end{split}
\end{align}
Now writing
\begin{align}
\delta^2\bigg(\beta+\alpha-\frac{\sqrt\kappa\,dw_0}{\kappa\,dt}\bigg)=(\kappa\,dt)^2\delta^2\big(\sqrt\kappa\,dw_0-\kappa\,dt(\beta+\alpha)\big)\,,
\end{align}
we do the integral over the initial Wiener increment $dw_0$,
\begin{align}\label{deltax1Ddtx}
D_{dt}(x)=
\delta(\phi)\,\delta\Big(\ell+\smallfrac12\kappa\,dt|\beta+\alpha|^2\Big)\,
\frac{1}{\sinh^2\!r}\delta(r-2\kappa\,dt)\,
2\pi\frac{\kappa\,dt}{\pi}e^{-\kappa\,dt|\beta+\alpha|^2}\,2\pi\delta^2(\beta-\alpha)\,.
\end{align}
It is easy to see that this integrates to 1 over the Haar measure~\ref{d7x}.

It is understood that in defining $\delta(x,1)$, we take the limit $dt\rightarrow0$.  We might as well take these limits, thus writing
\begin{align}
\delta(x,1)&=
\delta(\phi)\,\delta(\ell)\,\frac{1}{\sinh^2\!r}\delta(r)\bigg(2\pi\lim_{dt\to0}\frac{\kappa\,dt}{\pi}e^{-\kappa\,dt|\beta+\alpha|^2}\bigg)\,2\pi\delta^2(\beta-\alpha)\\
&=\delta(\phi)\,\delta(\ell)\,\frac{1}{\sinh^2\!r}\delta(r)\bigg(\pi\lim_{dt\to0}\frac{4\kappa\,dt}{\pi}e^{-4\kappa\,dt|\alpha|^2}\bigg)\,\pi\delta^2(\beta-\alpha)\,.
\end{align}
This has singular behavior in the ruler $r$, but this singularity goes away when integrating against the Haar measure.  More important is the Gaussian in $|\beta+\alpha|^2$, which limits to an infinitely wide, normalized Gaussian as $dt\rightarrow0$.  As discussed in appendix~\ref{app:CoordTrans}, the identity 1 is represented by $\phi=\ell=r=0$ and $\beta=\alpha$, with $\beta+\alpha$ free to take on any complex value.  The infinitely wide Gaussian in $\beta+\alpha$ and the $\delta$-function $\delta(\beta-\alpha)$ express this in $\delta(x,1)$.

Now let's transform the expression~\ref{deltax1Ddtx} to Harish-Chandra co\"ordinates.  With $r=2\kappa\,dt$, the sum and difference phase-space variables are related by
\begin{align}
\beta+\alpha&=\frac{\nu+\mu}{2\kappa\,dt}\,,\\
\beta-\alpha&=\frac{\nu-\mu}{2}\,,
\end{align}
and the Harish-Chandra center variables are given by, when $\alpha=\beta$,
\begin{align}
s&=\ell+f=\ell+\smallfrac12\kappa\,dt|\beta+\alpha|^2\,,\\
\psi&=\phi+\xi=\phi\,.
\end{align}
Substituting into equation~\ref{deltax1Ddtx} gives
\begin{align}\label{DdtHC}
D_{dt}(x)=
\delta(\phi)\,\delta(s)\,\delta(r-2\kappa\,dt)\,
4\pi\frac{1}{\pi 4\kappa\,dt}e^{-|\nu+\mu|^2/4\kappa\,dt}\,4\pi\delta^2(\nu-\mu)\,.
\end{align}
The wide Gaussian in $\beta+\alpha$ becomes a narrow Gaussian in $\nu+\mu$, so when we take the limit $dt\rightarrow0$, we get
\begin{align}
\delta(x,1)=
\delta(\phi)\,\delta(s)\,\delta(r)\,4\pi\delta^2(\nu+\mu)\,4\pi\delta^2(\nu-\mu)
=\delta(\phi)\,\delta(s)\,\delta(r)\,2\pi\delta^2(\nu)\,2\pi\delta^2(\mu)\,,
\end{align}
in agreement with equation~\ref{deltax1HC} and consistent with the fact that in Harish-Chandra co\"ordinates, the identity is represented uniquely by $\phi=s=r=0$ and $\nu=\mu=0$.

It is now straightforward to integrate over the center variables in either co\"ordinate system to find the $\delta$-function on~$G/Z$,
\begin{align}\label{deltaZxZ1Cartan}
\delta(Zx,Z1)&=\frac{1}{\sinh^2\!r}\delta(r)\,
\bigg(\pi\lim_{dt\to0}\frac{4\kappa\,dt}{\pi}e^{-4\kappa\,dt|\alpha|^2}\bigg)\,\pi\delta^2(\beta-\alpha)\\
&=\delta(r)\,2\pi\delta^2(\nu)\,2\pi\delta^2(\mu)\,.\label{deltaZxZ1HC}
\end{align}

\vfill\pagebreak

\section{Riccati Equations for the Three Moments}
\label{momentRiccatis}

In this appendix we derive th Ricatti equations~\ref{nTRiccati}--\ref{qTRiccati} for the three second moments, $n_T$, $m_T$, and $q_T$, from their expressions in equations~\ref{nuTsquaredM}, \ref{muTsquaredM}, and~\ref{nuTmuTM}.  When we work in terms of the modified path measure~$\sD\mu_M\!\left[dw_{[0,T)}\right]$ of equation~\ref{DmuM}, the correlation of two outcome increments, at times $t$ and $s$ within the interval $[0,T)$, depends on the overall time $T$, as noted in equation~\ref{dwdwM}.  Hence it seems natural here to reserve $t$ to denote times within the interval $[0,T)$ and to increment the overall time from $T$ to $T+dT$.  As a consequence, unlike anywhere else in the paper, we denote the duration of the increments by $dT$ instead of $dt$.

It is quite convenient to introduce a bra-ket notation and to define vectors
\begin{align}
\ket{\plus_T}&\equiv\sqrt{dT}\sum_{k=0}^{N-1}e^{-2\kappa\,dT\,(N-k)}\ket k\,,\\
\ket{\minus_T}&\equiv\sqrt{dT}\sum_{k=0}^{N-1}e^{-2\kappa\,dT\,k}\ket k\,,
\end{align}
where the kets $\ket k$ are orthonormal.  The matrix elements of $M_T$ are
\begin{align}
(M_T)_{kl}=\bra k M_T\ket{l}=\delta_{kl}-\kappa\,dT\,e^{-2\kappa\,dT\,|k-l|}\,,\quad k,l=0,\ldots,N-1.
\end{align}
The utility of the bra-ket notation becomes apparent when we write the three moments as matrix elements of $M_T$,
\begin{align}
n_T&=\kappa{}\bra{\plus_T}M_T^{-1}\ket{\plus_T}\,,\label{nTbk}\\
m_T&=\kappa\bra{\minus_T}M_T^{-1}\ket{\minus_T}\,,\label{mTbk}\\
q_T&=\kappa\bra{\plus_T}M_T^{-1}\ket{\minus_T}=\kappa\bra{\minus_T}M_T^{-1}\ket{\plus_T}\,.\label{qTbk}
\end{align}
Replacing $M_T$ and $M_t^\inv$ with the unit matrix gives the moments with respect to the original Wiener measure; these moments are the inner products of the kets $\ket{\pm_T}$.

Now increment $M_T$ forward one step in time, from $T$ to $T+dT$.  What happens is really nothing,
\begin{align}
\bra{k}M_{T+dT}\ket{l}=\delta_{kl}-\kappa\,dT\,e^{-2\kappa\,dT\,|k-l|}\,,\quad k,l=0,\ldots,N,
\end{align}
except that an outside row and column, labeled by the index $N$, are added,
\begin{align}
\vphantom{\Big(}k,l=0,\ldots,N-1:\quad&\bra{k}M_{T+dT}\ket{l}=\bra{k}M_T\ket{l}=\delta_{kl}-\kappa\,dT\,e^{-2\kappa\,dT\,|k-l|}\,,\\
\vphantom{\Big(}k=0,\ldots,N-1:\quad&\bra{k}M_{T+dT}\ket{N}=-\kappa\,dT\,e^{-2\kappa\,dT\,(N-k)}=-\kappa\sqrt{dT}\,\iprod{k}{\plus_T}\,,\\
\vphantom{\Big(}l=0,\ldots,N-1:\quad&\bra{N}M_{T+dT}\ket{l}=-\kappa\,dT\,e^{-2\kappa\,dT\,(N-k)}=-\kappa\sqrt{dT}\,\iprod{\plus_T}{l}\,,\\
\vphantom{\Big(}&\bra{N}M_{T+dT}\ket{N}=1-\kappa\,dT\,.
\end{align}
The incremented matrix can then be written in block form relative to the outer row and column,
\begin{align}\label{MTplusdT}
M_{T+dT}=M_T-\kappa\sqrt{dT}\big(\oprod{\plus_T}{N}+\oprod{N}{\plus_T}\big)+(1-\kappa\,dT)\oprod{N}{N}\,.
\end{align}
We also need to increment the kets $\ket{\pm_T}$:
\begin{align}
\ket{\plus_{T+dT}}&=\sqrt{dT}\,\sum_{k=0}^{N}e^{-2\kappa\,dT\,(N+1-k)}\ket k=e^{-2\kappa\,dT}\big(\ket{\plus_T}+\sqrt{dT}\,\ket N\big)\,,\\
\ket{\minus_{T+dT}}&=\sqrt{dT}\,\sum_{k=0}^{N}e^{-2\kappa\,dT\,k}\ket k=\ket{\minus_T}+\sqrt{dT}\,e^{-2\kappa T}\ket N\,.
\end{align}

The harder task, incrementing $M_T^\inv$, is done using the Schur complement.  Generally, for any matrix written in block~form,
\begin{align}\label{Mblock}
M
=
\begin{pmatrix}
A&B\\C&D
\end{pmatrix}\,,
\end{align}
deriving a formal expression for the inverse begins by a block Gaussian elimination that writes the matrix as
\begin{align}
M=
\begin{pmatrix}
I&0\\CA^\inv&I
\end{pmatrix}
\begin{pmatrix}
A&0\\0&M/A
\end{pmatrix}
\begin{pmatrix}
I&A^\inv B\\0&I
\end{pmatrix}\,,
\end{align}
where
\begin{align}
M/A\equiv D-CA^\inv B
\end{align}
is called the Schur complement.  A first thing to notice is that
\begin{align}
\det M=\det A\,\det(M/A)\,,
\end{align}
and the second thing is that the inverse of $M$ is
\begin{align}\label{SchurMinv}
M^\inv=
\begin{pmatrix}
\vphantom{\Big(}A^\inv+A^\inv B(M/A)^\inv CA^\inv&\quad&-A^\inv B(M/A)^\inv\\
\vphantom{\Big(}-(M/A)^\inv CA^\inv &\quad&(M/A)^\inv
\end{pmatrix}\,.
\end{align}

In the case at hand, the block form~\ref{MTplusdT} of $M_{T+dT}$ relative to the outer row and column can be written in the matrix form~\ref{Mblock} by identifying
\begin{align}\label{eq:MTplusdT2}
A&=M_T\,,\\
B&=-\kappa\sqrt{dT}\,\oprod{\plus_T}{N}\,,\\
C&=-\kappa\sqrt{dT}\,\oprod{N}{\plus_T}\,,\\
D&=(1-\kappa\,dT)\oprod{N}{N}\,.
\end{align}
The Schur complement is the $1\times1$ matrix
\begin{align}
M_{T+dT}/M_T=\big(1-\kappa\,dT-\kappa^2\,dT\,\bra{\plus_T}M_T^{-1}\ket{\plus_T}\big)\oprod{N}{N}=\big[1-\kappa\,dT(1+n_T)\big]\oprod{N}{N}\,.
\end{align}
One has immediately how the determinant advances by one increment,
\begin{align}\label{detMT}
\det M_{T+dT}=\det M_T\big(1-\kappa\,dT(1+n_T)\big)\,,
\end{align}
and thus the ODE
\begin{align}\label{detMTODE}
\frac{1}{\kappa}\frac{d\ln\det M_T}{dT}=-(1+n_T)\,,
\end{align}
which agrees with the differential equation~\ref{dlnNt} for the normalization factor $\sN_T=1/\det M_T$.  Further, one can work out from equation~\ref{SchurMinv} how the inverse increments,
\begin{align}\label{MTplusdTinv}
\begin{split}
M_{T+dT}^\inv
&=M_T^\inv +\kappa^2\,dT\,M_T^\inv \oprod{\plus_T}{\plus_T}M_T^\inv\\
&\qquad+\kappa\sqrt{dT}\,\big(M_T^\inv\ket{\plus_T}\bra N+\ket N\bra{\plus_T}M_T^\inv \big)\\
&\qquad+\big[1+\kappa\,dT\big(1+\kappa\bra{\plus_T}M_T^\inv \ket{\plus_T}\big)\big]\oprod{N}{N}\,.
\end{split}
\end{align}
where we keep only the leading-order term in $dT$ in each block.

Everything is set now to determine how the three moments increment,
\begin{align}
n_{T+dT}
&=\kappa\bra{\plus_{T+dT}}M_{T+dT}^\inv\ket{\plus_{T+dT}}
=n_T+\kappa\,dT\,(1-n_T)^2\,,\\
m_{T+dT}
&=\kappa\bra{\minus_{T+dT}}M_{T+dT}^\inv\ket{\minus_{T+dT}}
=m_T+\kappa\,dT\,\big(q_T+e^{-2\kappa T}\big)^2\,,\\
q_{T+dT}
&=\kappa\bra{\minus_{T+dT}}M_{T+dT}^\inv\ket{\plus_{T+dT}}
=q_T+\kappa\,dT\,\big({-}q_T(1-n_T)+e^{-2\kappa T}(1+n_T)\big)\,,
\end{align}
and these lead immediately to the three (coupled) Riccati equations,
\begin{align}
\frac{1}{\kappa}\frac{dn_T}{dT}&=(1-n_T)^2\,,\label{nTRiccatiapp}\\
\frac{1}{\kappa}\frac{dm_T}{dT}&=\big(q_T+e^{-2\kappa T}\big)^2\,,\label{mTRiccatiapp}\\
\frac{1}{\kappa}\frac{dq_T}{dT}&=-q_T(1-n_T)+e^{-2\kappa T}(1+n_T)\,,\label{qTRiccatiapp}
\end{align}
which are repeated in the main text as equations~\ref{nTRiccati}--\ref{qTRiccati}.

\pagebreak

\end{document}